\documentclass[conference]{IEEEtran}
\usepackage{amsmath}
	\usepackage{cancel}
\usepackage{mathtools}
\usepackage{graphicx} % can better scale and rotate graphics than graphic package
\usepackage{url}
\usepackage{stmaryrd} % for symbol \llceil
\usepackage{upgreek} % upright Greek letters; together with package "bm" -> upright bold letters
\usepackage{bbding} % for the \CheckmarkBold symbol

\usepackage{listings} % to get sql etc nicely formatted (source code listings)
\usepackage{tabularx} % table environment that allows to span a whole line
\usepackage{booktabs} % \toprule
\usepackage{multirow}
\usepackage{hhline} % cannot be overwritten by cellcolor
\usepackage{diagbox}
\usepackage{pgfplotstable} % powerful numerical table macros

\usepackage{xcolor,colortbl}    % !!! for some reason, overleaf needs xcolor separately, although that should already have been called by colortbl (10/2019)

\usepackage{tikz} % drawing figures
\usetikzlibrary{matrix}
\usetikzlibrary{calc}
\usetikzlibrary{math}
\usetikzlibrary{arrows.meta,positioning}
\usetikzlibrary{intersections, pgfplots.fillbetween}

\usepackage{easybmat} % \begin{BMAT} alternative syntax for math tabulars and also includes the command addpath which allows to draw a path between matrix elements.

\usepackage{adjustbox}
\def\eox{\unskip\kern 10pt{\unitlength1pt\linethickness{.4pt}$\diamondsuit${}}} % "\eox" command for end of example

\usepackage{rotating}		% for rotated text in tables, \begin{turn}{75}Accuracy\end{turn}

\usepackage{xspace} % for macros

\usepackage{subcaption} % recommended new package for subfigures (package subfig deprecated)
\newcommand{\hide}[1]{}

\usepackage{hyperref}
\usepackage[capitalise,nameinlink]{cleveref} %nameinlink makes whole word colored

\crefname{section}{Sec.}{Secs.}
\crefname{example}{Ex.}{Exes.}

\usepackage{aliascnt}  	
\newtheorem{theorem}{Theorem} % ...[section]

\newaliascnt{corollary}{theorem}

\aliascntresetthe{corollary}

\newaliascnt{example}{theorem}
\newtheorem{example}[example]{Example}
\aliascntresetthe{example}

\newaliascnt{definition}{theorem}
\newtheorem{definition}[definition]{Definition}
\aliascntresetthe{definition}

\newaliascnt{proposition}{theorem}

\aliascntresetthe{proposition}

\newaliascnt{lemma}{theorem}
\newtheorem{lemma}[lemma]{Lemma}
\aliascntresetthe{lemma}

\newaliascnt{conjecture}{theorem}

\aliascntresetthe{conjecture}

\newtheorem{questionW}{Question}
\newtheorem{resultW}{Result}

{\end{itshape}
}
\usepackage{tcolorbox}		% also allows to have shaded backgrounds around environments
\tcbuselibrary{breakable,skins}		% allow to break boxes, allow to use "enhanced jigsaw"

\tcbset{examplestyle/.style={
		enhanced jigsaw,	% if box is broken, don't show border at continuation
		colback=blue!08,	% !20	gray!10
		colframe=blue!08,	% !40
		arc=2mm,
		boxrule=0pt,		% 0pt
		left=1mm,
		right=1mm,
		left skip= 0mm,  % breaks ACM template unless specified for specific style (4/2019)
		right skip= 0mm, % breaks ACM template unless specified for specific style (4/2019)
		top=1mm,		% 0mm, -1mm
		bottom=1mm,		% problem in package with eq at end of env. just add: \tcbset{bottom=2mm} \tcbset{bottom=0mm}
		breakable,		% allows page break
		parbox = false,		% restores normal text behavior. ELSE: paragraphs are formatted slightly different as the main text. DOWNSIDE: unwanted side effects
		before={\par\pagebreak[0]\vspace{1mm}\parindent=0pt},		
		after={\par\pagebreak[0]\vspace{1mm}\parindent=0pt},				
		bottomrule = 0mm,
		boxsep = 0mm,					% common padding of hlengthi between the text content and the frame of the box, added to left, right, top, ...
		topsep at break=0pt,			% Additional vertical space at the top of middle and last parts in a break sequence
		bottomsep at break=0pt,			% Additional vertical space at the first and middle parts in a break sequence		
		pad at break=0mm,
		pad before break=1mm,		
		pad after break=1mm,		
		bottomrule at break=0mm,
		toprule at break=0mm,		
		}}

\usepackage{footnote}

\tcolorboxenvironment{example}{examplestyle}
\makeatletter
\DeclareRobustCommand*\uell{\mathpalette\@uell\relax}
\newcommand*\@uell[2]{
  \setbox0=\hbox{$#1\ell$}
  \setbox1=\hbox{\rotatebox{10}{$#1\ell$}}
  \dimen0=\wd0 \advance\dimen0 by -\wd1 \divide\dimen0 by 2
  \mathord{\lower 0.1ex \hbox{\kern\dimen0\unhbox1\kern\dimen0}}
}
\makeatother

\usepackage{marginnote}

\renewcommand{\epsilon}{\varepsilon} % nicer epsilon symbol
\usepackage{scalerel}
\usepackage{tikz}
\usetikzlibrary{svg.path}

\definecolor{orcidlogocol}{HTML}{A6CE39}
\tikzset{
  orcidlogo/.pic={
    \fill[orcidlogocol] svg{M256,128c0,70.7-57.3,128-128,128C57.3,256,0,198.7,0,128C0,57.3,57.3,0,128,0C198.7,0,256,57.3,256,128z};
    \fill[white] svg{M86.3,186.2H70.9V79.1h15.4v48.4V186.2z}
                 svg{M108.9,79.1h41.6c39.6,0,57,28.3,57,53.6c0,27.5-21.5,53.6-56.8,53.6h-41.8V79.1z M124.3,172.4h24.5c34.9,0,42.9-26.5,42.9-39.7c0-21.5-13.7-39.7-43.7-39.7h-23.7V172.4z}
                 svg{M88.7,56.8c0,5.5-4.5,10.1-10.1,10.1c-5.6,0-10.1-4.6-10.1-10.1c0-5.6,4.5-10.1,10.1-10.1C84.2,46.7,88.7,51.3,88.7,56.8z};
  }
}

\RequirePackage{etoolbox}
\DeclareRobustCommand\orcidicon[1]{\href{https://orcid.org/#1}{\mbox{\scalerel*{
\begin{tikzpicture}[yscale=-1,transform shape]
\pic{orcidlogo};
\end{tikzpicture}
}{|}}}}

\definecolor{ForestGreen}{rgb}{0.0, 0.66, 0.47}

\newcommand{\technicalReport}[1]{{{\color{red}{--Commented-out section to-be-moved to a Technical Report--}}}}

\newcommand{\bigT}{\mathcal{T}}

\def\ojoin{\setbox0=\hbox{$\bowtie$}%
  \rule[-.02ex]{.25em}{.4pt}\llap{\rule[\ht0]{.25em}{.4pt}}}
\def\leftouterjoin{\mathbin{\ojoin\mkern-5.8mu\bowtie}}

\def\fullouterjoin{\mathbin{\ojoin\mkern-5.8mu\bowtie\mkern-5.8mu\ojoin}}
\newcommand{\cc}{olive}
\newcommand{\algocomment}[1]{\textcolor{\cc}{{//#1}}}
\newcommand{\revision}[1]{{\color{black}{{#1}}}}

\newcommand{\name}{\textsf{Gen-T}\xspace}
\newcommand{\aliteps}{\textsf{ALITE-PS}\xspace}
\newcommand{\alite}{\textsf{ALITE}\xspace}
\newcommand{\aPip}{\textsf{Auto-Pipeline*}\xspace}
\newcommand{\ver}{\textsf{Ver}\xspace}
\newcommand{\tpch}{\textsf{TP-TR}\xspace}
\newcommand{\tpchSmallBench}{\textsf{TP-TR Small}\xspace}
\newcommand{\tpchMedBench}{\textsf{TP-TR Med}\xspace}
\newcommand{\tpchLargeBench}{\textsf{TP-TR Large}\xspace}
\newcommand{\tdBench}{\textsf{T2D Gold}\xspace}
\newcommand{\santosLargeBench}{\textsf{SANTOS Large}\xspace}
\newcommand{\wdc}{\textsf{WDC Sample}\xspace}

\newcommand{\dkl}{D$_{\text{KL}}$\xspace}

\newcommand{\ise}{EIS\xspace} % used to be Value Similarity
\usepackage{cite}
\usepackage{amsmath,amssymb,amsfonts}
\usepackage{algorithmic}
\usepackage{graphicx}
\usepackage{textcomp}
\usepackage{xcolor}
\def\BibTeX{{\rm B\kern-.05em{\sc i\kern-.025em b}\kern-.08em
    T\kern-.1667em\lower.7ex\hbox{E}\kern-.125emX}}

\usepackage{url}
\usepackage{verbatim}
\usepackage{multirow}
\usepackage{balance}
\usepackage{lscape}
\usepackage{float}
\usepackage{paralist}
\usepackage{booktabs}
\usepackage{xspace}
\usepackage{listings}
\usepackage{lipsum}
\usepackage{subcaption}
\usepackage{dsfont}

\usepackage[ruled,noend,linesnumbered]{algorithm2e} % boxruled, linesnumbered,ruled

\usepackage{setspace} % for setting line spacing

\DontPrintSemicolon     % does not print ";" at end of each line
\SetNlSty{}{}{}                % style of line numbers "\Setnlsty{small}{}{:}"
\SetAlgoInsideSkip{smallskip}   % distance between caption rule and beginning of algorithm
\SetAlFnt{\small}			% size of actual algorithm text\scriptsize\sf\rm
\SetAlCapFnt{\small}		% size of name of algorithm
\SetAlCapNameFnt{\small}

\begin{document}

\title{Gen-T: Table Reclamation in Data Lakes}

\author{\IEEEauthorblockN{Grace Fan}
\IEEEauthorblockA{\textit{Northeastern University} \\
Boston, United States \\
fan.gr@northeastern.edu}
\and
\IEEEauthorblockN{Roee Shraga}
\IEEEauthorblockA{\textit{Worcester Polytechnic Institute} \\
Worcester, United States \\
rshraga@wpi.edu}
\and
\IEEEauthorblockN{Renée J. Miller}
\IEEEauthorblockA{\textit{Northeastern University} \\
Boston, United States \\
miller@northeastern.edu}
}

\maketitle
\thispagestyle{plain}
\pagestyle{plain}

\begin{abstract}
We introduce the problem of Table Reclamation.  Given a Source Table and a large table repository, reclamation finds a set of tables that, when integrated, reproduce the source table as closely as possible.  Unlike query discovery problems like Query-by-Example or by-Target, Table Reclamation focuses on reclaiming the data in the Source Table as fully as possible using real tables that may be incomplete or inconsistent.
To do this, we define a new measure of table similarity, called %rjm don't use acronyms before they are defined
%\ise, 
error-aware instance similarity, to measure how close a reclaimed table is to a Source Table, a measure grounded in instance similarity used in data exchange.
Our search covers not only \textsc{Select-Project-Join} queries, but integration queries with unions, outerjoins, and the unary operators subsumption and complementation  that have been shown to be important in data integration and %data 
fusion.  Using reclamation, a data scientist can understand if any tables in a repository can be used to exactly reclaim a tuple in the Source.  If not, one can understand if this is due to differences in values or to incompleteness in the data.
%errors or inconsistencies in null values.
%We present a solution for Table Reclamation named \name. 
Our solution, \name, performs table discovery to retrieve a set of candidate tables from the table repository, filters these down to a set of {\em originating tables}, then integrates these tables to reclaim the Source as closely as possible.  We show that our solution, while approximate, is accurate, efficient and scalable in the size of the table repository with experiments on real data lakes containing up to 15K tables, where the average number of tuples varies from small (web tables) to extremely large (open data tables) up to 1M tuples.  
\end{abstract}
\setcounter{page}{1}

% \begin{IEEEkeywords}
% component, formatting, style, styling, insert
% \end{IEEEkeywords}

\section{Introduction}\label{sec:intro}
%As more tables in data lakes become openly available, users have easier access to more government, academic, and enterprise datasets. 
%However, there may be concerns around the origins of the tables that may need to be addressed before they are used for further tasks~\cite{DBLP:journals/cacm/AbadiAABBBBCCDD22}. 
%Considering the rise of generative AI models, such as ChatGPT~\cite{chatgpt}, there is an increasing need to verify the AI outputs, for example, using values from data lake tables~\cite{DBLP:journals/corr/abs-2307-02796}.
We introduce the problem of {\em Table Reclamation} where we are given a source table and seek to find a set of tables from a data lake (a large table repository) that, when integrated, reproduce the source table as  closely as possible.  We begin with an example showing how table reclamation can be used.

\begin{figure}[h]
	\centering
	\includegraphics[width=0.48\textwidth]{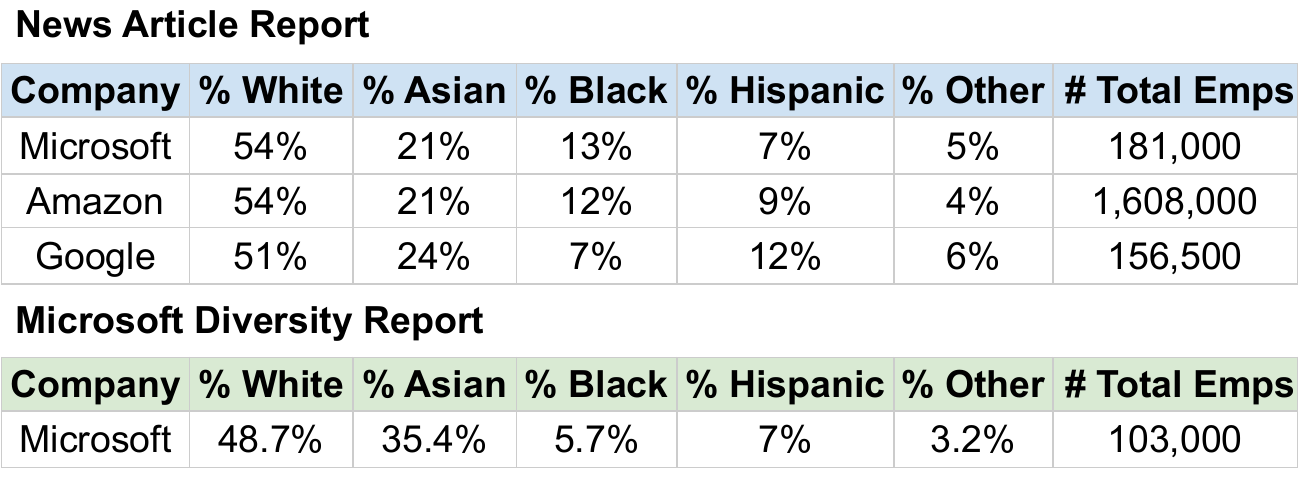}
	\caption{A news article reports the top blue table.  A user has access to Microsoft's diversity report, which seems to contradict the article (bottom green table).
 }
	\label{fig:gpt_motivating_ex}
\end{figure}

%\todo{Change example~\ref{ex:gpt_company_diversity} and Figure~\ref{fig:gpt_motivating_ex} to journalism}
\begin{example}\label{ex:gpt_company_diversity}
% Suppose a user prompts ChatGPT\footnote{To create this and the next example we used the free research preview of GPT 3.5.} to ``Show demographics of employees in Top US tech companies in 2021'', for which ChatGPT returns two tables for gender and racial demographics 
Suppose a user is reading a news article that reports the demographics of employees in Top US tech companies in 2021 (top blue table in Figure~\ref{fig:gpt_motivating_ex}).  The user has access to 2021 Microsoft's Diversity Report~\cite{ms_diversity_report} which seems to contradict the numbers in the news article (bottom green table in Figure~\ref{fig:gpt_motivating_ex}).
Using table reclamation and her data lake (which may include wikitables, NYTimes Data and other public and private datasets), the user can ask if there is a set of tables that, when integrated, recreates the data in the news article (blue table).  
%However, after consulting Microsoft's Diversity Report from 2021~\cite{ms_diversity_report}, the user finds that ChatGPT's returned statistics for Microsoft (first row of top two tables) that contradict those reported by Microsoft itself (bottom green tables). %rjm - they can be hallucinated.
%Thus, data values corresponding to ``Microsoft'' in ChatGPT's output tables must be derived from a source different than the company's published report.
Table reclamation is able  to reproduce this table using a number of tables including {\tt World\_MS\_Ethnicity} and {\tt World\_MS\_Employees} (which, after being joined and a selection on 2021 is applied, produce the first tuple).   We call the tables used in reclamation, originating tables.   Other tables can be unioned with the Microsoft tables to reclaim the other tuples.  
From this (the originating tables including their meta-data and data), a user can understand that while her table (green) is reporting US statistics numbers, the article is reporting international numbers.
% From this, the user can understand that while her table (green) is reporting international numbers, the article is reporting only US statistics.  
\end{example}

%\todo{Remove Figure~\ref{fig:biased_gpt_motivating_ex} and Example~\ref{ex:gpt_biased_output}}
%\begin{figure}[h]
%	\centering
%	\includegraphics[width=0.48\textwidth]{fig/gent_motivation_example_gpt_2.pdf}
%	\caption{
%            Two prompts provide ChatGPT with two applicants' information that differ in names and cities, and ask for a list of starting salaries at companies from which they can receive offers. For ``Malik Brown'' in Table 2, his starting salaries are \$20-35K less than those of ``James Smith'' in Table 1. 
% }
%	\label{fig:biased_gpt_motivating_ex}
%\end{figure}

\begin{figure*}[!t]
	\centering
	\includegraphics[width=0.8\textwidth]{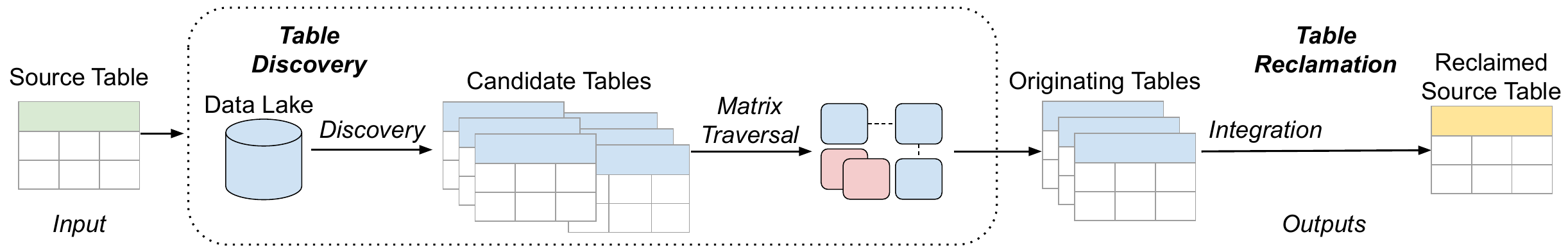}
 \caption{\name Architecture. Given a Source Table, \name finds a set of originating tables (Table Discovery), produces a reclaimed Source Table from the (Table Reclamation), and returns the originating tables and the reclaimed Source Table.}
 \label{fig:pipeline}
\end{figure*}

%rjm:  move to related work?
%A similar line of work is Table Discovery, in which existing systems find relevant tables to the user's table~\cite{DBLP:journals/pvldb/Miller18,DBLP:conf/sigmod/SarmaFGHLWXY12,DBLP:conf/sigmod/ZhangI20,DBLP:journals/debu/MillerNZCPA18}. Specifically, joinable and unionable table search find tables that can join or union with the given table, respectively~\cite{DBLP:conf/sigmod/ZhuDNM19,DBLP:journals/pvldb/ZhuNPM16,DBLP:journals/pvldb/NargesianZPM18}. However, retrieved tables in these searches may primarily share semantics with the columns in the given table~\cite{starmie22, santos23}, or only have overlapping sets of values in individual columns. Instead, we aim to find originating tables whose values across columns for each tuple are consistent with those in the given table.
% to its origins in order to perform verifications. 
Unlike the well-known problem of Data Provenance~\cite{DBLP:journals/ftdb/CheneyCT09,DBLP:reference/db/CheneyT18}, we do not have prior knowledge of the query or tables that were originally used to create a Source Table. Instead, we focus on recovering possible  tables that, when integrated, confirm the  data values and facts in a Source Table.
Table reclamation is related to the common Query-By-Example (QBE) or Query-By-Target (QBT) that discover a query over input tables that produces an instance-equivalent table to the given example output table~\cite{DBLP:journals/tbd/KoehlerABCMKFKL21,DBLP:conf/pldi/WangCB17,DBLP:conf/sigmod/ShenCCDN14,DBLP:conf/icde/DeutchG16,DBLP:journals/pacmpl/BavishiLFSS19,DBLP:journals/pvldb/YangHC21}. 
In order to generalize to a data lake setting, we do not assume we know a complete and correct set of input tables.  Rather, we use an additional step of finding candidate tables within or across data lakes that may contribute to the Source Table (i.e., that may be originating tables). 
Also, existing QBE/QBT systems focus primarily on discovering \textsc{(Select)-Project-Join} queries over (largely complete) relational tables~\cite{DBLP:conf/sigmod/ZhangEPS13, DBLP:conf/sigmod/KalashnikovLS18,DBLP:journals/pvldb/OrvalhoTVMM20,DBLP:conf/sigmod/TranCP09,DBLP:conf/vldb/BonifatiCLS14}, with some using both the data values and the schema of the tables. Due to the noise and heterogeneity of data lake tables, these queries may not be sufficient to fully integrate data lake tables to produce a given Source table. So, we aim to recover \textsc{Select-Project-Join-Union} queries using only the data values, since the metadata of data lake tables may be missing or inconsistent~\cite{DBLP:journals/pvldb/NargesianZPM18,DBLP:journals/pvldb/AdelfioS13,DBLP:conf/sigmod/FaridRIHC16,DBLP:journals/pvldb/NargesianZMPA19}.  We also consider operations that have proven to be important in data integration and data fusion of incomplete data, namely subsumption and complementation~\cite{DBLP:journals/csur/BleiholderN08}. 
%Both operations have been used to integrate incomplete data.  In addition, Rajaraman and Ullman describe the role of subsumption in computing outerjoins over complete data in a way that "preserves all possible connections among facts"~\cite{DBLP:conf/pods/RajaramanU96}.
%rjm repeated at end of next paragraph
%Also note that our goal is table reclamation not query discovery {\em per se}, so our focus is on reclaiming the Source Table as fully as possible.

%Reclamation focuses on a Source Table and does not assume we know the originating tables.  Hence, it can be used to verify tuples. 
Our goal is to reclaim the source table completely, but this may not always be possible. 

\begin{example}\label{ex:partial}
Continuing our example, it is possible that the best reclamation we can find has null values for the percentages of  Hispanic employees for Google and a different number of Asian employees  (20\% instead of 24\%).  These differences indicate that the source data about Google was not completely found within the data lake.
The user can analyze the originating tables returned by our approach to understand these differences.
As an example, it may be that the originating tables for the Google data in the repository are European in origin and do not report values for all categories like Hispanic employees as this is a protected category under US, but not European law.  
\end{example}

If certain tuples cannot be reclaimed, a data scientist would know these are not derivable from her 
data lake. 
%Reclamation can also be used to verify values.  For example, if the reclaimed table contains nulls where the Source Table has a value, the data scientist can look to see if value imputation might account for the values in her table and can examine whether  value imputation is appropriate for her task.  Similarly, 
If the reclaimed table contains different values from the source, a scientist can investigate whether the source values %may be
are wrong or %alternatively 
if they %may be 
are valid corrections to errors in the originating tables.  Unlike traditional QBE and QBT approaches, our focus is on the data (rather than the query) and on understanding what data in a Source Table (and only data in a Source Table) can be reclaimed. We make the following contributions.

% \noindent {\bf Contributions}
%  We present a data-driven technique for Table Reclamation in data lakes, named \name, in which we aim to discover a set of originating data lake tables whose integration can reclaim (or reproduce) a Source Table given by the user.

\noindent$\bullet$ We define the novel problem of {\em Table Reclamation} -- finding a set of originating tables that, when integrated, can 
reproduce a Source Table as closely as possible.

\noindent$\bullet$ 
%We present a set of properties that a metric for determining how close a reclaimed table is to a Source Table should possess.  And based on this we evaluate our methods using tuple-based precision and recall which consider only how many tuples are exactly reclaimed,
To evaluate how close a reclaimed table is to a Source Table, we define a new 
%rjm don't use acronym before defining it\ise 
error-aware instance similarity (\ise) score that is a principled extension of instance similarity used in  data exchange~\cite{DBLP:journals/vldb/AlexeHPT12}), and show how it can be computed efficiently. %in our setting.
%a new metric that uses conditional KL-divergence~\cite{DBLP:books/daglib/0016881}.

\noindent$\bullet$ We present an approximate Table Reclamation solution 
named \name 
%(Figure~\ref{fig:pipeline}) 
that performs table discovery to retrieve a set of candidate tables, and filters out poor candidates using a novel table representation that simulates table integration without performing expensive integration operations.  The remaining originating tables are integrated to produce a table whose values are as close as possible to the Source Table.

\noindent$\bullet$
    % \renee{this bullet doesn't make sense - we haven't described base lines and what is significance of 13?}\grace{resolved}
    % We conduct extensive effectiveness experiments, showing that \name is able to perfectly reclaim up to 13 Source Tables on benchmarks for which the baselines only perfectly reclaim at most 1 Source Table.
    We conduct extensive %effectiveness 
    experiments on real and synthetic data lakes,
    showing that \name outperforms all baseline methods.
    %when reclaiming Source Table.  
    \name reclaims 5X more values from Source Tables than the best-performing baseline.  We perform an ablation study on the sensitivity of \name to erroneous data (data that cannot be reclaimed) and to incomplete data.
    
\noindent$\bullet$ 
%\renee{I think intro need just a bit more intuition/motivation on search process for this to make sense} 
    % We show the efficiency of our solution as the data lake size scales up to 11K tables. Our method that uses table representations to narrow down the space of candidate tables shows to be 5X faster in runtime than the state-of-the-art table integration algorithm.
    We show that our solution is efficient and scalable to the size of the data lake with experiments on real data lakes containing up to 15K tables, where the average table size (number of tuples) varies from small (web tables) to extremely large (open data tables) with on average over 1M tuples.  
    In addition, our solution is scalable to large source tables, with experiments on source tables containing up to 22 columns and 1K rows.
    
% \noindent$\bullet$ We show that \name is robust when generalizing to different real-world application setting.
% \rjm{make this one more specific}\grace{Since the previous bullet includes web tables, we can remove this bullet}
% \section{\name Architecture and Preliminaries}\label{sec:overview}
\section{Overview}\label{sec:overview}
% We first provide %give 
% an overview of \name and its architecture. We then discuss how we evaluate reclaimed tables.
%a possible reclaimed Source Table for the problem of Source Table Reclamation.

% \subsection{Overview of \name}\label{subsec:architecture}
A data scientist provides a \textit{Source Table} that she would like to reclaim by understanding if it can be produced by integrating any combination of tables within a data lake.
%rjm Given a table by the user, denoted as a \textit{Source Table}, that has been generated from some subset of data lake tables, our goal is reverse engineer its creation. %Specifically, we aim to find a set of tables (termed \emph{originating tables}) such that, if integrated, reclaim the Source Table. W
Specifically, we aim to 
%rjm pinpoint the 
determine a set of tables from which the Source Table's values may originate (termed \textit{originating tables}), and use them to reclaim (regenerate) the Source Table. % from the originating table. 
Given our data lake setting where tables can be changed autonomously, we formulate the problem as an approximate search of finding a set of tables that can best be used to reclaim the Source, {\em as closely as possible}.

% \roee{I would also adopt a bit here and make blue:} 
% While existing Query-By-Example methods~\cite{Gong2023Ver,DBLP:journals/corr/abs-1911-11876} consider semantic similarity between instances (e.g., home addresses vs. work addresses), they rely on a human-in-the-loop to disambiguate the instances. In our current problem setting, \name aims to reproduce all values from the Source Table, as is. To disambiguate instances, we currently only consider syntactic similarity between instances in the data lake and those in the Source Table. We believe that if the reproduced Source Table contains the same values as those in the Source Table, then it must share the same semantics as the Source Table.

\revision{Unlike many existing \textit{Query-by-Example}~\cite{DBLP:conf/pldi/WangCB17,DBLP:conf/sigmod/ZhangEPS13, DBLP:conf/sigmod/KalashnikovLS18,DBLP:journals/pvldb/OrvalhoTVMM20,DBLP:conf/sigmod/TranCP09,DBLP:conf/vldb/BonifatiCLS14,DBLP:conf/sigmod/ShenCCDN14,DBLP:conf/icde/DeutchG16,DBLP:journals/pacmpl/BavishiLFSS19} or \textit{by-target}~\cite{DBLP:journals/pvldb/YangHC21} approaches, we do not assume 
%rjm to have 
that we know 
the exact set of input tables whose values first formed the Source Table or even {\bf if} 
%rjm have prior knowledge of whether 
the Source Table can be reclaimed. 
% \roee{Here is a good place to emphasize our assumptions:} 
In addition, while we do not assume tables in a data lake to have keys or any foreign key relationships,
we assume the Source Table to have a (possibly multi-attribute) key, which can be found using existing mining techniques~\cite{DBLP:journals/jiis/JiangN20,DBLP:conf/www/BornemannBKNS20}. This is a restriction, but it is made to make the instance comparison (which is done often in the algorithm) efficient.  Without the source table having a key, instance similarity requires homomorphism checks which is NP-hard~\cite{DBLP:journals/vldb/AlexeHPT12}.  
In general, we do not assume that metadata is available for any tables (column names are included in examples only for clarity).
}

To solve the problem of table reclamation, we 
%rj require 
use 
a two-step solution. 
First, we 
% \roee{although I like the term "retrieve", let's be consistent and use "discover"} \grace{resolved}
discover %a subset of 
tables from the data lake that share values with the Source Table and therefore may have created portions of it
%the Source
, we call these {\em candidate tables}. 
Then, we search for ways of combining subsets of these tables to regenerate the Source Table. 
\begin{figure}[!t]
	\centering
	\includegraphics[width=0.48\textwidth]{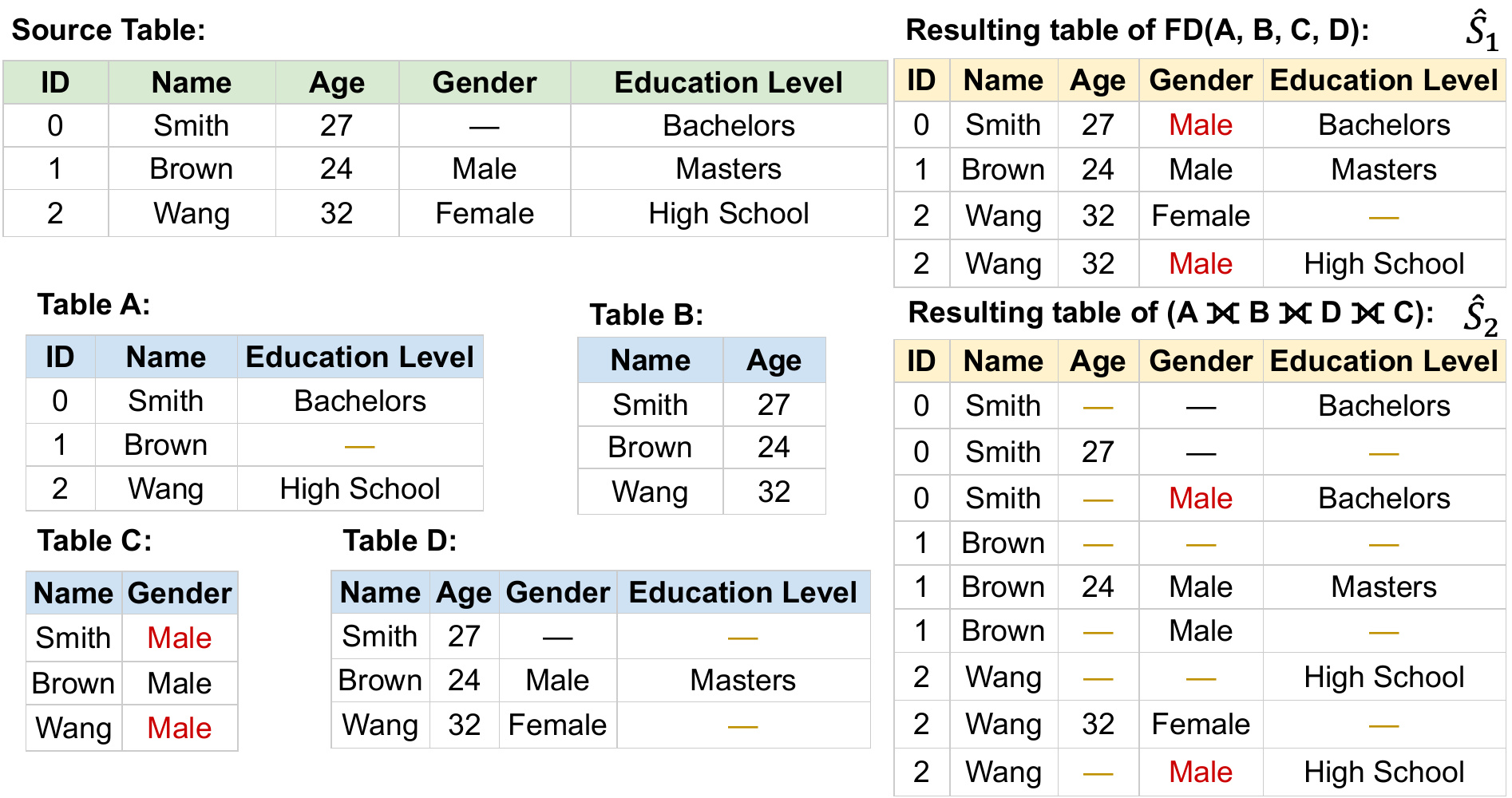}
	\caption{Source Table (in 
 green) contains applicants' information, such as ID, Name, Age, Gender, and Education Level. Tables A, B, C, D (in 
 blue) are possible tables from which the Source Table's instances originated. 
    % All common columns with the Source Table are in blue. 
    Missing values and inconsistent values w.r.t. Source Table are depicted in yellow (`---') and red, respectively. Tables on the right (in 
    yellow) are possible integrations of % the data lake 
    tables 
    % (with only shared columns with the Source Table) 
    resulting from %state-of-the-art 
    integration methods using Full Disjunction (FD) and outer join ($\fullouterjoin$). 
 }
	\label{fig:raw_integ}
\end{figure}
Figure~\ref{fig:pipeline} shows the  pipeline of \name.  The input is
%rjm the user's table as our 
a
Source Table and the output a % (perhaps partially) 
Reclaimed Source Table and its originating tables. 
% First, i
In the Table Discovery phase (Section~\ref{sec:discovery}), \name discovers 
%rjm this is an example of where "the vs a" makes a big difference
%the 
a
set of candidate tables whose values may have contributed to the  creation of the Source Table. Then, we apply our novel solution of representing tables as matrices in order to simulate table integration via matrix traversal (Section~\ref{subsec:matrix_traversal},~\ref{subsec:ternary-matrix}). The goal of this step is to refine the set of candidate tables to a set of originating tables, and essentially filter out %any 
%rjm misleading or noisy 
%tables from the set of 
candidate tables that are not needed before performing table integration.
% \name %thus 
% uses an approximate algorithm to search for and integrate a set of originating tables. 
 To efficiently retrieve a set of candidate tables, \name uses an existing, data-driven table discovery method that has no guarantees for this problem setting. \name then prunes the candidate tables to a set of originating tables by computing each candidate table's similarity with a given Source Table and simulating table integration.
 %Given this set of candidate tables, \name then prunes it to a set of originating tables by approximating each candidate table's similarity with a given Source Table and simulating table integration. 
% \begin{figure}[!t]
% 	\centering
% 	\includegraphics[width=0.48\textwidth]{fig/motivating_integration.pdf}
% 	\caption{Source Table about applicants' data, and the possible integrations of the data lake tables resulting from state-of-the-art integration method using Full Disjunction (FD), and an outer join $\fullouterjoin$ ordering. 
%  }
% 	\label{fig:raw_integ}
% \end{figure}
% \roee{ let's say why we provide this example -- ``We now return to our running example to illustrate the necessity of pinpointing the originating tables''.}
Once Matrix Traversal pinpoints 
%rjm the set of Originating tables,
a set of originating tables,
%rjm it's excludes not discludes :-)
%rjm that disclude any misleading tables, 
we integrate these originating tables in the Table Reclamation phase (Section~\ref{subsec:integration}) and produce a reclaimed Source Table.

\begin{example}\label{ex:running_example}
Suppose a user has the top left, green table in Figure~\ref{fig:raw_integ} as a Source Table. %As this table contains instances for applicants, with sensitive values for age, gender, and education level, it is crucial to verify them. 
To reclaim this table, we
%we trace its values to 
use table discovery to find a subset of tables in the data lake with overlapping values -- in this example, tables A, B, C, and D. However, Table C contains contradicting non-null values in the ``Gender'' column compared to values in the same tuples of the ``Gender'' column in the Source Table. 
% Thus, to correctly verify every cell value in the Source Table, we will look for other tables in a discovered set of candidate tables that may more faithfully reclaim these tuples.
    We demonstrate the consequence of directly integrating all these tables, including Table C.
    % , by first projecting out columns that do not appear in the Source Table. 
    The top right, yellow table is the integration result from the state-of-the-art full disjunction (FD) method~\cite{DBLP:conf/sigmod/Galindo-Legaria94,Khatiwada2022IntegratingDL} and the bottom right table shows the result using one possible outerjoin order that may be learned by Auto-Pipeline (a by-target approach)~\cite{DBLP:journals/pvldb/YangHC21}.
The resulting tables
 contain different values in the Gender column (in red) with respect to the 
corresponding value in the Source Table.
These values originate from Table C. When possible, we need to refine the set of candidate tables to filter out tables like Table C that  produce integrated tables with erroneous values 
that do not make the Source Table.  Note in this example, the integration of Tables A, B and D alone produces a better reclamation than using all four candidates.
\end{example} 

% The remainder of the paper is outlined as follows: we present related work in Section~\ref{sec:related}. 
% We now describe \emph{Value Similarity Score}, allowing us to compare reclaimed tables with source tables and to define the problem of \emph{Source Table Reclamation}. Then, in Section~\ref{sec:related} we present related work. Sections~\ref{sec:discovery}-\ref{sec:reclamation} describe our proposed solution \name in detail and the experiments in Section~\ref{sec:experiments} show its effectiveness, scalability, and generalizability.

% Going into the solution pipeline of \name, we first discuss the Table Discovery phase in Section~\ref{sec:discovery}, specifically the Matrix Traversal solution that finds a set of originating table. Next, we integrate these tables in the Table Reclamation phase in Section~\ref{sec:reclamation}. Finally, the experiments in Section~\ref{sec:experiments} show the effectiveness, scalability, and generalizability of \name.

% \grace{Should we remove this outline? Awkward placement here...}
%rjm:  I've shortened
%The remainder of the paper is outlined as follows: w
We present related work next in Section~\ref{sec:related}. 
%Before discussing the solution pipeline of \name, w
We define the problem of \emph{Table Reclamation} and preliminaries for our solution in Section~\ref{sec:preliminaries}. Then, we discuss the two steps in \name -- Table Discovery phase (Section~\ref{sec:discovery}) 
%that includes the Matrix Traversal solution that finds a set of originating table, 
and the Table Reclamation phase (Section~\ref{subsec:integration}).
%in which we integrate the tables.
% Going into the solution pipeline of \name, we first discuss the Table Discovery phase in Section~\ref{sec:discovery}, specifically the Matrix Traversal solution that finds a set of originating table. Next, we integrate these tables in the Table Reclamation phase in Section~\ref{sec:reclamation}. 
Finally, the experiments in Section~\ref{sec:experiments} show the effectiveness, scalability, and generalizability of \name and we conclude with open problems and exciting directions for this new area (Section~\ref{sec:conclusion}).
% Outline:
    % Table Discovery
    % Table Integration
    % By-Target Synthesis
\section{Related Work}\label{sec:related}
% In this section, we discuss related work to the two phases of \name: Table Discovery (Section~\ref{subsec:related-discovery}) and Table Integration (Section~\ref{subsec:related-integ}). We conclude with  work related to 
% %our problem of 
% finding the origins of a table, referred to as By-Example or By-Target approaches in the literature (Section~\ref{subsec:related-qbe}).

% \roee{Trimming suggestion: paragraphs/bold instead on subsections.}

% \subsection{Table Discovery}\label{subsec:related-discovery}

% We now discuss related work to Table Discovery and Integration. We then discuss work related to finding the origins of a table, referred to as By-Example or By-Target approaches in the literature.

We now discuss related work on Table Discovery and Integration and work related to finding the origins of tables.%, referred to as By-Example or By-Target approaches in the literature.

\vspace{.05cm}\noindent\textbf{Table Discovery:} Table Discovery 
has a rich literature, specifically keyword search over tables, unionable table search, and joinable table search. 
Early work such as Octopus~\cite{DBLP:journals/pvldb/CafarellaHK09} and %along 
Google Dataset Search~\cite{DBLP:conf/www/BrickleyBN19}, support keyword search over the metadata of tables~\cite{DBLP:journals/pvldb/AdelfioS13,DBLP:journals/pvldb/LimayeSC10} and smaller scale web-tables~\cite{DBLP:conf/www/ShragaRFC20,DBLP:conf/sigir/ShragaRFC20}. %To support d
Data-driven table discovery systems~\cite{DBLP:conf/sigmod/SarmaFGHLWXY12, DBLP:journals/pvldb/ZhuNPM16, DBLP:conf/sigmod/ZhuDNM19, DBLP:conf/icde/FernandezMQEIMO18, DBLP:journals/pvldb/NargesianZPM18} were then developed to find schema complements,   entity complements, joinable tables, and unionable tables. 

For joinable table search, early systems %make 
use schema matching or syntactic similarities between tables' metadata, such as Jaccard similarity~\cite{DBLP:conf/sigmod/YakoutGCC12, DBLP:journals/ws/LehmbergRRMPB15}. LSH Ensemble~\cite{DBLP:journals/pvldb/ZhuNPM16} makes use of approximate set containment between column values and supports set-containment search using LSH indexing. JOSIE~\cite{DBLP:conf/sigmod/ZhuDNM19} uses exact set containment to retrieve joinable tables that can be equi-joined with a column in the user's table. MATE~\cite{DBLP:journals/pvldb/EsmailoghliQA22} supports multi-attribute join with a user's table. 
DeepJoin~\cite{DBLP:journals/pvldb/Dong0NEO23} leverages a deep learning model to retrieve equi-joinable and semantically joinable tables.
These systems can be used to retrieve a set of candidate tables that have high set similarity with a given user's table.

For table union search, early systems also used schema similarity %was first supported by systems that leverage schema similarity to retrieve unionable table
~\cite{DBLP:conf/sigmod/SarmaFGHLWXY12, DBLP:conf/ijcai/LingH0Y13}.  Using data (rather than metadata), a formal problem statement for unionability was first defined by %TUS
Nargesian et al.~\cite{DBLP:journals/pvldb/NargesianZPM18} who presented a data-driven solution  %system 
that leverages syntactic, semantic, and natural language measures. This problem was refined by SANTOS~\cite{santos23} to consider  relationship semantics as well as column semantics.
%when retrieving semantically unionable tables. 
Most recently, Starmie~\cite{starmie22} offers a scalable solution to finding unionable tables that leverages the entire table context to encode its semantics. Although our method also retrieves relevant tables to a user's table, we aim to retrieve tables for a specific task -- reclaiming the user's table. 
Finally, other recent work~\cite{DBLP:conf/icde/GalhotraGF23} presents a goal-oriented discovery for specific downstream tasks, aiming to augment columns. We tailor table discovery towards the goal of reclaiming the Source Table. 

% \roee{Why this is in table discovery? Maybe add another subsection with others or rename this or next as X and Provenance? Actually maybe it fits better under by-example} Our problem setting of tracing the Source table's values back to its origins can be related to Data Provenance~\cite{DBLP:journals/ftdb/CheneyCT09}, which given the query and its output table, explains where the tuples originates from, why and how they was produced. However in our problem setting, we need to recover the tables and the query used to create the user's table, and thus do not know the query that was originally used to create the user's table.

% More similar to our problem setting, in which we retrieve relevant tables to the Source table such that we can reclaim the Source Table, METAM~\cite{DBLP:journals/corr/abs-2304-09068} offers a generic framework by querying the downstream task in order to drive the table discovery and augmentation. Unlike METAM, we only aim to create a system that tailors the table discovery and augmentation towards the goal of reclaiming the Source Table.
% \grace{Should we remove the emphasis? - more vague about similarity}

% \subsection{Table Integration}\label{subsec:related-integ}

\vspace{.05cm}\noindent\textbf{Table Integration:} 
%rjmOnce we have a set of candidate tables retrieved from table discovery, we continue to Table Integration. 
% \roee{Schema/Entity matching discussion: Change some of the discussion here as well to reflect that we have updated it based on the reviews (in other words, make it look blue ;))} 
 Lehmberg et al.~\cite{DBLP:journals/pvldb/LehmbergB17} stitches unionable tables together, but does not support join augmentation of tables. %More r
 Recently, \alite~\cite{Khatiwada2022IntegratingDL} performs full disjunction (FD)~\cite{DBLP:conf/sigmod/Galindo-Legaria94} to maximally combine tuples from a set of tables (intuitively, full disjunction is a commutative and associative form of full outer join).
Our goal is to reproduce the given Source Table, which may contain incomplete tuples, so we do not aim to maximally combine tuples if it produces a table that is not identical to the Source Table. Nonetheless, \alite is a candidate baseline for \name, as it offers a state-of-the-art integration solution.

\revision{Preceding the table integration process, there are pre-integration tasks to find alignments between table elements. First, instance-based schema matching 
%rjm corresponds 
determines how the schemas of two tables align to prepare for integration~\cite{DBLP:journals/vldb/RahmB01,DBLP:journals/pvldb/ShragaGR20,DBLP:conf/icde/KoutrasSIPBFLBK21,DBLP:conf/vldb/DoR02,DBLP:conf/icde/MelnikGR02,DBLP:journals/debu/ChenGHTD18}. Our solution, \name, aligns schemas implicitly by
%after retrieving a set of originating tables.
%rjm to a user's table, and 
renaming columns in the retrieved tables with the column of the source table that best matches.
%to align with the column headers of 
%rjm the user's table.
%the Source Table so we can apply joins and unions.
Entity matching~\cite{DBLP:conf/sigmod/CappuzzoPT20, DBLP:journals/pvldb/0001LSDT20, DBLP:journals/jdiq/LiLSWHT21,DBLP:journals/corr/abs-1905-06397,DBLP:journals/pvldb/GetoorM12,DBLP:conf/sigmod/MudgalLRDPKDAR18,DBLP:conf/www/ZhangNWST20,DBLP:conf/www/ZhaoH19,DBLP:conf/cikm/Gurajada0QS19} is another common pre-integration task, aiming to align tuples for cleaning or joining tables. In our context, since we assume that the Source Table has a key, tuples can be aligned by matching using equality on the key.} %In our solution, we assume the Source Table has a key, and thus align tuples by matching using equality on the key. 

% In addition, works proposed for entity matching work to align tuples in preparation for join tasks between tables, for example~\cite{DBLP:conf/sigmod/CappuzzoPT20, DBLP:journals/pvldb/0001LSDT20, DBLP:journals/jdiq/LiLSWHT21,DBLP:journals/corr/abs-1905-06397,DBLP:journals/pvldb/GetoorM12,DBLP:conf/sigmod/MudgalLRDPKDAR18,DBLP:conf/www/ZhangNWST20,DBLP:conf/www/ZhaoH19,DBLP:conf/cikm/Gurajada0QS19}. In our solution, we require tables to have keys, and thus align entities via their keys.

% \subsection{By-Example and By-Target}\label{subsec:related-qbe}
% \subsection{Finding Origins of Tables}\label{subsec:related-qbe}
\vspace{.05cm}\noindent\textbf{Finding Origins of Tables:} Our problem setting of tracing a Source Table's values back to its origins can be related to Data Provenance~\cite{DBLP:journals/ftdb/CheneyCT09,DBLP:reference/db/CheneyT18}, which given a query and its output table, explains from where the (values or) tuples originate, why and how they were produced. However, in our problem setting, we 
%need to recover the tables and the integration required to reproduce the 
%the Source Table and thus 
do not know the query or originating tables that were originally used to create the Source Table.

%rjm There have been a lot of work on 
% \roee{Change some of the discussion here as well to reflect that we have updated it based on the reviews (in other words, make it look blue ;))} \grace{done}
\revision{Query-By-Example (QBE) is a popular approach.  The original QBE was a language allowing nonexpert users to query a database~\cite{DBLP:conf/afips/Zloof75}.  More recently this term has been used for methods that are given a pair of matching input and output tables, and the task is to synthesize a query from the input to the output\cite{DBLP:conf/pldi/WangCB17}. %Specifically, s
%Systems, such as SQL-by-example~\cite{DBLP:conf/pldi/WangCB17}, synthesize a SQL query to produce an output table, given an input table.
For this task, some systems only consider Project and Join operators~\cite{DBLP:conf/sigmod/ZhangEPS13, DBLP:conf/sigmod/KalashnikovLS18,DBLP:journals/pvldb/OrvalhoTVMM20,Gong2023Ver}, whereas others also consider the Select operator~\cite{DBLP:conf/sigmod/TranCP09,DBLP:conf/vldb/BonifatiCLS14}.
% For example, Ver~\cite{Gong2023Ver} aims to find Project-Join views over large tables in which the join path is not known. 
Others output a set of queries %rather than one query 
that could reproduce the example output table, given the input table~\cite{DBLP:conf/sigmod/ShenCCDN14,DBLP:conf/icde/DeutchG16}. AutoPandas~\cite{DBLP:journals/pacmpl/BavishiLFSS19} performs transformation-by-example by synthesizing Pandas programs rather than SQL queries.
More recently, proposed techniques relax the assumption that a set of tables from which the query table can be generated is provided~\cite{Gong2023Ver, DBLP:journals/pvldb/RezigBFPVGS21, DBLP:journals/corr/abs-1911-11876}. Instead, they discover a set of tables that, when integrated, produce a table that contains the query table. These methods often expect only a \emph{partial} query table with a small  set of attributes and possibly a set of tuple examples (for example, Ver~\cite{Gong2023Ver} uses queries that are tables of 2 columns and 3 rows). Their goal is to generate an output table that \emph{completes} this query table by returning a table that contains many additional tuples in addition to those in the query table. 
However in our problem, our goal is to reproduce all and only tuples from the source table.
Nonetheless, since Ver~\cite{Gong2023Ver} is the state-of-the-art Query-by-Example method, we use it as a baseline for \name.
}
%rjm , for the same inputs and goal.

Auto-Pipeline~\cite{DBLP:journals/pvldb/YangHC21} defines %a similar problem, 
Query-By-Target, %, 
with the similar goal of synthesizing the pipeline used to create the target table, given the target table and a set of input tables. Using the synthesized pipeline on the input tables, it then produces a table that %is 
``schematically'' %aligned
aligns with the input target table. As the state-of-the-art in this line of work, it is a baseline for our approach. In both By-Example and By-Target paradigms, 
many systems assume that
the set of input tables on which the system synthesizes a query to generate the example or target table is 
%rjm minimal.
known and perhaps more importantly are known
to contain the tuples and columns needed to reproduce the output table. %However, i
In our problem, we do not assume 
%rj such 
this 
is the case.
%rjm redundant?
%We are only given the 
%Source Table and a data lake as input. From the data lake, we need to search for and filter a set of candidate tables such that 
%by integrating the filtered set of candidate tables (which we call the originating tables), 
%we can reclaim the Source Table as close as possible.

% Another similar work to ours is Ver~\cite{Gong2023Ver}, a system that finds \textsc{Project-Join} views over large tables in which the join path is not known. %This framework is similar to ours, but w
% Different from their setup, we assume that our given Source table was created by any variation of \textsc{Select-Project-Join-Union} query.
% \grace{Should we remove the emphasis?}
\section{Problem Definition and Preliminaries}\label{sec:preliminaries}
We first discuss how we evaluate a possible reclaimed Source
Table and then define the problem of \emph{Table Reclamation}.
% \emph{Value Similarity Score}, allowing us to compare reclaimed tables with source tables and to define the problem of Source Table Reclamation.
Then, we describe preliminaries for our solution, specifically the set of table operators with which we perform table integration to produce a possible reclaimed Source Table. 

\subsection{\ise Score and Problem Definition}\label{subsec:tuple_alignment}
%Given a 
To evaluate a possible reclaimed table, we compare it with the Source Table to see how close they are. 
% \roee{Can we attach refs to each of the following statements here? I think we may even be able to cover them with the existing refs.} 
The problem of comparing database instances is prevalent in many applications such as analyzing how a dataset has evolved over time (e.g., data versioning)~\cite{DBLP:journals/pvldb/ShragaM23}, evaluating data cleaning solutions (e.g., compare a clean instance produced by a data repair algorithm against a gold standard)~\cite{DBLP:journals/pvldb/MahdaviA20}, or comparing solutions generated by data exchange or transformation systems~\cite{DBLP:journals/tcs/FaginKMP05, DBLP:journals/vldb/AlexeHPT12}. 
\revision{This \emph{similarity score} requires the computation of a mapping between the tuples across instances, which can be used to explain the result.  The most general measures rely on homomorphism checking and are NP-hard~\cite{FKNP08}. Since data lake tables do not have keys, integrating them can produce multiple copies of tuples from the source table.  We will align data lake tuples with a single source tuple where the lake and source tuple share the same key value (i.e., are {\em aligned tuples} iff they share the same values on key attributes).   Hence, multiple lake tuples may align with the same source tuple, and some will align with no source tuple.  But a lake tuple will align with at most one source tuple and because of this, we can do the mapping  efficiently. 
%However, since the source table has a key, we will assume that tuples map (i.e., are {\em aligned tuples}) iff they share the same values on key attributes.
% Note that while some recent Query-By-Example methods~\cite{Gong2023Ver,DBLP:journals/corr/abs-1911-11876} consider semantic similarity between values, we aim to reproduce all values from the Source Table, as is.
%rjm I found the following confusing as it says "share the same semantics" which sounds like semantic
%, i.e., the reproduced Source Table must share the same semantics as the Source Table.
}

%Our 
We %now 
propose an {\em error-aware instance similarity score} %is 
as a generalization of {\em instance similarity} defined by Alexe et al.~\cite{DBLP:journals/vldb/AlexeHPT12}, which has been widely used in the integration literature~\cite{10.1007/978-3-031-42941-5_37,DBLP:journals/pvldb/ArocenaGCM15,DBLP:conf/ssdbm/MaziluPFK19,DBLP:journals/debu/ArocenaGMMPS16}. 
% \roee{Any ref indicating the usage?}  
Instance similarity %was developed to 
quantifies the preservation of data associations when source data is exchanged into target database~\cite{DBLP:journals/vldb/AlexeHPT12}. %Instance similarity 
It relies on computing homomorphisms (as it is designed for data exchange where keys are not assumed).   Considering two relations with the same schema,  if two tuples are mapped, they define the {\em tuple similarity} as the ratio of the number of values that are shared over the size (cardinality) of the tuples. Note that %since this is 
in data exchange two tuples cannot be mapped if they disagree on any non-null values. In our setting, we can map tuples that differ on their non-null values.  Hence, we define an {\em error-aware tuple similarity} that penalizes %for 
mismatching (erroneous) values. 

\begin{definition}\label{dfn:tuple-sim}
Given two tuples $s$ and $t$ with the same schema 
%and same number of non-key attributes, 
containing $n$ non-key attributes, where $s$ and $t$ share the same key value.  Let $\alpha(s,t)$ be the number of non-key attributes on which $s$ and $t$ share the same value and $\delta(s,t)$ be the number of non-key attributes on which $s$ and $t$ have different values and $t$ is a non-null value.  Then the {\bf error-aware tuple similarity} is\footnote{Note the tuple similarity defined by Alexe et al.~\cite{DBLP:journals/vldb/AlexeHPT12} is $\alpha(s,t)/n$}:
\begin{equation}\label{eq:error-tuple-sim}
    E(s,t) = (\alpha(s,t) - \delta(s,t)) / n 
\end{equation} 
% \footnote{Note the {\bf tuple similarity} defined by Alexe et al.~\cite{DBLP:journals/vldb/AlexeHPT12} is $\alpha(s,t)/n$}
\end{definition}

Since a tuple can map to more than one tuple, Alexe et al. define {\em instance similarity} using the maximum tuple similarity score and we do the same in defining {\em error-aware instance similarity}, but we use error-aware tuple similarity rather than tuple similarity.  For completeness (and since in our experiments we use both measures), we now define both.

\begin{definition}
\label{dfn:instance-sim}
Let $S$ be a source table (with key attributes $K$) and $T$ a possible reclaimed table  with the same schema that has $n$ non-key attributes.  Note that $T$ does not have to satisfy the key constraint.
For a tuple $s \in S$, let $m(s) = \{ t \in T | s[K] = t[K] \}$.  
 Then the {\bf instance similarity} of $S$ and $T$ is:
\begin{equation}\label{equ:instance-similarity}
    % \text{Instance Similarity} = 
    \dfrac{\sum_{s\in S} \text{max}_{t \in m(s)} (\alpha(s,t)/n) }{|S|}
\end{equation}
The {\bf Error-Aware Instance Similarity (\ise)} of $S$ and $T$ is (for a normalized score in range[0,1]):
\begin{equation}\label{eq:normalized_ise}
    % \dfrac{
    % 0.5\cdot\sum_{s\in S} \text{max}_{t \in m(s)} (\alpha(s,t) - \delta(s,t)+n)/n}{|S|}
    \dfrac{
    0.5\cdot\sum_{s\in S} \text{max}_{t \in m(s)} (1+E(s,t))}{|S|}
\end{equation}
\end{definition}

\begin{figure}[ht]
    \vspace{-.1in}
    \centering
    \includegraphics[width=0.45\textwidth]{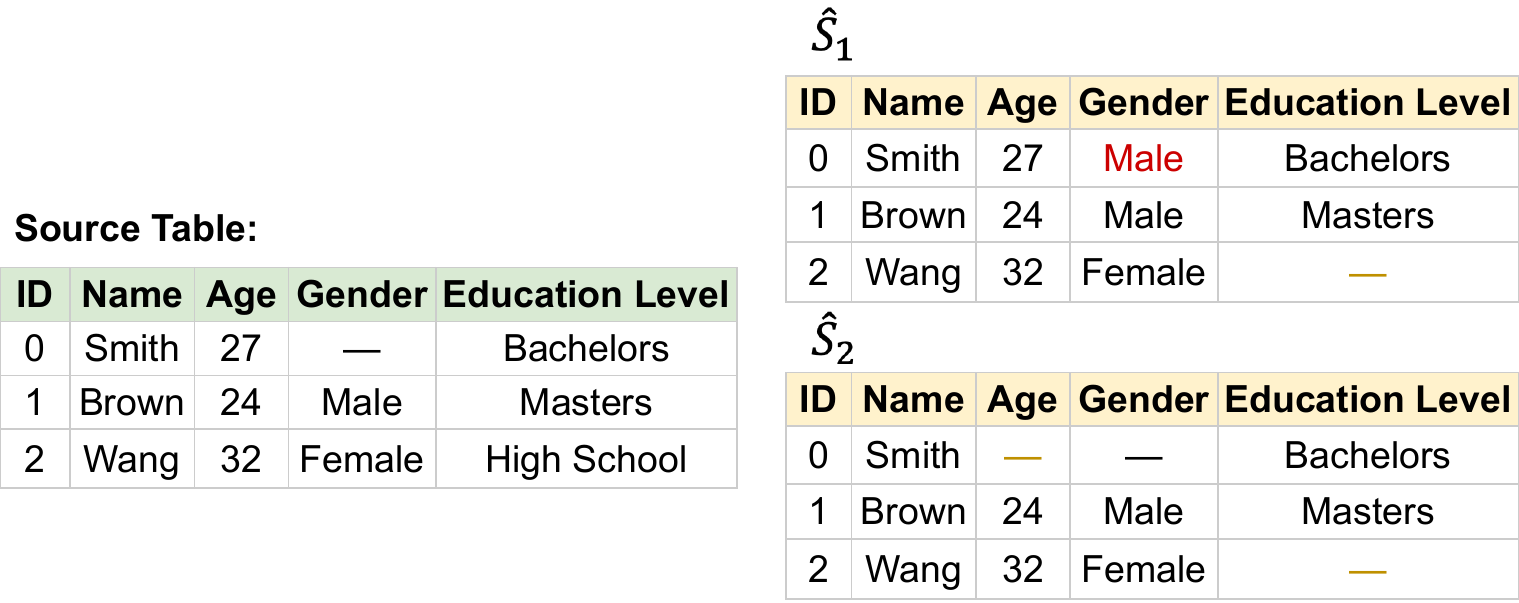}
    \caption{Aligned tuples between a Source Table (left green table) and two possible reclaimed tables (right yellow tables) from Figure~\ref{fig:raw_integ}, aligned based on key column `ID'.}
    \label{fig:vss_example}
\end{figure}

We illustrate these scores in an example based on Figure~\ref{fig:raw_integ}.

\begin{example}\label{ex:similarity}
Consider Source Table $S$ (with key column ``ID'') from Figure~\ref{fig:raw_integ} (top-left green table), and two possible reclaimed tables, $\hat{S}_1$ (top-right yellow table) and $\hat{S}_2$ (bottom-right yellow table). Their aligned tuples are shown in Figure~\ref{fig:vss_example}.
%If there were multiple tuples for a single key, we would find the one with the highest (error-aware) tuple similarity.
Notice the instance similarity score (counting only matching values) of 
$S$ and $\hat{S}_1$ is higher than of $S$ and $\hat{S}_2$.

$\hat{S}_1: t_0 = 3/4, t_1 = 4/4, t_2 = 3/4 \rightarrow 0.833$
% \rightarrow (\frac{3}{4}+\frac{4}{4}+\frac{3}{4})/3 = 0.833$$

$\hat{S}_2: t_0 = 2/4, t_1 = 4/4, t_2 = 3/4 \rightarrow  0.75$
% \rightarrow (\frac{2}{4}+\frac{4}{4}+\frac{3}{4})/3 = 0.75$$

However, we want to favor $\hat{S}_2$ that contains nullified (unknown) values (one tuple matching the source correctly and two tuples missing values), over $\hat{S}_1$, which 
has reclaimed a possibly erroneous value for a source null. Using \ise score, $\hat{S}_2$ has a higher similarity with $S$ than $\hat{S}_1$.

$\hat{S}_1: t_0 = (3-1)/4, t_1 = 4/4, t_2 = 3/4 \rightarrow 0.875$

$\hat{S}_2: t_0 = 3/4, t_1 = 4/4, t_2 = 3/4 \rightarrow  0.917$
% \begin{align*}
% & \hat{S}_1: t_0 = \frac{(3-1)}{4}, t_1 = \frac{4}{4}, t_2 = \frac{3}{4} \\ & \rightarrow \dfrac{\frac{1}{2}*(\frac{(3-1+4)}{4} + (\frac{(4+4)}{4})+(\frac{(3+4)}{4}))}{3} = 0.875
% \end{align*}
% \begin{align*}
% & \hat{S}_2: t_0 = \frac{3}{4}, t_1 = \frac{4}{4}, t_2 = \frac{3}{4} \\& \rightarrow \dfrac{\frac{1}{2}*(\frac{(3+4)}{4} + (\frac{(4+4)}{4})+(\frac{(3+4)}{4}))}{3} = 0.917
% \end{align*}
\label{example:vss}
\end{example}

% Recall that we evaluate Source Table $S$'s similarity (Algorithm~\ref{alg:table_integ} Lines~\ref{line:integ_comp_coverage}, ~\ref{line:integ_subsump_coverage}) before and after applying Complementation($\kappa$) and Subsumption($\beta$) to see if the resulting table contains fewer values and tuples shared with $S$. With \ise, we can evaluate the resulting table with respect to $S$ by measuring how similar the values in the resulting table are to $S$'s values. 

Using \ise Score as a similarity measure to compare a possible reclaimed table $\hat{S}$ and Source Table $S$, we aim to solve the following problem:

\begin{definition}[Table Reclamation] \label{def-qd}
    Given a collection of tables $\bigT$ and a Source Table $S$, %rjm important!  I removed this as S does not have to be generated from T!!!!
    %generated from a set of tables $\bigT^{*} \subseteq \bigT$, 
    find a set of originating tables $\hat{\bigT} \subseteq \bigT$ such that its integration produces a reclaimed table $\hat{S}$ with the maximum \ise Score to $S$.
\end{definition}

\subsection{Preliminaries: Integration Operators}\label{subsec:table-operators}
We now present the operators we use to produce a possible reclaimed Source Table.   We will show that they are sufficient for integration, inspired by recent work on data lake integration~\cite{Khatiwada2022IntegratingDL}.  First the unary operators.
%Consider the following table operators performed on one table.

\noindent$\bullet$ Projection($\pi$): Project on specified columns of the table.\\
$\bullet$ Selection($\sigma$): Select tuples that satisfy a specified condition.\\
$\bullet$ Subsumption($\beta$)~\cite{DBLP:conf/sigmod/Galindo-Legaria94}: Given tuples $t_1, t_2$ 
%in the same table,
with the same schema, $t_1$ subsumes $t_2$ if for every attribute on which they are both non-null they have the same value, and
%some non-null column value(s)
$t_1$ contains one or more attributes with a non-null value where $t_2$ has nulls. Applying subsumption we remove $t_2$. Applying $\beta$ on a table involves repeatedly applying subsumption and discarding the subsumed tuples ($t_2$).\\
$\bullet$ Complementation($\kappa$)~\cite{DBLP:journals/csur/BleiholderN08,DBLP:conf/icde/BleiholderSHN10}: Given tuples $t_1, t_2$ 
%in the same table, 
with the same schema, $t_1$ complements $t_2$ if they share at least one non-null column value, and $t_1$ contains some non-null values where $t_2$ has nulls while $t_2$ contains some non-null values where $t_1$ has nulls. The tuples must agree on all values on which they are both non-null.
%non-null values. 
Applying $\kappa$ on $t_1$ and $t_2$ produces a single tuple that contains all non-null values of either (both) tuples and is null only if both $t_1$ and $t_2$ are null. Applying $\kappa$ on a table produces a table with no complementing tuples and involves repeatedly applying complementation to pairs of tuples. %that complement each other.\\  

% Schemas of tables have been aligned (Section~\ref{subsec:candidate_table_retrieval}), so joinable (unionable) columns share the same name and we use the natural version of the Outer Union operator~\cite{cowbook}. 
We use a single  binary operator natural Outer Union, and  we assume the schemas of the tables have been aligned so unionable columns share the same name~\cite{cowbook}. 

%\noindent$\bullet$ Inner Union($\cup$): Union two tables that share the same schema. This operator is commutative and associative.\\
\noindent$\bullet$ Outer Union($\uplus$)~\cite{DBLP:conf/sigmod/Codd79}: Union two tables, even if their schemas are not equal. The result %ing table 
contains the union of the columns from both tables. If a column $C$ is missing from one table ($T$), but appears in the other table ($S$), then in the result, the tuples of $S$ contain a null ($\bot$) in their $C$ column. This operator is commutative and associative. Note that when applied to tables with the same schema, outer union is the same as inner union.

\noindent To make reclamation search more efficient, we %will make use of 
use the fact that
Outer Union and the set of unary operators above can be used to represent any SPJU query.  Using this result, our search will focus on outer union and the unary operators.

\begin{theorem}[Representative Operators] \label{thm-ops}
Given two tables that contain no duplicate tuples, and no tuples that can be subsumed or complemented, for all SPJU queries, there exists an equivalent query consisting of only Outer Union and the four unary operators (selection, projection, complementation, and subsumption).\footnote{The proof is included in 
% Appendix~\ref{app:ops_proof}
our technical report~\cite{gen-t_technical_report}.
}
\end{theorem}

% \rjm{Yes - remove from here and include this in V-B}
% \grace{should we move this to Section~\ref{subsec:integration}?}
% Using this set of operators $\bigoplus = \{\uplus,\sigma, \pi, \kappa, \beta\}$, 
% we present an efficient algorithm to  explore the integration space to reclaim a Source Table. Instead of traversing through all possible join paths
% in a space consisting of all tables, as in existing by-example and by-target work, we apply Outer Union and unary operators with conditions based on a Source Table. 

% \section{Table Discovery}\label{sec:discovery}
% Outline:
% Table Discovery:
%     We need to retrieve the set of candidate tables to perform table reclamation (continue running example)
%     Starmie (SotA) + Set Similarity
%     Matrix Traversal
\section{Table Reclamation using \name}\label{sec:reclaim}
\revision{%We first describe Table Discovery (first step in Figure~\ref{fig:pipeline}, Section~\ref{sec:discovery}) of \name,  which aims to find a good set of originating tables with which we can effectively reclaim a source table. With %the set of originating tables this set, we enter the Table Reclamation step (second step in Figure~\ref{fig:pipeline}, Section~\ref{subsec:integration}), in which we perform table integration to produce a reclaimed source table.
}
We describe Table Discovery (Section~\ref{sec:discovery}),  
which finds a set of originating tables that we can effectively integrate to reclaim a source table %in the Table Reclamation step 
(Section~\ref{subsec:integration}).

\subsection{Table Discovery}\label{sec:discovery}
We first discover a set of candidate tables%(Section~\ref{subsec:candidate_table_retrieval})
, after which we discuss a novel methodology, termed \emph{Matrix Traversal}, to refine this set into a set of originating tables. %(Sections~\ref{subsec:matrix_traversal}-\ref{subsec:ternary-matrix}).
% based on Theorem~\ref{thm-ops}. %The method, \emph{Matrix Traversal}, is based on Theorem~\ref{thm-ops}. 
% Accordingly, instead of traversing through all possible join paths in a space consisting of all tables, as in existing by-example and by-target work (see Section~\ref{sec:related}), we apply Outer Union and unary operators with conditions based on a Source Table. 

\subsubsection{Candidate Table Retrieval}\label{subsec:candidate_table_retrieval}
Discovering a set of candidate tables requires discovering tables that share some of the same values as the Source Table in an efficient manner. In the context of data lakes, where metadata is inconsistent or missing, searching using schema names is unreliable~\cite{DBLP:journals/pvldb/AdelfioS13,DBLP:conf/sigmod/FaridRIHC16,DBLP:journals/pvldb/NargesianZMPA19,DBLP:conf/sigmod/ZhangI20}. 
% So, we leverage a data-driven, unsupervised approach to table discovery. 
% However, we can use any existing data-driven %unsupervised 
% approach to table discovery that is scalable in a data lake setting.
However, we can use any existing data-driven %unsupervised 
table discovery approach that is scalable in a data lake setting. %\revision{Note that schema matching is applied here implicitly (high value overlap) and we rename the data lake tables' columns with the Source Table's column name, see Section~\ref{subsec:table-operators}.}\grace{moved to end of (1). Why see Section~\ref{subsec:table-operators}?}
% More importantly, we aim to use a table discovery system that is scalable to the size of data lakes and their tables. This way, it can discover a set of candidate tables for the Source Table among a large data lake. In our experiments (Section~\ref{subsec:accuracy-exp}), we make use of a state-of-the-art system for 
% %rjm unionable 
% table search called Starmie~\cite{starmie22}. 

With 
%rjm the
% \renee{I've changed a few dozen "the" to "a" -- consider this as you write}
a set of top-$k$
tables returned as relevant to the Source Table, we need to verify the set similarities of their values with the Source Table.
%, especially if they were discovered primarily using table semantics.  
To do so, we retrieve candidate tables among the previously discovered tables using a set similarity algorithm. This could be done efficiently with a system like JOSIE~\cite{DBLP:conf/sigmod/ZhuDNM19} that computes exact set containment or MATE~\cite{DBLP:journals/pvldb/EsmailoghliQA22} that supports multi-attribute joins. 
% \rjm{I don't understand the next sentence.  What is the list, here we have only talked about sets of tables.}
In addition to finding candidate tables containing columns that have high set similarity with a Source Table, we also \emph{diversify} the set of candidate tables such that each candidate table has minimal overlap with %its previous 
other candidates.
This is especially important in public data lakes, which tend to have multiple versions of the same tables~\cite{DBLP:journals/pvldb/ShragaM23,koch2023duplicate} and a large percentage of duplicate column sets~\cite{DBLP:conf/sigmod/ZhuDNM19}.
By diversifying candidates, each candidate may overlap with different values in $S$, as illustrated in the following example.

\begin{example}
    Suppose we have the Source Table from %our running example in 
    Figure~\ref{fig:raw_integ}. In addition to data lake tables A, B, C, D, we also have Table E, an exact duplicate of Table D. If we only rank these tables using set overlap with the Source Table, Tables D and (its duplicate) Table E become top candidates, since all their columns have high set overlap with those in the Source Table. However, Table E does not add any new information when integrated with Table D. Thus, diversifying the set of candidate tables decreases Table E's score, pushing other tables such as Table A higher in the top-$k$ ranking.
\end{example}

With a diverse set of candidates found for each column in a Source Table $S$, we ensure that each candidate table still has high set overlap with the Source Table across related columns. To do so, we find all tuples in a candidate table that share column values with $S$. We verify that for each column that has high set overlap with a column in $S$, it still has high set overlap within these tuples. 
\revision{
For columns from the candidate table that have high value overlap with columns in $S$, we rename them with names of corresponding columns in $S$. This way, we implicitly perform schema matching between each candidate table and $S$.
}
% \roee{We haven't introduced ``subsumption" yet: Can we remove it from the discussion here?} 
Finally, we check for and remove any candidate table whose columns and column values are subsumed by other candidate tables.
% , specifically if their columns and column values are contained in other tables. 
% If so, we remove the subsumed tables from the returned set of candidate tables.

\subsubsection{Matrix Traversal}\label{subsec:matrix_traversal}
% \roee{I think this puts too much highlight on two-valued. I would use one Matrix Traversal subsection, start with the more ``natural'' solution with two-valued (potentially a sub-sub-section), explain why it is flawed and then more to three-valued} \roee{We can also make this distinction using a good example that shows (1) how matrix traversal works, (2) why three-valued is better than two-valued and (3) what would have happened if we were to use graph traversal (which will be introduced in Related Work)}\grace{I added Figure~\ref{fig:raw_integ} for what would happen if we directly integrated the tables from~\ref{subsec:candidate_table_retrieval}. Should I move it here?} \roee{let's discuss pros/cons. We can definitely refer back to it here.}

With a set of diversified candidate tables,
%found previously, 
we could potentially enter the table integration phase (Section~\ref{subsec:integration}).\revision{%using all candidate tables as an originating set of tables.
}
However, it may be
computationally expensive to 
\revision{% use all candidate tables to perform table integration.
% To minimize the integration cost, 
}
directly integrate all candidate tables.
Thus, we need to refine the set of candidate tables to a set of
%the 
originating tables containing a maximum set of aligned tuples with respect to the Source Table. To do so, we %can 
emulate the table integration process 
\revision{% without performing the expensive integration computations 
}
and see what candidate tables are 
necessary to reclaim our Source Table. 
\revision{%By simulating the alignment of candidate tables' tuples with each other, we can uncover contradicting (erroneous) 
}
By simulating tuple alignment, we can uncover erroneous aligned tuples with respect to the Source Table, and discard tables that could decrease the \ise Score.

First, we need to align tuples in candidate tables to tuples in a Source Table, based on shared values with the key attribute from the Source Table.
To do so, we need to ensure that each candidate table contains a key column of the Source Table. If it does not, we
greedily find a best way to join it with candidates that include the source key. We use standard join cardinality estimation to find a path that covers the most source key values and is as close to functional as possible (this procedure is denoted as Expand()).\footnote{The Expand algorithm is included in 
% Appendix~\ref{app:expand}
the technical report~\cite{gen-t_technical_report}.
}  
 % Note that this step could be replaced with outer union (as per Theorem~\ref{thm-ops}), but this would make our similarity computation (Section~\ref{subsec:tuple_alignment}) expensive (potentially exponential).
 % first join it with a copy of another candidate table with which it has common column values, and that has a key from the Source Table. 
 This way, all tables can align its tuples with the Source Table using key values. 

\revision{%To capture tuple alignment, we represent each candidate table and its aligned tuples with the Source Table in the form of a matrix.For each table, we encode its aligned tuples with respect to the tuples in the Source Table, for columns that the table shares with the Source Table. 
}
We represent aligned tuples and shared columns from a candidate table in the form of a matrix. 
To encode aligned tuples, we initialize the matrices to have the 
% \renee{what does "same shape" mean?}\grace{resolved}
same dimensions as the Source Table $S$, \revision{%containing the same number of rows and columns as $S$,
}such that the matrix indices represent the Source Table's indices. For each key value and its associated column values in the Source Table, check if the value appears in the candidate table 
% \renee{hmm, you mean candidate table?}\grace{resolved} 
at the corresponding column and key value. If so, then the matrix has 1 at the same index as the value's index in Source Table, and 0 otherwise.  
% \renee{I'm confused as this doesn't account for candidate tables that don't contain the source key value.}
% \grace{added a line in set similarity (Section~\ref{subsec:candidate_table_retrieval}) about retrieving candidate tables with shared keys with $S$}

% \begin{figure*}[!ht]
%     \centering 
%     \includegraphics[width=\textwidth]{fig/matrix_integration.pdf}
%     \caption{Simulate table integration of two tables by taking the logical {\tt OR} of two matrix representations. \grace{Added back in (not in original paper -- in response to R2D5)}
%     % \roee{Trimming suggestion: I think this figure is not a must in the main text.}
%     }
%     \label{fig:matrix_integ}
% \end{figure*} 

\begin{figure*}[!ht]
\centering
    \includegraphics[width=.9\linewidth]{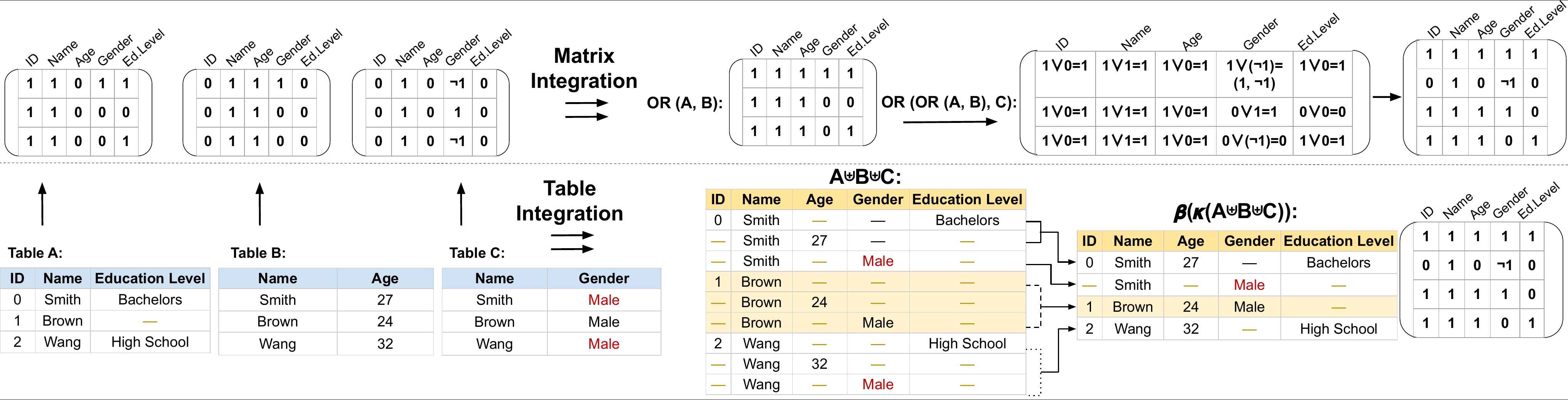}
    % } \roee{I suggest we keep only (b) for the main text and refer to (a) in words + provide it in the TR.}}
    \caption{
    Matrix initialization and integration of tables A, B, C given the Source Table from Figure~\ref{fig:raw_integ} simulate their table integration. The result of matrix integration is equivalent to the matrix representation of the table integration result. 
    }
    \label{fig:matrix_integ}
\end{figure*}

\revision{
Next, we simulate table integration by applying the logical {\tt OR} on the matrices, which takes the maximum value at each position. 
Suppose we have tuples $t_1, t_2$ from matrices $m_1, m_2$, respectively, at the same row index $i$. We want to combine values from $t_1, t_2$ at column $j$. Assuming that $ S[i, j] \neq \bot$, the produced tuple $m_r$ contains the following value at position ($i, j$): $m_r[j] = max(t_1[j], t_2[j])$.
If $t_1[j]$ is 1 and $t_2[j]$ is 0, for example, $max(t_1[j], t_2[j])$ is 1, which is equivalent to the logical {\tt OR}. 
}
This is comparable to applying the Outer Union ($\uplus$) of two tables, and Subsumption ($\beta$) and Complementation ($\kappa$) on the resulting table (refer to Theorem~\ref{thm-ops}).
\revision{%Taking the Outer Union result of two tables, we apply $\beta, \kappa$ such that 
}
In this table integration, for tuples $t_1, t_2$ that share a non-null value at the same column, the resulting tuple $t_r$ is formed such that for every column $j$, $t_r[j] = \not\perp \text{if} \, t_1[j] \neq \bot \text{ or } t_2[j]\neq \bot \text{ and } t_r[j] =\bot \text{ otherwise}$.
% \begin{equation}
%     \centering
%     t_r[j] = 
%     \begin{cases}
%         \not\perp &\text{if} \, t_1[j] \neq \bot \vee t_2[j]\neq \bot\\
%         \bot \quad &\text{otherwise}
%  \end{cases}
% \end{equation}
% In matrix representations, we combine two matrices by combining tuples at the same row index, since all tuples are encoded based on the order of the Source Table's keys. Recall that we encode a 1 in the matrix if the corresponding table shares the same value as the Source Table at the same index, and a 0 if it has a null where the Source Table contains a non-null value at the same index. Then, integrating tuples $t_1, t_2$ can be simulated with respective matrix tuples $m_1, m_2$. %, both at row index $i$. %, that encode $t_1, t_2$ based on Source Table $S$, respectively. 
% %$m_1$ and a different matrix's tuple $m_2$, both at row index $i$, that encode $t_1, t_2$ based on Source Table $S$, respectively. 
% This way, when we combine the values in $m_1, m_2$ at column $j$, assuming that $ S[i, j] \neq \bot$, the produced tuple $m_r$ contains the following value at position ($i, j$): $m_r[j] = max(t_1[j], t_2[j])$.
% Suppose $ S[i, j] \neq \bot$:
% \begin{equation}
%     \centering
%     m_r[j] = max(t_1[j], t_2[j])
% \end{equation}
% Thus, matrix integration results in a 1 if there is a matching, non-null value to the Source Table's value. Similarly, table integration results in a non-null value if there exists a non-null value in the element-wise integration. 
Thus, both table and matrix integrations maximally combine tuples such that non-null values replace a null value at the same index.

% In both integrations, we maximally combine the tuples such that non-null values can replace a null value at the same index.
%\grace{Is this clear? If not, we can add the Figure fig/matrix\_integration.pdf back in}

% Putting it all together, 

\begin{algorithm}[h]
    \caption{Matrix Traversal}\label{alg:matrix}
    % \setstretch{0.95} % sets line height
    % \SetInd{1}{1}
    \footnotesize
    \SetAlgoLined
    \LinesNumbered
    \textbf{Input}: $\bigT= \{T_1, \dots T_n\}$: set of candidate tables; $S$: Source Table \label{line:mt_input}\\
    % \textbf{Input}: Set of candidate tables $\bigT=\{T_1, \dots, T_n\}$; Source Table $S$ \label{line:mt_input}\\
    \textbf{Output}: $T_{\mathsf{orig}} = \{T_1, T_2, \dots T_i\}$: refined set of originating tables \label{line:mt_output}\\ 
    $\bigT \gets$ Expand($\bigT, S$)    \label{line:join}\algocomment{join tables without source key}\;
    $\mathcal{M} \gets $ MatrixInitialization($\bigT$),    \label{line:mt_initialize}\algocomment{Initialize Matrices of $S$ shape}\; 
    $T_{\mathsf{start}} \gets$ GetStartTable($\mathcal{M}$)\; \label{line:mt_getStart}
    prevCorrect = mostCorrect $\gets \mathsf{evaluateSimilarity}(T_{\mathsf{start}})$  \label{line:mt_evalStart}\;
    $T_{\mathsf{orig}} \gets []$\; 
    \While {$|T_{\mathsf{orig}}| < |\bigT|$} { \label{line:mt_whileCondition}
        \If{$T_{\mathsf{orig}}$} {
            $M_c \gets \mathsf{Combine}(T_{\mathsf{orig}})$    \label{line:mt_combine}\algocomment{Iteratively combine each pair of consecutive matrices}\; 
        }
        prevCorrect = mostCorrect; nextTable = $\bot$\;
        \For {all tables $T \in \bigT s.t. T\notin T_{\mathsf{orig}}$}{
            $M_c \gets \mathsf{Combine}(\mathsf{M_c}, T)$\; 
            percentCorrectVals $\gets \mathsf{evaluateSimilarity}(M_c)$\; \label{line:mt_pickNext}
            \If{percentCorrectVals $>$ mostCorrect}{
                mostCorrect $\gets$ percentCorrectVals\;
                nextTable $\gets T$\;
            }
        }
        \If{mostCorrect = prevCorrect}{ \label{line:mt_converge}
            Exit,    \algocomment{Integration did not find more of $S$'s values}\; 
        }  \label{line:mt_endConverge}
        $T_{\mathsf{orig}} = T_{\mathsf{orig}}\cup$ nextTable\; 
    }

	\Return $T_{\mathsf{orig}}$;
\end{algorithm}
% \vspace{-.5in}

We demonstrate matrix initialization and traversal in Algorithm~\ref{alg:matrix}. Given a set of candidate tables $\bigT$ and Source Table $S$, 
we first ensure that each candidate table contains a key column from $S$ by joining candidate tables that do not share a key column with $S$ with those that do (Expand() on line~\ref{line:join}).
This way, we align tuples in each candidate table with respect to $S$ and initialize each candidate table's 
% \renee{you mean each candidate table?  what if candidate table does not contain Source Table key and so does not have any aligned tuples?} 
matrix representation (Line~\ref{line:mt_initialize}). Then, we traverse over the matrices and perform the logical {\tt OR} operator to combine a pair of matrices in $\mathsf{Combine()}$ (Line~\ref{line:mt_combine}). To evaluate the resulting matrix, we check the fraction of 1's in the matrix, which represent the number of values in the resulting table integration found in the Source Table, thus evaluating the \ise score.
% its similarity with $S$ or \ise Score with respect to $S$.
%\renee{Remember to define coverage -- I still don't understand what it means.} \grace{resolved} 
At each step of the matrix traversal (Line~\ref{line:mt_pickNext}), including the start (Lines \ref{line:mt_getStart}-\ref{line:mt_evalStart}), we choose the matrix that results in a matrix containing the most 1's in $\mathsf{evaluateSimilarity()}$. This traversal ends when either all matrices have been traversed (Line~\ref{line:mt_whileCondition}), or the 
percentage of 1's in the resulting matrix 
% (equivalent to the number of Source Table's values found in the resulting table)
% Source Table's coverage 
converges (Lines \ref{line:mt_converge}-\ref{line:mt_endConverge}). 
% Note that early convergence can be achieved if the lower-ranked candidate tables do not add new information (can replace 0's with 1's) to the resulting matrix. 
We return the set of tables used in the final traversal as the set of originating tables to perform table integration.
If a candidate table that was joined with other candidates to contain a key column from $S$ (from Expand() on line~\ref{line:join}) becomes an originating table, we include its expanded form in the returned set.

\subsubsection{Three-Valued Matrices}\label{subsec:ternary-matrix}
Previously, we use matrices populated with binary values to represent aligned tuples with respect to the Source Table. However, this representation cannot distinguish between nullified and erroneous aligned tuples with respect to the Source Table. Specifically, it does not account for cases in which a tuple in the Source Table and an aligned tuple in a candidate table have different non-null values in the same column, and if a tuple in the Source Table has a null value while the aligned tuple has a non-null value at the same column.
Rather, it represents both types of values as 0 in the matrices.
% , as shown in Figure~\ref{fig:ternary_matrix}(a).
In actuality, when we apply Outer Union on two tables with aligned tuples containing different non-null values in the same column, we keep the tuples separate. 
% \begin{figure*}[!ht]
%     \subfloat[Two-Valued Matrix Representation]{
%     \begin{minipage}[t]{0.47\linewidth}
%     \includegraphics[width=\linewidth]{fig/ternary_motivate.pdf}
%     \end{minipage}
%     }
%     \subfloat[Three-Valued Matrix Representation]{
%     \begin{minipage}[t]{0.49\linewidth}
%     \includegraphics[width=\linewidth]{fig/matrix_ternary_initialization.pdf}
%     \end{minipage}
%     }
%     % \caption{Suppose we have a new Source Table about applicants' data and candidate tables A, B, C, D. Figure (a) shows two-valued matrix representations, which does not distinguish between nullified and erroneous values. Figure (b) shows three-valued matrices to make this distinction.}
%     \caption{A new Source Table %about applicants' data 
%     and candidate tables A, B, C, D. Figure (a) shows two-valued matrix representations, which does not distinguish between nullified and erroneous values. Figure (b) shows three-valued matrices to make this distinction.}
%     \label{fig:ternary_matrix}
% \end{figure*}
% \renee{add example that requires join?  all examples so far just use union.}\grace{updated the motivating example and especially Figure~\ref{fig:ternary_matrix}}
Thus, we need to distinguish between nullified and erroneous aligned tuples in the matrix representation (Line~\ref{line:mt_initialize} in Algorithm \ref{alg:matrix}). To do so, we %make 
use %of 
three-valued matrices, in which we encode a 1 if a candidate table shares the same value with the Source Table at the same index in an aligned tuple, 0 if a candidate table contains a null where the Source Table has a non-null value at the same index, and -1 if a candidate table contains a non-null value that contradicts with the Source Table's value at the same index (shown in Figure~\ref{fig:matrix_integ}).
Formally, given Source Table $S$ and candidate table $T$, we populate position $(i, j)$ for each aligned tuple $t_{\text{Align}} \in T$ in matrix $M$ as:
\begin{equation}
    M[i, j] = 
    \begin{cases}
        1 &\text{if} \, S[i,j] = T[i,j]\\
        0 & \text{elif} \, S[i,j] \neq \bot \wedge T[i,j] = \bot \\
        -1 \quad &\text{otherwise}
 \end{cases}
\end{equation}

% Now that we have 
\revision{%With the amended matrix representations, we now discuss how to combine them during matrix traversal. With three-valued matrices,  the logical {\tt OR} over two matrices takes the maximum of two truth-values at each index. 
}
Combining the amended three-valued matrix representations with the logical {\tt OR} takes the maximum of two truth-values at each index.
Specifically, if we have two tuples from two matrices that contain a 1 and -1 at the same position, applying logical {\tt OR} would choose the 1~\cite{bolc2013many}. However, in practice when applying Outer Union on two tuples with contradicting non-null values at the same index, the resulting integration would contain both tuples. Thus, we need to keep both tuples from the matrices if they contain different non-0 values at the same index. We re-define $\mathsf{Combine}$()  (Line~\ref{line:mt_combine}) between two matrices, given tuples $t_1, t_2$ at the same row index accordingly.
% \todo{Is there a way to make this look better?} \roee{let me know what you think}
% \begin{equation}\label{eq:combine_ternary}
%     \mathsf{Combine}(t_1, t_2) = 
%     \begin{cases}
%         \text{Return } t_1, t_2 \quad &\text{if} \, t_1[j] \neq t_2[j] \neq 0 \\& \text{ for column }j\\
%         \text{OR} (t_1, t_2) \quad &\text{otherwise}
%  \end{cases}
% \end{equation}
\begin{equation}\label{eq:combine_ternary}
    \mathsf{Combine}(t_1, t_2) = 
    \begin{cases}
        t_1, t_2 \quad &\text{if } \exists j:\, t_1[j] \neq t_2[j] \neq 0 \\
        \text{OR} (t_1, t_2) \quad &\text{otherwise}
 \end{cases}
\end{equation}
% \begin{equation}
%     \centering
%     OR(\text{tuples }t_1, t_2) = 
%     \begin{cases}
%         Max(v_i, w_i) \quad &\text{if} \, v_i \neq w_i \neq 0\\
%         & \text{ for } v_i \in t_1, w_i \in t_2\\
%         \text{Return } t_1, t_2 \quad &\text{otherwise}
%  \end{cases}
% \end{equation}
This way, we keep %two 
tuples separated if they contain contradicting, non-0 values at the same position. %Otherwise 
Else we apply logical {\tt OR} and take the maximum of truth values element-wise. The %amended 
new Combine() function is illustrated in Example~\ref{ex:merge_three_values}.

%As expected, this 
The new $\mathsf{Combine}$() could result in matrices with more rows than in the Source Table. %To account for this, 
Thus, we encode each matrix as a dictionary, with each key value in the Source Table as a dictionary key, and the list of aligned tuples in the resulting matrix with respect to a tuple in the Source Table as values.
\begin{example}\label{ex:merge_three_values}
    Given the Source Table from Figure~\ref{fig:raw_integ}, Figure~\ref{fig:matrix_integ} shows the result of integrating tables A, B, and C and their matrix representations. We start with matrix A with the largest number of correct values. Integrating matrices A and B produces more correct values after taking logical {\tt OR}'s of 0's and 1's (from the Combine() function). When combining its resulting matrix with matrix C, we find a (1) and ($\neg1$) in the first tuple for column ``Gender''. In this case, we keep both tuples from {\tt OR}(A, B) and C. For all other value-pairs, we take the logical {\tt OR}. When we integrate tables A, B, C, we also find tuples for ``Name''=Smith to be separate, since we check if we over-combine tuples and replace correct nulls (Algorithm~\ref{alg:table_integ}). As a result, the matrix representation of this table integration result is exactly the output of the matrix integration, and so 
    table integration is simulated by integration of matrix representations. 
    % Note that in table integration, over-combination of tuples is checked to ensure that correct nulls are not replaced, and that erroneous values are not chosen (see Algorithm~\ref{alg:table_integ}).
\end{example}

For $\mathsf{evaluateSimilarity()}$ of the start (Lines \ref{line:mt_getStart}-\ref{line:mt_evalStart}) and resulting matrices (Line~\ref{line:mt_pickNext}), we evaluate the \ise Score by taking the aligned tuple with the 
% \renee{largest, not most} 
largest number of aligned values to its corresponding tuple in the Source Table, or the largest number of 1's. 
% Then, we check the number of correct values (1's) vs.~the number of erroneous values (-1's), out of the Source Table size. 
To find the \ise score (Equation~\ref{eq:normalized_ise}) between a tuple $t$ in a resulting matrix that shares a key value with a Source tuple $s$, we set
$\alpha(s,t)$ to be the number of non-key attributes for which tuple $t$ has (1)'s, representing shared values between $t$ and $s$. For $\delta(s,t)$, we take the number of non-key attributes for which $t$ has (-1)'s, representing different, non-null values.
% \begin{equation}
%     % \dfrac{\sum_{s\in S} \text{max}_{t_{\text{Align}} \in \hat{S}} (1 + \frac{(\sum_{i=1}^n t_{\text{Align}_i}=1) - (\sum_{i=1}^n t_{\text{Align}_i}=-1)}{n})*\frac{1}{2}}{|S|}
%     \dfrac{\sum_{s\in S} max_{t \in m(s)} (\frac{((\sum_{i=1}^n t_{i}=1) - (\sum_{i=1}^n t_{i}=-1)+1)*\frac{1}{2}}{n})}{|S|}
% \end{equation} 
Thus, we treat correct, nullified, and erroneous aligned tuples with respect to the Source Table in different manners, and combine their matrix representations depending on the behavior of applying Outer Union and unary operators.

% \begin{example}\label{ex:combine_ternary_matrix}
%     Consider the matrix integration results from Figure~\ref{fig:matrix_integ}. The integration of tables A, B, C does not result in a perfect matrix consisting only of 1's, and thus not a perfect reclamation of the Source Table. However, matrix integration of tables A and D (also from Figure~\ref{fig:raw_integ}) does result in a perfect matrix, correctly representing perfect reclamation after table integration. Thus, Tables A and D become the set of originating tables. 
%     Consider another example, in which combining tables A and D creates a tuple of "1 0 0 0 0" for tuple at ID=0, and combining tables A, B, and C creates a tuple of "1 1 0 -1 -1" for tuple at ID=0 (with key column ``ID'' as the first column). In this case, the first tuple with one 1 and four 0's has a \ise score of 
%     $\dfrac{\frac{1}{2}*((0-0+4)/4)}{1}=\frac{1}{2}=0.5$. The second tuple with two 1's, one 0, and two (-1)'s has a \ise score of  $\dfrac{\frac{1}{2}*((1-2+4)/4)}{1}=\frac{3}{8}=0.375$. Thus, the first tuple, even though it has fewer 1's, has a higher \ise score than the second tuple due to erroneous values in the second tuple.
% \end{example}
% \section{Table Reclamation}\label{sec:reclamation}

% With the set of originating tables found from Table Discovery, we now enter the Table Reclamation step. Given the set of Table Operators defined in Section~\ref{subsec:table-operators}, we perform table integration to produce a reclaimed source table.

\subsection{Table Reclamation via Integration}\label{subsec:integration}
%Table Integration takes 
With the set of discovered originating tables, %returned from Table Discovery, 
we now integrate them to reproduce the Source Table as closely as possible. Using the set of representative operators $\bigoplus = \{\uplus,\sigma, \pi, \kappa, \beta\}$ (Theorem~\ref{thm-ops})%from Section~\ref{subsec:table-operators})
, we present an efficient algorithm to  explore the integration space to reclaim the Source Table. 
\revision{%Instead of traversing through all possible join paths in a space consisting of all tables, as in existing by-example and by-target work, we apply Outer Union and unary operators with conditions based on a Source Table. \grace{after revisions, add back in!!}
}

Given a set of originating tables ($\bigT$), a Source Table ($S$) and using $\bigoplus$, \name outputs a table ($T_{\mathsf{result}}$) that reclaims $S$ as best as possible. 
%Note that from the previous discovery phase, each table in $\bigT$ adopts their column names from $S$ for every column that shares values with a column in $S$. 
Our table integration method is depicted in Algorithm~\ref{alg:table_integ}.
%as inputs (Line \ref{line:integ_input}), and produces the integration result as a candidate table that ideally reclaims all of $S$'s values (Line \ref{line:integ_output}). 
%\grace{Key Discussion}
From Section~\ref{subsec:candidate_table_retrieval}, we ensure that all candidate tables (and thus %its subset of 
originating tables) contain $S$'s columns.
% We currently assume that all originating tables given as input to the Table Integration step have been integrated to contain $S$'s columns. Later in Section~\ref{sec:discovery}, we describe Table Integration's previous step in the pipeline, Table Discovery. There, when we discuss how a set of candidate tables is refined to a set of originating tables, we will describe how we achieve this goal of ensuring that all originating tables contain $S$'s columns.

% \noindent{\bf Preprocessing:} 
\subsubsection{Preprocessing} First, we project out columns not in $S$ ($\pi$), and 
% for tables that contain a key column from $S$,
select tuples whose values are in the source key column  ($\sigma$). 
Hence, we only keep columns and tuples that overlap with $S$ (ProjectSelect(), Line~\ref{line:integ_ps}).
% Then, for efficiency, we take any candidate table without the source join key and greedily find a best way to join it with candidates that include the source key.  We use standard join cardinality estimation to find a path that covers the most source key values and is a close to functional as possible (Expand() on Line~\ref{line:join}).  Note that this step could be replaced with outer union (as per Theorem~\ref{thm-ops}), but this would make our similarity computation (Section~\ref{subsec:tuple_alignment}) expensive (potentially exponential).\footnote{The Expand algorithm is included in the technical report~\cite{gen-t_technical_report}.}
%Next, w
We then union all originating tables that share the same schema (InnerUnion(), Line~\ref{line:integ_iu}) to reduce the space of tables we need to explore. %Recall outer union on such tables is the same as inner union.
To prevent over-combining tuples that share nulls with tuples in $S$, for each table $T \in \bigT$, we find tuples in $S$ that share key values and contain nulls in the same columns, and replace these nulls in $T$ with unique labeled non-null values
% label these nulls with a dummy non-null value 
(LabelSourceNulls(), Line~\ref{line:integ_labelNull}). 
% \renee{How do we know which values these are?  I think you've skipped some explanation.  We are assuming we know how all attributes align with each other and with $S$.  How is this done?}  \grace{revised the explanation, and added a sentence towards the beginning about aligning column names}
% Before integration,
%applying our set of integration operators $\bigoplus$, 
Finally,
we remove duplicate tuples, subsumed tuples ($\beta$), and take the resulting tuples of complementation ($\kappa$) (TakeMinimalForm() on Line~\ref{line:integ_minimalForm}).

\subsubsection{Integration}
% \noindent{\bf Integration:}
%In the integration (Lines \ref{line:integ}-\ref{line:subsump}),
We integrate all resulting tables $\bigT_{\cup}$, all of which contain the source key column.
\revision{% For each $T_i \in \bigT_{\cup}$, we outer union it with $T_{\mathsf{result}}$. Next, we check if applying Complementation($\kappa$) and Subsumption($\beta$) on $T_{\mathsf{result}}$ results in a table that is more similar to $S$ in the evaluateSimilarity() function. To do so, we find the \ise score and evaluate how similar the values in the resulting table are to $S$'s values.
 }
At each iteration of $T_i \in \bigT_{\cup}$, we outer union $T_i$ with the integrated result so far, $T_{\uplus}$.
Next, we check if applying Complementation($\kappa$) and Subsumption($\beta$) on $T_{\uplus}$ results in a table that has a higher \ise score in the evaluateSimilarity() function.
This lets us check if these operators are over-combining tuples (e.g., removing a subsumed tuple that is identical to a tuple in $S$) and decreasing the number of values shared with $S$.
\revision{%We do this iteratively for all tables from the input set.
}
After iterating through all tables from the input set, we revert the previous labeling of shared nulls with $S$ (RemoveLabeledNulls() on Line~\ref{line:integ_revertNull}).
To ensure that the resulting integrated table ($T_{\mathsf{result}}$) has the same schema as $S$, we add null columns in $T_{\mathsf{result}}$ for every column it does not share with $S$ 
(Line~\ref{line:padNulls}). 
Finally, we return the resulting integration as a 
possible reclaimed table. 

\begin{algorithm}[t]
    % \setstretch{0.95} % sets line height
    % \small
    \footnotesize
    \SetAlgoLined
    \LinesNumbered
    \textbf{Input}: $\bigT= \{T_1, \dots, T_n\}$: tables to integrate; $S$: the Source Table \label{line:integ_input}\\
    \textbf{Output}: $T_{\mathsf{result}}$: integration result \label{line:integ_output}\\ 
    % $\bigT \gets$ AlignColumns($\bigT, S$) \label{line:integ_renameCols}\algocomment{align columns in $\bigT$ with $S$}\;
    $\bigT \gets$ ProjectSelect($\bigT, S$)    \label{line:integ_ps}\algocomment{$\sigma, \pi$ $(T \in \bigT)$ on columns, keys in $S$}\;
    % $\bigT \gets$ Expand($\bigT, S$)    \label{line:join}\algocomment{join tables without source key}\;
    $\bigT_{\cup} \gets $ InnerUnion($\bigT$)    \label{line:integ_iu}\algocomment{Inner Union tables with shared schemas}\; 
    $\bigT_{\cup} \gets$ LabelSourceNulls($\bigT_{\cup}$)     \label{line:integ_labelNull}\algocomment{Label Nulls shared with Source Table}\; 
    % $T_{\uplus} \gets T_1 \uplus T_2 \uplus \dots \uplus T_n\qquad\qquad\quad\,$    \algocomment{Apply outer union $\uplus$}\; 
    $\bigT_{\cup} \gets$ TakeMinimalForm($\bigT_{\cup}$)     \label{line:integ_minimalForm}\algocomment{Apply $\beta, \kappa$ on each table}\; 
    $T_{\uplus} \gets \emptyset$\;
    \For {$T_i \in \bigT_{\cup}$}{ \label{line:integ}
        $T_{\uplus} \gets T_{\uplus} \uplus T_i$ \algocomment{Apply outer union $\uplus$}\;
		% \If{$S$ similarity does not decrease}{\label{line:integ_comp_coverage}
            \If{$\mathsf{evaluateSimilarity}(\kappa(T_{\uplus}), S) \geq \mathsf{evaluateSimilarity}(T_{\uplus}, S)$}{\label{line:integ_comp_coverage}
            $T_{\uplus} \gets \kappa(T_{\uplus})\qquad\qquad\qquad$ 	\algocomment{Apply complementation $\kappa$}\; 
        }
        % \If{$S$ similarity does not decrease}{\label{line:integ_subsump_coverage}
        \If{$\mathsf{evaluateSimilarity}(\beta(T_{\uplus}), S) \geq \mathsf{evaluateSimilarity}(T_{\uplus}, S)$}{\label{line:integ_subsump_coverage}
            $T_{\uplus} \gets \mathcal \beta(T_{\uplus})\qquad\qquad\qquad$ 	\algocomment{Apply subsumption $\beta$}\; \label{line:subsump}
        }
	}
    
    $T_{\mathsf{result}} \gets$
    RemoveLabeledNulls($T_{\uplus}$) \label{line:integ_revertNull}\;

    \If{$T_{\mathsf{result}}$ has fewer columns than $S$}{
        add null columns in $T_{\mathsf{result}}$ for each column $\in S \setminus T_{\mathsf{result}}$\label{line:padNulls}\;
    }
    Output $T_{\mathsf{result}}$ \;
	\caption{Table Integration}\label{alg:table_integ}
\end{algorithm}

\section{Experiments}\label{sec:experiments}
% We now present the evaluations, showing the effectiveness in Section~\ref{subsec:accuracy-exp}, scalability in Section~\ref{subsec:scalability}, and generalizability in Section~\ref{subsec:generalizability} of \name. \todo{show concrete numbers}\grace{I would like to show that \name outperforms baseline by XX\%. Which methods / benchmarks? Also for scalability}
% We now present evaluations on benchmarks with tables containing real instances (Source Tables) from the well-known TPC-H Benchmark~\cite{TPC} and  also the \tdBench~\cite{WDC} Benchmark.
% %\renee{what does it mean for these to be embedded in a data lake - you need to explain}\grace{resolved}, 
% We use these benchmarks
% %as well as benchmarks with these tables 
% along with tables from real data lakes, specifically 
% the large SANTOS benchmark~\cite{santos23} and a sample of WDC~\cite{DBLP:conf/www/LehmbergRMB16}. 
% There are no existing solutions for reclamation.  Hence, our baselines are versions of related techniques that we modify to solve the reclamation problem. Effectiveness experiments in Section~\ref{subsec:accuracy-exp} show that \name can perfectly reclaim 15-17 Source Tables, whereas most baselines only fully reclaim 1-3 Source Table across benchmarks. Scalability experiments in Section~\ref{subsec:scalability} show that \name is 5X faster than the next-fastest baseline on a large, real data lake. Finally, Section~\ref{subsec:generalizability} shows that \name generalizes to a different real-world application. 

\revision{We evaluate \name on benchmarks with tables containing real instances from well-known Benchmarks~\cite{TPC,WDC} along with tables from real data lakes~\cite{santos23,DBLP:conf/www/LehmbergRMB16}. Our baselines %are versions of related techniques that we 
are modified related techniques to solve the reclamation problem.} Section~\ref{subsec:accuracy-exp} shows that \name can perfectly reclaim 15-17 Source Tables, whereas most baselines only fully reclaim 1-3 Source Table across benchmarks. Section~\ref{subsec:scalability} shows that \name is 5X faster than the next-fastest baseline on a large %, real 
data lake. Finally, Section~\ref{subsec:generalizability} shows that \name generalizes to a different real-world application. 

\subsection{Experimental Setup}\label{subsec:exp_setup}
%\subsubsection{Environment}
We implement \name in Python on a CentOS server with Intel(R) Xeon(R) Gold 5218 CPU @ 2.30GHz processor.
\revision{
%The c
Code for \name and all the baselines is publicly available%on our repository
~\cite{gen-t_repo}.%\footnote{\url{https://github.com/northeastern-datalab/gen-t}}.
}
% \rjm{why not use citation 10 here as well for consistency and to save space?}
%\subsubsection{Datasets}
\label{subsubsec:datasets}
% To evaluate both effectiveness and scalability of \name, we use 6 benchmarks whose statistics are outlined in Table~\ref{tab:bench_stats}.%, generated from three main sources.
We evaluate \name on 6 benchmarks (%statistics shown in
see Table~\ref{tab:bench_stats}).

% \begin{table}[h]
%     \small
%     % \begin{subtable}{\linewidth}
%     \scalebox{0.95}{
%     \begin{tabular}{lcccccc}\\ \toprule
%     Benchmark & \# Tables & \# Cols & Avg Rows   & Size (MB) \\ \midrule
%     \tpchSmallBench    & 32   & 244 & 782 & 3 \\
%     % \tpchMedBench    & 32   & 244 & 10,827 & 40 \\
%     \tpchMedBench    & 32   & 244 & 10.8K & 40 \\
%     % \quad+\santosLargeBench  & 11,118   & 122,040 & 7,683 & 11K \\
%     \santosLargeBench+\tpchMedBench  & 11K   & 122K & 7.7K & 11K \\
%     % \tpchLargeBench   & 32   & 244 & 1,082,653 & 3.9K \\
%     \tpchLargeBench   & 32   & 244 & 1M & 3.9K \\
%         \midrule
%     \tdBench  & 515   & 2,147 & 74 & 4 \\ 
%     \wdc+\tdBench  & 15K   & 75K & 14 & 66 \\\bottomrule
%     \end{tabular}}
%     \caption{Statistics on Data lakes of each benchmark}\label{subtab:bench_stats_dl}
%     % \end{subtable}

%     % \begin{subtable}{\linewidth}
%     % \scalebox{0.9}{
%     % \begin{tabular}{lcccc}\\ \toprule
%     % Benchmark & \# Reclaimable Sources & Avg Cols & Avg Rows & Avg Vals\\\midrule
%     % \tpchSmallBench    & 26 & 9 & 27 & 211\\
%     % \tpchMedBench    & 26  & 9 & 1,171 & 8,303\\
%     % \tpchLargeBench   & 26  & 9 & 1,171 & 8,303\\
%     %  \midrule
%     % \tdBench  & 0 & --- & --- & --- \\ \bottomrule
%     % \end{tabular}}
%     % \caption{Statistics on Source Tables of each benchmark \roee{Since this is very repetitive, maybe we can just mention it in the text?}}\label{subtab:bench_stats_st}
%     % \end{subtable}
%     % \caption{Benchmark Statistics}
%     \label{tab:bench_stats}
% \end{table}

% \vspace{.05in}
\begin{table}[ht]
    \small
    % \vspace{-.15in}
    % \vspace{-.1in}
    \scalebox{0.8}{
    \begin{tabular}{lcccccc}\\ \toprule
    Benchmark & \# Tables & \# Cols & Avg Rows   & Size (MB) \\ \midrule
    \tpchSmallBench    & 32   & 244 & 782 & 3 \\
    \tpchMedBench    & 32   & 244 & 10.8K & 40 \\
    \tpchLargeBench   & 32   & 244 & 1M & 3.9K \\
        \santosLargeBench+\tpchMedBench  & 11K   & 122K & 7.7K & 11K \\
        \midrule
    \tdBench  & 515   & 2,147 & 74 & 4 \\ 
    \wdc+\tdBench  & 15K   & 75K & 14 & 66 \\\bottomrule
    \end{tabular}}
    \caption{Statistics on Data lakes of each benchmark}\label{subtab:bench_stats_dl}
    % \vspace{-.1in}
    \label{tab:bench_stats}
\end{table}
\noindent\textbf{\tpch Benchmarks:} We use the 8 tables from the TPC-H benchmark~\cite{TPC}, which contain business information including customers, products, suppliers, nations, etc. Using these tables, we create three versions of a benchmark suite titled TP-Table Reclamation (\tpch). 
\tpchLargeBench has TPC-H tables with original table sizes. % using scale factor of 1. 
%resolved
%\rjm{using scale factor of 1}, \renee{Hmm, the point of TPC-H is to test scalability so there is not a single "original size".  Perhaps state which scale factor you are using?  Looks like the smallest or scale factor of 1 - where lineitem has 6M tuples and other dimension tables are smaller?  Confirm or correct.} \grace{We create 3 \tpch benchmarks such that the largest has a scale factor of 1 (e.g. lineitem has 6M tuples) and the smallest scales down to 1000th of its size (e.g. linitem has 6K tuples)}
\tpchMedBench has TPC-H tables that are each 1/100 of its original table's rows, 
and \tpchSmallBench has TPC-H tables that are each $\sim$1/1000 of its original table's rows. 
In each,  we take each of the 8 tables and create 4 versions of the same table -- creating 32 tables in total.
%  (detailed in Table~\ref{subtab:bench_stats_dl})
For two versions, we randomly nullify different subsets of values, and for the other two versions, we randomly inject different non-null (erroneous) values in different subsets of values.  
For the majority of the experiments, the number of nulls (respectively,  erroneous values) is 50\% meaning that we randomly take 50\% of each table's values and replace them with nulls (respectively, replace them with different new strings).  In an ablation study (Section~\ref{subsec:accuracy-exp} last two paragraphs), we vary (independently) the number of nulls/erroneous values from 10\% to 90\%.   Our goal in the table discovery phase is then to filter out originating tables with injected non-null noise, so that resulting reclaimed tables do not contain erroneous values. Thus, we 
seek to verify that our approach uses the nullified versions rather than the erroneous versions so that combining them can %maximally 
reproduce the Source Table.

%{\bf Source Tables for TPC:}
The Source Tables for the \tpch benchmarks are created using all 8 of the original (unmodified) TPC-H tables over which we randomly generated 26 queries 
%queries that produce the Source Tables that we aim to reclaim, from the 8 original tables (without injected nulls and noisy values) of each benchmark. To ensure variations of SPJU queries with no aggregations or string-transformations involved, we create 26 queries, 
each having a subset of operators \{$\pi, \sigma, \bowtie, \leftouterjoin, \fullouterjoin, \cup, \uplus$\}. In these 26 queries, the number of operations ranges from 2 (just $\pi, \sigma$), to 9, such that the query with the maximum number of 
unions contains 4 unioned tables, and the query with the maximum number of joins joins 3 tables. We ran the same queries on each \tpch benchmark to create
26 Source Tables for the \tpchSmallBench benchmark containing an average of 9 columns and 27 rows, and 26 Source Tables for the \tpchMedBench and \tpchLargeBench benchmarks that have an average of 9 columns and 1K rows.

% \vspace{.05in}
\noindent\textbf{\santosLargeBench+\tpchMedBench Benchmark:}
In addition, to further assess the effectiveness and scalability of our table discovery method, we embed \tpchMedBench into a real, large data lake \santosLargeBench~\cite{santos23}. 
In doing so, we evaluate how well \name prunes a potentially large set of candidate tables retrieved from a large data lake to a smaller set of originating tables that can more accurately reclaim a Source Table when integrated.  We use the same Source Tables as for \tpchMedBench. 
\noindent\textbf{\tdBench Benchmark:} In addition, we explore the real-world application of our method with the \tdBench Benchmark~\cite{WDC}, which takes web tables and matches them to properties from DBpedia. This benchmark was not originally created for the problem of Table Reclamation, so we test the generalizability of \name by seeing if it can reclaim any of this benchmark's tables.
We take 515 raw tables that contain some non-numerical columns and a key column. We do not have prior knowledge of whether or not any of these 515 tables can be ``reclaimed'' as a Source.
%no Source is known to be able to be perfectly reclaimed from a subset of tables in the benchmark (0 Reclaimable Sources). 
Thus, we iterate through each of the 515 tables as potential sources. 

% \vspace{.05in}
\noindent\textbf{\wdc+\tdBench:} 
To further assess the effectiveness of \name, we embed \tdBench tables into a sample of the WDC web table corpus~\cite{DBLP:conf/www/LehmbergRMB16}, which contains ~15K relational web tables.
This way, we can examine how well \name prunes a large set of candidate tables found from a large table corpus to a small set of originating tables that can be integrated to reclaim a Source Table.
\subsubsection{Baselines}\label{subsubsec:baselines}
% For all experiments, w
\revision{
We compare \name %against 
to the %current 
state-of-the-art for \textit{by-target synthesis}, 
Auto-Pipeline~\cite{DBLP:journals/pvldb/YangHC21}, 
the state-of-the-art for \textit{Query-by-example}, Ver~\cite{Gong2023Ver},
and the state-of-the-art for table integration, \alite~\cite{Khatiwada2022IntegratingDL} which were %ere we have 
modified %them to be solutions 
for the reclamation problem.}
% \revision{We provide code for our baselines in our repository~\cite{gen-t_repo}.}

Auto-Pipeline has a similar framework to our problem in discovering the integration (query or pipeline) that reclaims a Source Table. However, \name does not assume to have the perfect set of input tables from which we can synthesize the query that reproduces the Source Table. 
Auto-Pipeline has both query-search and deep reinforcement learning approaches, but since we propose an unsupervised approach, we use the query-search variation as our baseline. Auto-Pipeline's code implementation is not openly available, so we 
adopted an open re-implementation of their search approach~\cite{DBLP:journals/pvldb/ShragaM23}, which adapts the framework in Foofah~\cite{DBLP:conf/sigmod/JinACJ17}, 
and, for a fair comparison, revised their set of table operators to only contain table operators that \name considers (\{$\sigma, \pi, \cup, \bowtie, \leftouterjoin, \fullouterjoin$\}).
We call this re-implemented, adapted baseline \textbf{\aPip}. Since Auto-Pipeline's benchmarks contain small tables, and most of their operators are string-transformation operators, we do not consider their benchmarks for our experiments.

\textbf{\ver} has a similar objective of discovering and integrating tables to produce a table that contains the source table and other similar tuples. In contrast, \name aims to reproduce only the tuples from the source table. 
\ver takes small source tables as input (e.g., source tables with 2 columns and 3 rows~\cite{Gong2023Ver}), so we query \ver with two columns from the Source Tables. We evaluate the output table for each run, and aggregate the results to evaluate the entire source table. 

% To show 
We validate the need for our Matrix Traversal rather than directly integrating the set of candidate tables returned from Set Similarity (Section~\ref{subsec:candidate_table_retrieval}), 
%\renee{Have you stated what Set Similarity is?  So far I think you only mentioned Starmie in Sec 4.}\grace{added reference} 
%we also compare 
by comparing against \textbf{\alite} provided with the set of candidate tables from Set Similarity as input. %Also, since \name first projects and selects on the Source Table's columns and keys before performing integration, we 
We also compare with a variation of \alite, which we call \textbf{\aliteps}, that, similar to \name, first performs projection and selection to match the Source Table before performing table integration.
% Hence, \aliteps, also includes unary operators to make its performance more comparable with \name.
\alite without project %ion
 and select %ion 
  is much slower as it creates a %much 
  larger integration result.%, hence \aliteps is a fairer comparison %from a performance point of view.}
% in terms of performance.

For all
baselines on the \tpch benchmarks, we create another variant in which we give each method a specific integrating set (int. set) 
of tables as input, rather than the full set of candidate tables returned from Set Similarity. %Since we know what subset of tables from the 8 original tables were used to create the 26 Source Tables, we know that a perfect reclamation of each Source Table contains variants of these tables. 
We know what subset of tables from the 8 original tables were used to create the 26 Source Tables, so, we know that a perfect reclamation %of each Source Table 
contains variants of these tables. Thus, for all original tables used to create each Source Table, the integrating set includes all variations (2 nullified and 2 erroneous versions of each original table) of these tables.
% This set includes all variations of expected tables \renee{expected table?  different from expected candidate?}, or tables used in the original creation of the Source Tables. It contains variants of the tables with nullified values, and tables with erroneous values with respect to the Source Table.

\subsubsection{Metrics}\label{subsubsec:metrics}
%\roee{I changed to \cap{S} instead of T for consistence.}
For effectiveness, we evaluate how many values in a Source Table have been reclaimed, or how similar the values in the reclaimed table are to those of the Source Table. Thus, the Source Tables are essentially our ground truth in that we see how many of its values we can reproduce. 
In an aligned tuple in a reclaimed table with respect to a Source Table, it contains an erroneous value if there is a non-null, different value at a given column compared to a value in the same column in the Source Table. Similarly, it contains a nullified value if it contains a null value in a column where the Source tuple contains a non-null value (see Section~\ref{subsec:tuple_alignment}).

\begin{table*}[!ht]
\centering
\scalebox{0.9}{
% \hspace{.3in} 
\begin{tabular}{l|cccc|cccc|cccc}\hline
 & \multicolumn{4}{c|}{\tpchMedBench} & \multicolumn{4}{c|}{\santosLargeBench+\tpchMedBench} & \multicolumn{4}{c}{\tpchLargeBench}\\
% Method & Recall & Precision & Recall & Precision & Recall & Precision \\  \hline \hline
Method & Rec & Pre & Inst-Div.& \dkl & Rec & Pre & Inst-Div.& \dkl & Rec & Pre & Inst-Div.& \dkl \\  \hline \hline
\alite &    0.662 & 0.202 & 0.100 & 35.831 & --- & --- & --- & ---& --- & --- & --- & ---\\
\alite w/ int. set &  0.694 & 0.202 &  0.085 & 36.348 & 0.694 & 0.202 & 0.085 & 36.348 & --- & --- & --- & ---\\
\aliteps &  0.880 & 0.556 & 0.009 & 3.524 & 0.842 & 0.554 & 0.011 & 4.629 & 0.775 & 0.521 & 0.049 & 21.978\\ 
\aliteps w/ int. set &  0.880 & 0.569 &  0.009 & 3.524 & 0.880 & 0.569&  0.009 & 3.524 & 0.880 & 0.569 & 0.009 & 3.524\\ \midrule
\name & \textbf{0.976} & \textbf{0.867} & \textbf{0.004} & \textbf{1.326} & \textbf{0.976} & \textbf{0.867} & \textbf{0.004} & \textbf{1.326} & \textbf{0.971} & \textbf{0.807} & \textbf{0.004} & \textbf{1.490} \\\bottomrule
\end{tabular}}
% \newline
% \vspace*{.05in}
% \newline
% \scalebox{0.9}{\begin{tabular}{l|cc|cc|cc}\hline
%  & \multicolumn{2}{c|}{\tpchMedBench} & \multicolumn{2}{c|}{\santosLargeBench+\tpchMedBench} & \multicolumn{2}{c}{\tpchLargeBench}\\
% Method & Inst-Div.& \dkl & Inst-Div.& \dkl & Inst-Div.& \dkl \\  \hline \hline
% \alite &    0.100 & 35.831 & --- & --- & --- & ---\\
% \alite w/ int. set &  0.085 & 36.348 & 0.085 & 36.348 & --- & ---\\
% \aliteps &  0.009 & 3.524 & 0.011 & 4.629 & 0.049 & 21.978\\ 
% \aliteps w/ int. set &  0.009 & 3.524 & 0.009 & 3.524 & 0.009 & 3.524\\\midrule
% \name & \textbf{0.004} & \textbf{1.326} & \textbf{0.004} & \textbf{1.326} & \textbf{0.004} & \textbf{1.490} \\\bottomrule
% \end{tabular}}
\caption{
Effectiveness
of \name and baselines \alite, \aliteps on the larger \tpch benchmarks. \alite, \aliteps, and \name are given the same set of candidate tables from Set Similarity, and \alite and \aliteps are also given an integrating set (``w/ int. set''). If there are no results for some method, then it timed out for most, if not all, Source Tables.}
\label{tab:effectiveness}
% \vspace{-.15in}
\end{table*}

\noindent\textbf{Precision and Recall:} Consider a Source Table $S$ and %resulting 
an output reclaimed table $\hat{S}$.
% , let $|S|$ be the expected output size (\# tuples in $S$), and $|\hat{S}|$ be the reclaimed table size of the method result (\# tuples in $\hat{S}$). 
From the measure Tuple Difference Ratio (TDR)~\cite{Khatiwada2022IntegratingDL}, we derive two similarity measures, Recall (Rec) and Precision (Pre), that measure the \# of tuples in the intersection of $S$ and $\hat{S}$ relative to the \# of tuples in each table. \\
$\text{Rec} = |S\cap{\hat{S}|}/{|S|}$ and $\text{Pre} = |S\cap{\hat{S}|}/{|\hat{S}|}$.
% \renee{hmm you can't take the intersection of two numbers.  I think in the numerator  the bars are in wrong place - we are intersecting the sets, not their cardinality.  $|S \cap \hat{S} |$}\grace{resolved}
% \begin{equation}
%     \text{Recall} = \dfrac{|S\cap{\hat{S}|}}{{|S|}},
%     \text{Precision} = \dfrac{|S\cap{\hat{S}|}}{{|\hat{S}|}}
% \end{equation}

In addition to metrics that measure the similarity between the tuples of
%rjm the resulting 
a reclaimed
table and a Source Table, we also include finer-grain metrics that measure the number of values that do not match within aligned tuples (tuples with the same key value). If there are multiple aligned tuples with respect to one tuple in the Source Table (multiple tuples in the reclaimed table with the same key value), then we consider the tuple that contains the largest number of column values shared with the corresponding tuple in the Source Table. This way, there is at most 1 aligned tuple in the reclaimed table for each tuple in the Source Table.
%how much the values in the resulting table diverge from those in the Source Table. 
In these measures, which we denote as \textit{divergence measures}, the ideal score is 0 (the reclaimed table is identical to the Source Table).
% \renee{see earlier comment -- what does instance-equivalent mean?  Earlier it was used for queries rather than tables -- for tables don't you just mean "identical" for tables}\grace{resolved}.
% Specifically, we measure how 
% %rjm much of the 
% many tuples from the Source Table are not found in the 
% %rjm resulting 
% reclaimed table (using \textit{Instance Divergence}), and how 
% % \renee{many, not much} 
% many values found in 
% %rj the resulting tables 
% reclaimed tuples differ from those in the Source Table (using \textit{Conditional KL-Divergence}). This 
% %rjm is in part 
% enables us
% to measure the nullified and erroneous values in the reclaimed table's aligned tuples, with respect to the Source Table (see Section~\ref{subsec:tuple_alignment}). 
%as defined in Section~\ref{subsec:tuple_alignment}.
Specifically, we introduce \textit{Instance Divergence} and \textit{Conditional KL-Divergence}, enabling us
to measure the nullified and erroneous values in the reclaimed table's aligned tuples, with respect to the Source Table (see Section~\ref{subsec:tuple_alignment}). 

\noindent\textbf{Instance Divergence:} We measure how many missing values there are in each aligned tuple, with respect to its corresponding tuple in the Source Table. 
To do so, we use the inverse of Instance Similarity~\cite{DBLP:journals/vldb/AlexeHPT12} (see Equation~\ref{equ:instance-similarity}) and introduce \emph{Instance Divergence:}
$ \text{Inst-Div.} = 1 - \text{Instance Similarity}$.

\noindent\textbf{Conditional KL-divergence:} %Finally, w
We aim to capture how many erroneous values there are in aligned tuples from a reclaimed table with respect to tuples in a Source Table.
%Thus, we also consider the Conditional KL-divergence of a reclaimed table, with respect to a Source Table, conditioned on the probability that the key values from the Source Table are found in the reclaimed table. 
We adopt the traditional definition of conditional KL-divergence~\cite{DBLP:books/daglib/0016881,DBLP:books/daglib/0021593}, and %also 
add a penalization for erroneous values, such that the score is higher (diverges more) for 
reclaimed tables containing erroneous values as opposed to nulls in their aligned tuples with the Source Table (see 
% Appendix~\ref{app:metrics}
the technical report~\cite{gen-t_technical_report}
for more details).
% \roee{Consider introducing the measures where we introduce the definition of Value Divergence Score}
% For all effectiveness metrics, we take the average scores across all Source Tables.
We report average scores over all Source Tables.

\noindent\textbf{Efficiency Measures:} For efficiency, we measure the average runtimes for all Source Tables, as well as the average ratio of the output 
%rjm resulting 
reclaimed table's size to the 
%rjmexpected output 
Source Table's size (large integrations can significantly increase runtimes). 
%resolved
%\renee{you've lost me, I don't understand the next sentence and why are we talking about timeouts before discussing any other times?}\grace{should I move this to the second paragraph in Section~\ref{subsec:accuracy-exp}?}

\subsection{Effectiveness}\label{subsec:accuracy-exp}

\begin{table}[!t]
\centering
\scalebox{0.9}{
\begin{tabular}{l|cc|ll}\hline
% Method & Recall & Precision  & Inst-Div.&\dkl \\  \hline \hline
Method & Rec & Pre  & Inst-Div.&\dkl \\  \hline \hline
\alite & 0.704 & 0.128  & 0.095 &1.332 \\
\alite w/ int. set & 0.745 & 0.133  & 0.086 &1.197 \\
\aliteps & 0.805 & 0.539  & 0.040 &0.655 \\ 
\aliteps w/ int. set & 0.833 & 0.552  & 0.037 &0.688 \\ 
\aPip & 0.674 & 0.272  & 0.158 &2.574 \\ 
\aPip w/ int. set & 0.683 & 0.289  & 0.133 &2.109 \\
  \revision{\ver w/ int. set} &  \revision{0.746} & \revision{0.351} & \revision{0.127} & \revision{10.393}\\
  \midrule
\name & \textbf{0.954} & \textbf{0.799}  & \textbf{0.015} &\textbf{0.165} \\\bottomrule
\end{tabular}}
% \newline
% \vspace*{.05in}
% \newline
% \scalebox{0.9}{}
\caption{\revision{
Similarity (Rec, Pre) and Divergence (Inst-Div., \dkl)
of \name and baselines on \tpchSmallBench benchmark.}}
\vspace{-.15in}
\label{tab:small-effectiveness}
% \end{table*}
\end{table}

% Tables~\ref{tab:small-effectiveness} and ~\ref{tab:effectiveness} report the similarity and divergence scores for all four \tpch benchmarks across all methods.
\revision{
Tables~\ref{tab:small-effectiveness} and~\ref{tab:effectiveness} report the results for all methods on the \tpchSmallBench benchmark and the larger \tpch benchmarks, respectively. 
}
For experiments
%resolved \renee{still don't know what an expected candidate is}\grace{removed} 
on \tpchSmallBench, \tpchMedBench, and \tpchLargeBench benchmarks, we input candidate tables discovered from just Set Similarity (Section~\ref{subsec:candidate_table_retrieval}). For experiments on \santosLargeBench+\tpchMedBench, we first discover relevant tables from the large data lake using Starmie~\cite{starmie22}, a state-of-the-art self-supervised system for scalable table discovery. %We choose to use Starmie in our experiments since it is  scalable to large data lakes.
%rjm and their tables. 
Hence, it can discover a set of candidate tables for the Source Table from a large data lake. Although the primary use case of Starmie was table union search, it %has been
was shown to apply to other search semantics such as table discovery to improve the performance of downstream machine learning tasks via feature discovery (join search) and column clustering. 
% After running Starmie, 
Following Starmie, we %then 
run Set Similarity to find syntactically similar tables among the returned tables from Starmie.
\revision{
We run baselines \aPip and \ver (w/ int. set) only on \tpchSmallBench benchmark (Table~\ref{tab:small-effectiveness}), since they timeout for most, if not all Source Tables in the larger benchmarks. \ver times out when given the entire data lake from \tpchSmallBench.
}
\alite times out only on \tpchLargeBench benchmark (Table~\ref{tab:effectiveness}). We discuss scalability and timeouts in Section~\ref{subsec:scalability}.

% Some baselines timeout for most Source Tables as the sizes and/or number of tables increase across the benchmarks. \aPip times out for every benchmark except for \tpchSmallBench, and \alite times out on \tpchLargeBench benchmark. We discuss scalability and timeouts in Section~\ref{subsec:scalability}.

Across all benchmarks, \name outperforms the baselines for all metrics, while perfectly reclaiming 15-17 Source Tables across all benchmarks. Baselines \aliteps and \aPip only perfectly reclaim 3 Source Tables and 1 Source Table, respectively, across the benchmarks on which they do not time out, and \alite and \ver does not perfectly reclaim any.
In fairness, \alite is an integration method that does not consider the Source Table (it is not "target-driven" like \aPip). Also, \ver is a QBE method whose goal is to produce source tuples along with many additional tuples.
In terms of similarity (Table~\ref{tab:effectiveness}), \name outperforms the top performing existing baseline method (\aliteps) by 11-25\% in Recall and by 48-56\% in Precision %; \name outperforms \aliteps by 10-19\% in Recall and by 56-57\% in Precision; and \name outperforms \aPip by 42\% in Recall and by 208\% in Precision 
across all \tpch benchmarks.
For the divergence measures, we see that \name produces tables that contain 
%rjm less -- less oxygen, fewer values :-)
%\rjm{
fewer %} 
nullified values in its aligned tuples with respect to the Source Table (Inst-Div.), as well as 
%rjm less 
fewer erroneous values in its aligned tuples, which is reflected in the lower \dkl scores than the baselines. 
% \grace{
% Notice that as the data lake and table sizes grow larger across %\tpchMedBench, \santosLargeBench+\tpchMedBench, and \tpchLargeBench 
% benchmarks (left to right in Table~\ref{tab:effectiveness}), \name's results remain fairly consistent whereas baseline \aliteps's similarity results decrease and divergence scores increase, especially the KL-divergence measure.
% }

Even compared to each baseline that is given specified integrating sets of tables rather than large sets of candidates (`w/ int. set'), \name performs much better. 
% For example in Table~\ref{tab:small-effectiveness}, \name outperforms Ver. Ver produces tables that contain source tuples as well as many additional tuples (hence lower precision), which is the goal of Query-By-Example. 
Thus, the matrix traversal method (Section~\ref{subsec:ternary-matrix}) used in \name to refine the set of originating tables works well in filtering out misleading tables that could be integrated to produce tables containing erroneous values. We provide %examples and 
benchmark samples to exemplify this in our repository~\cite{gen-t_repo}.%\footnote{\url{https://github.com/northeastern-datalab/gen-t}}
%We now provide two examples, illustrating interesting cases from the benchmark. Benchmark samples corresponding to these examples are provided in our repository.\footnote{\url{https://github.com/northeastern-datalab/gen-t}}
% \roee{Trimming suggestion: let's choose one of the examples to leave here and the other will be provided in the technical report. Which one is more "attractive"?}

\begin{figure}[t]
\centering
    % \vspace{-.15in}
    {
    \centering
    \begin{minipage}[t]{\linewidth}
    \includegraphics[width=\linewidth]{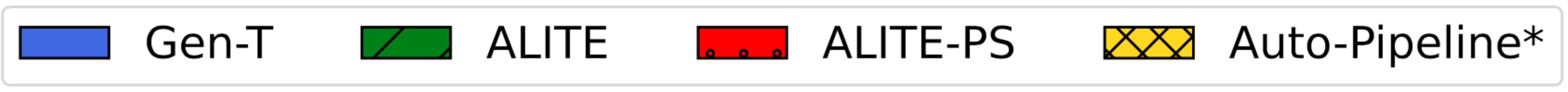}
    \end{minipage}
    }
    \subfloat[Recall on \tpchSmallBench]{
    \begin{minipage}[t]{0.4\linewidth}
    \includegraphics[width=\linewidth]{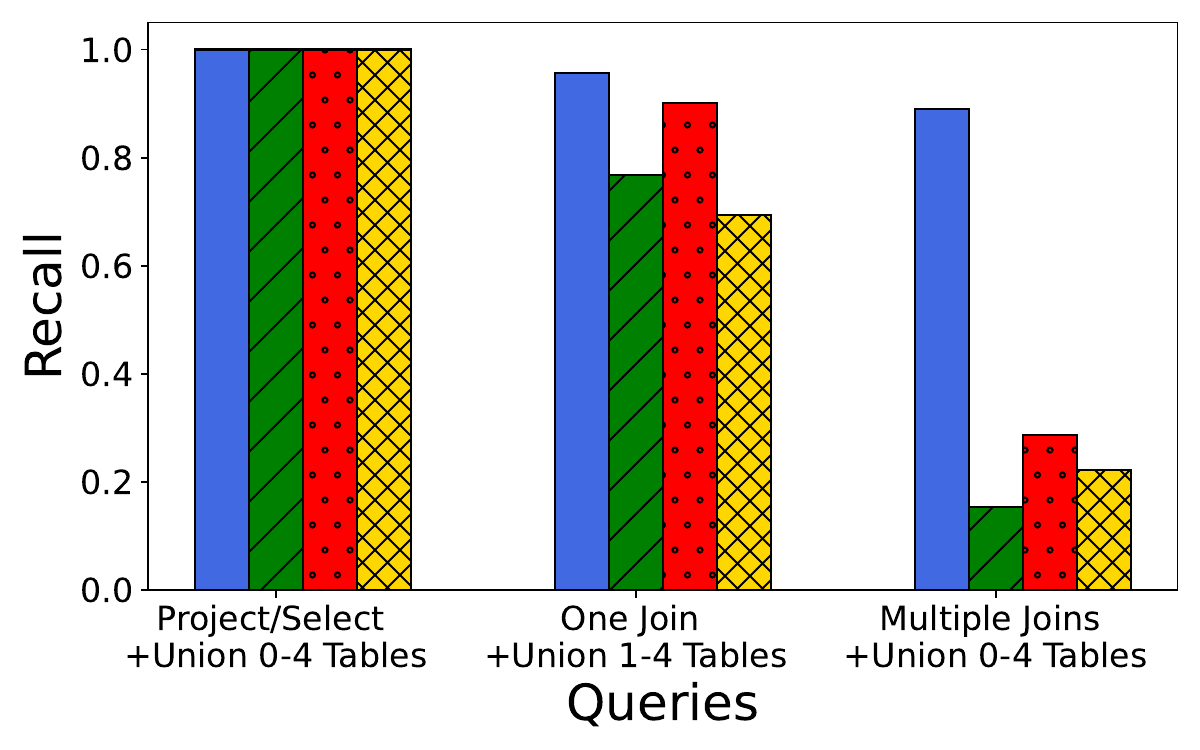}
    \end{minipage}
    }
    \subfloat[Prec. on \tpchSmallBench]{
    \begin{minipage}[t]{0.4\linewidth}
    \includegraphics[width=\linewidth]{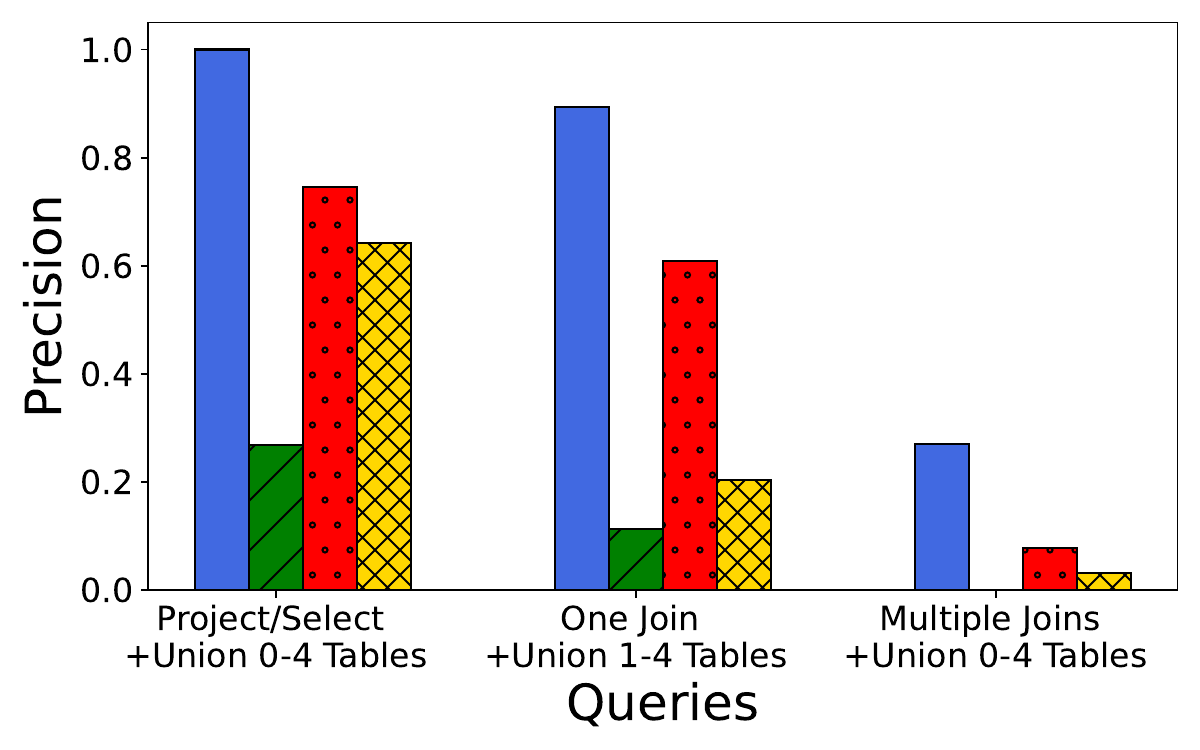}
    \end{minipage}
    }
    \hfill
    \subfloat[Recall on \tpchMedBench]{
    \begin{minipage}[t]{0.4\linewidth}
    \includegraphics[width=\linewidth]{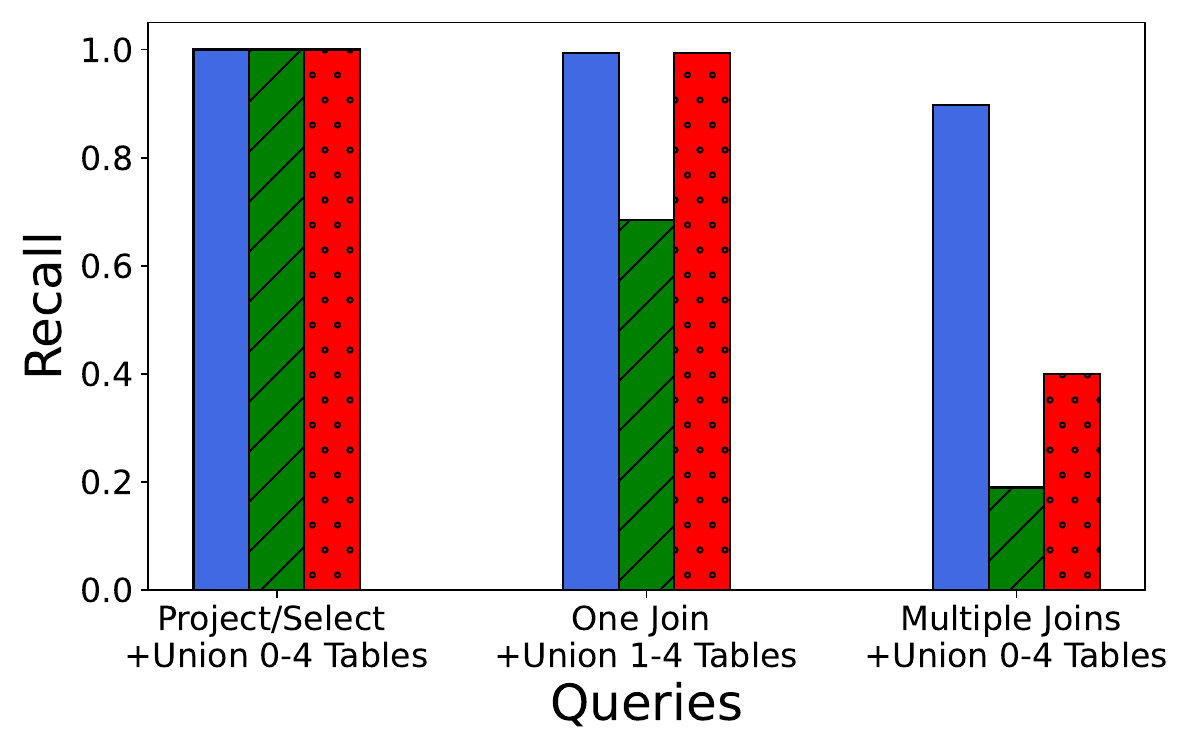}
    \end{minipage}
    }
    \subfloat[Prec. on \tpchMedBench]{
    \begin{minipage}[t]{0.4\linewidth}
    \includegraphics[width=\linewidth]{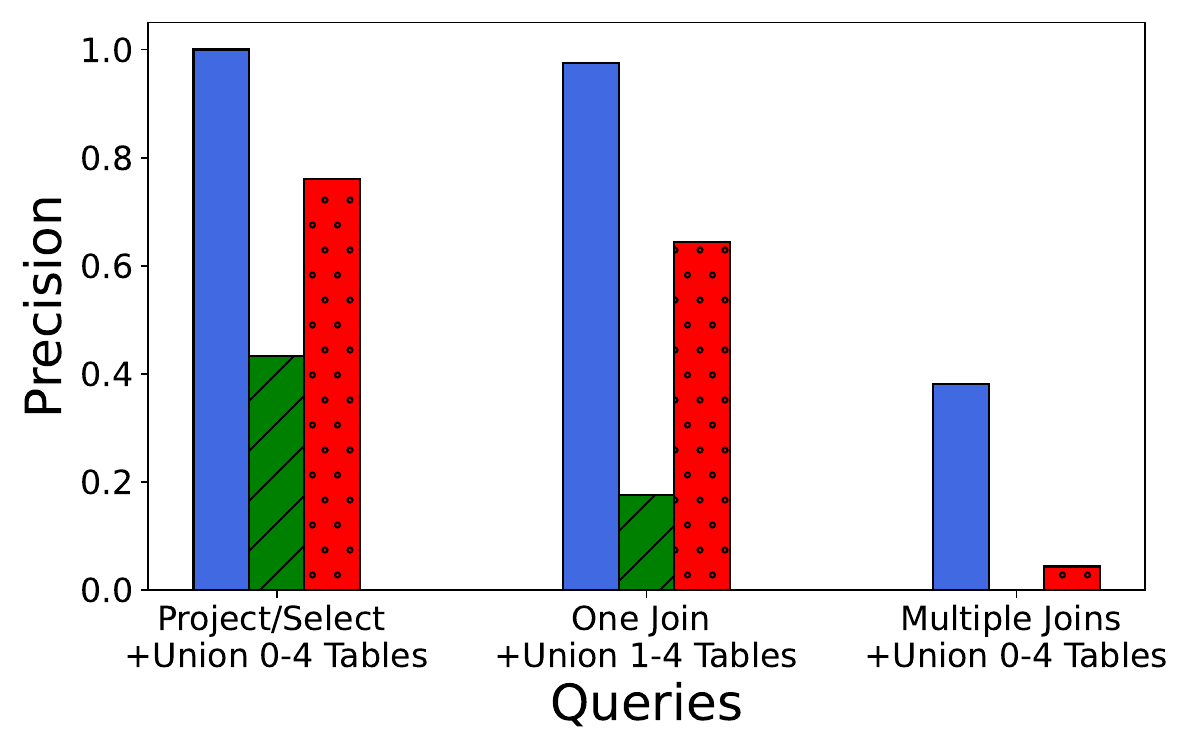}
    \end{minipage}
    }
    \hfill
    \subfloat[Recall on \tpchLargeBench]{
    \begin{minipage}[t]{0.4\linewidth}
    \includegraphics[width=\linewidth]{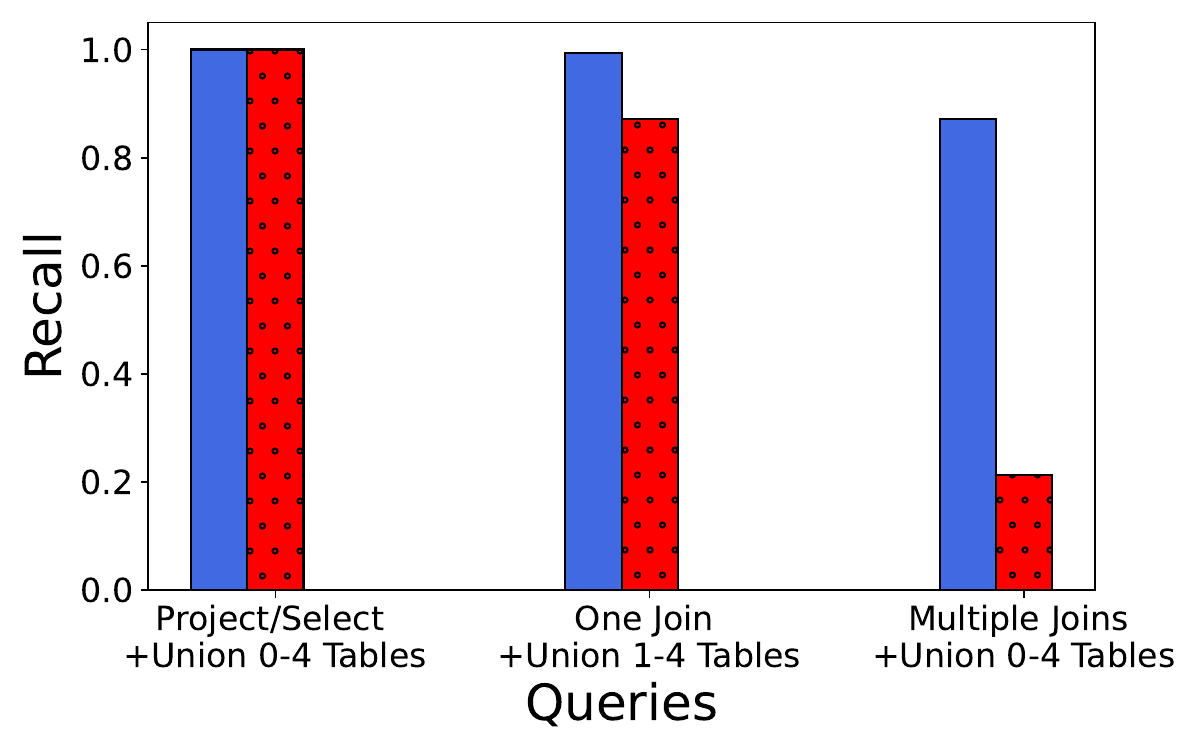}
    \end{minipage}
    }
    \subfloat[Prec. on \tpchLargeBench]{
    \begin{minipage}[t]{0.4\linewidth}
    \includegraphics[width=\linewidth]{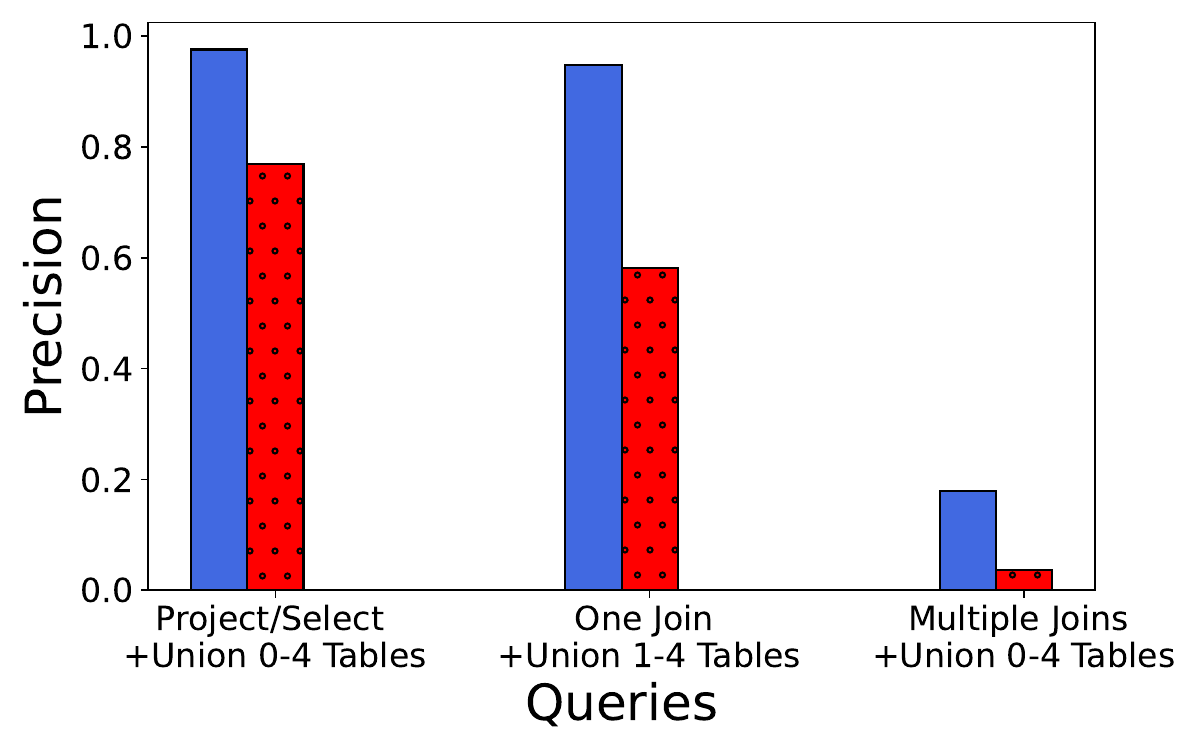}
    \end{minipage}
    }
    
    % \caption{Recall and Precision Analysis of all baselines on \tpchSmallBench, \tpchMedBench, \tpchLargeBench Benchmarks, for different types of queries that produce Source Tables.}
    \caption{Recall and Precision of different types of queries that produce Source Tables over the \tpch Benchmarks.}
    \label{fig:sim_plots}
    \vspace{-.15in}
\end{figure}

%To better understand the performance of the methods on different types of queries used to initially create the Source Tables, w
We now perform an analysis of the similarity measures for all methods on different types of queries used to form the Source Tables in \tpch %\tpchSmallBench, \tpchMedBench, and \tpchLargeBench 
benchmarks, shown in Figure~\ref{fig:sim_plots}. Ranging from simple queries (that just perform Projection, Selection, and Union) to more complex queries (joining up to 3 tables and unioning up to 4 tables), we see that \name outperforms the baselines on queries of all complexities used to initially create the Source Table. Thus, not only is the matrix traversal 
effective, but the set of operators used in table integration %(Section~\ref{subsec:table-operators}) works well in 
represents
%ing 
different types %and complexities 
of queries well.

\begin{figure}[t]
    \vspace{-.05in}
    \centering
    \includegraphics[width=0.7\linewidth]{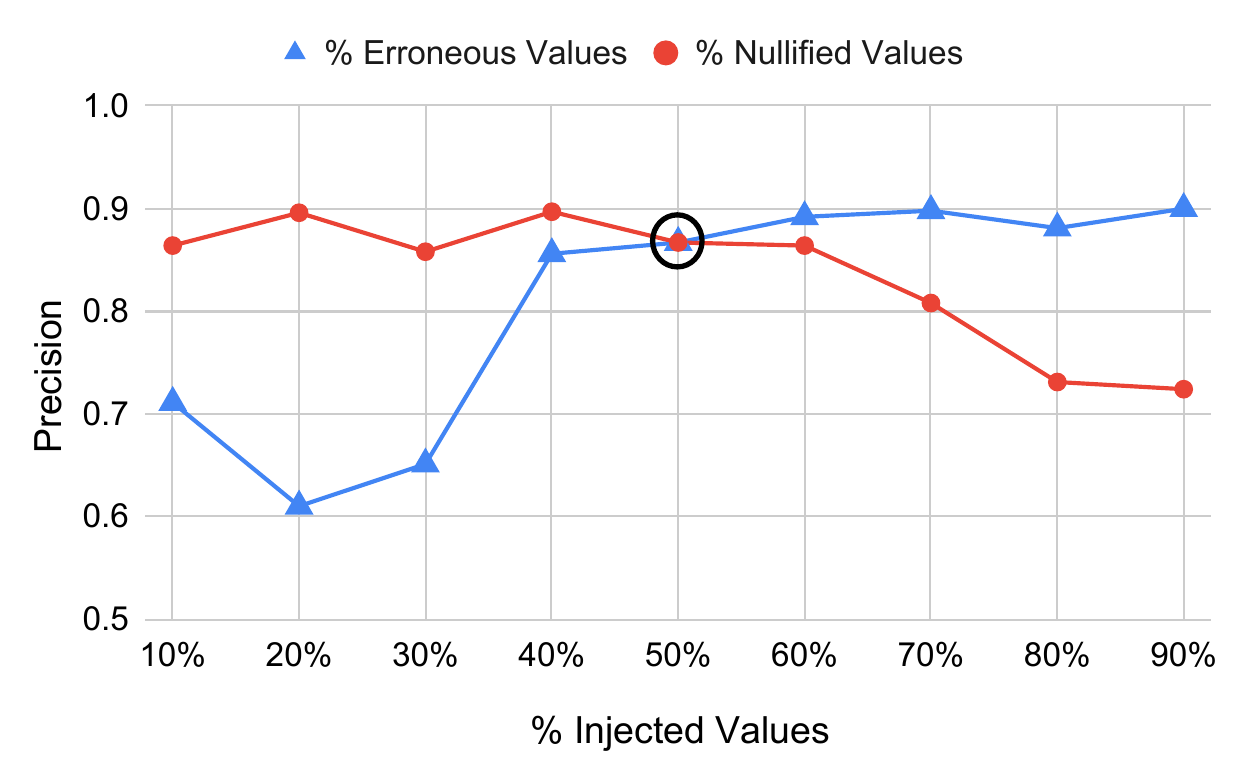}
    \caption{\name Precision as \tpchMedBench has different \% of Erroneous %Values 
    (blue %line with 
    triangles) and Nullified Values (red %line with 
    circles).}
    \label{fig:ablation_err_nulls}
    \vspace{-.1in}
\end{figure}

\noindent\textbf{Tuning \% Erroneous vs. Nullified Values:} 
%Lastly, w
We further analyze \name's performance %when run
on data lake tables with different number of erroneous and nullified values in \tpchMedBench tables (Figure~\ref{fig:ablation_err_nulls}). So far, \tpchMedBench tables have 50\% erroneous values in erroneous versions and 50\% nulls in nullified versions (intersection point on the graph where \name has 0.867 Precision). Now, we tune the percentage of values replaced with non-null, random strings (blue line in Figure~\ref{fig:ablation_err_nulls}) in erroneous versions, while the nullified versions always contain 50\% nulls. Similarly, we tune the percentage of values replaced with nulls (red line in Figure~\ref{fig:ablation_err_nulls}) while holding the erroneous versions constant. For \name to produce a perfect reclamation of a Source Table, it should only have originating tables with injected nulls so that these nulls can be replaced with correct values during table integration.

As data lake tables have more %values replaced with 
erroneous values (blue line), \name is more likely to contain tables with nullified tuples in its set of originating tables, which results in an integrated table with higher precision. On the other hand, as we tune the percentage of values replaced with nulls (red line), precision decreases. As more nulls are injected, these tables also have fewer correct values. \name is thus more inclined to have originating tables with 50\% erroneous values, or 50\% correct values, leading to a final integration with lower precision.

\subsection{Scalability}\label{subsec:scalability}
\begin{figure}[t]
\centering
    % \vspace{-.12in}
    {
    \centering
    \begin{minipage}[t]{0.4\textwidth}
    \includegraphics[width=\linewidth]{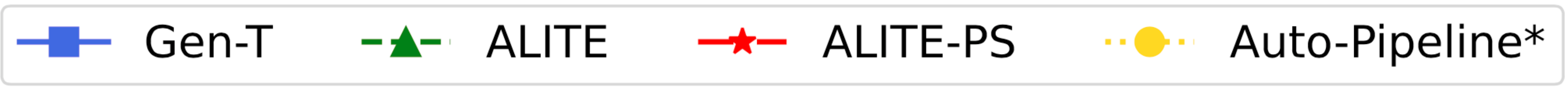}
    \end{minipage}
    }
    \subfloat[Average Runtimes in sec.]{
    \begin{minipage}[t]{0.48\linewidth}
    \includegraphics[width=\linewidth]{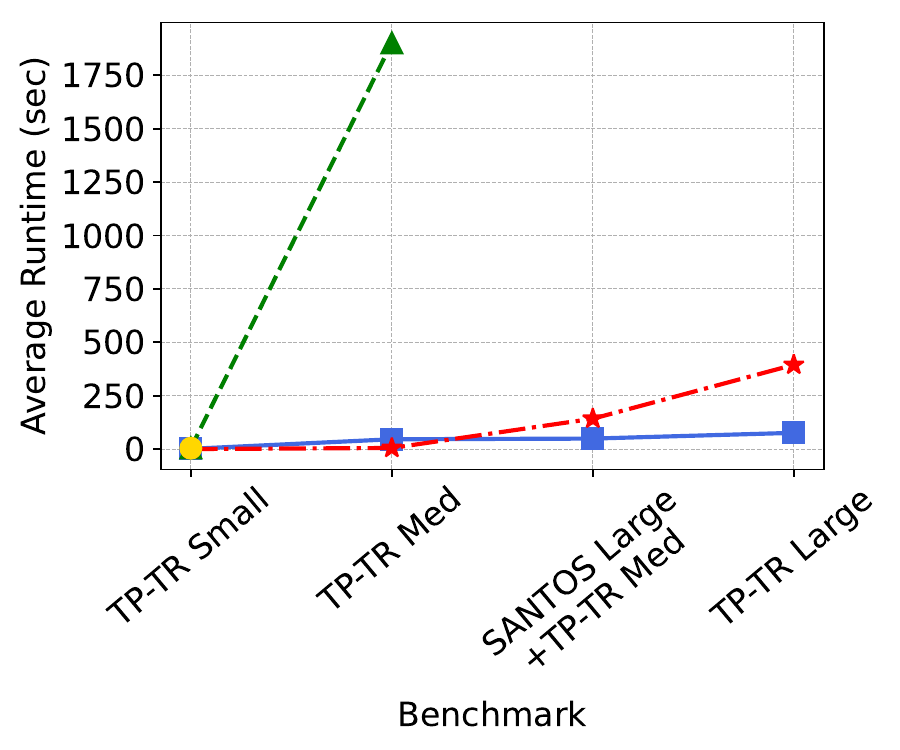}
    \end{minipage}
    }
    \subfloat[Average Ratio of Output Size
    % (total \# values) / Source Table Size
    % Average Output Sizes (total \# values), in tens of millions
    ]{
    \begin{minipage}[t]{0.48\linewidth}
    \includegraphics[width=\linewidth]{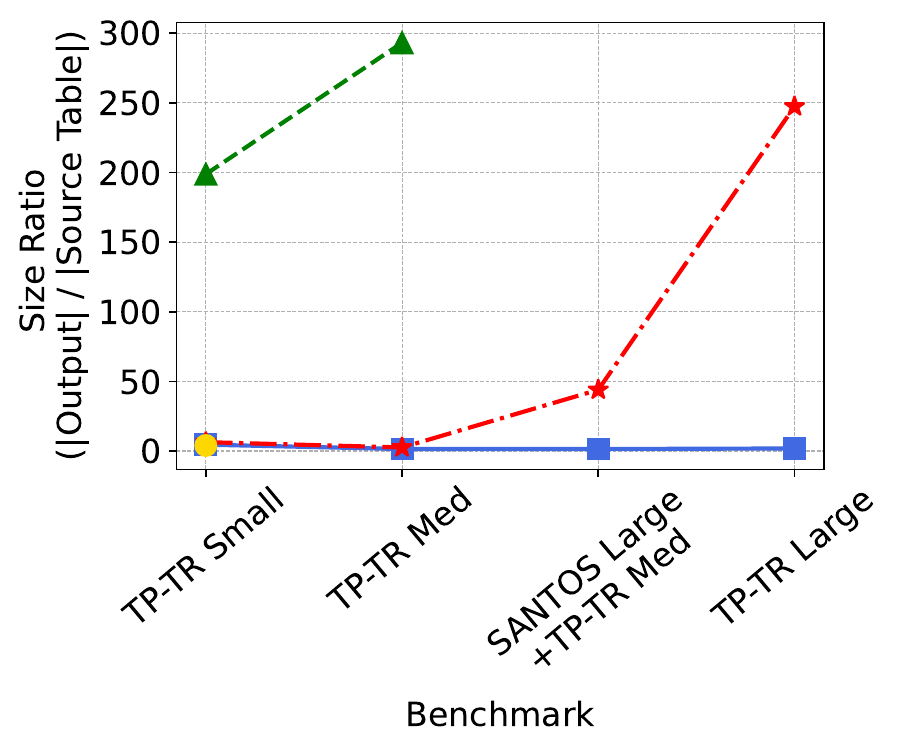}
    \end{minipage}
    }
    % \caption{Measuring Scalability via Average Runtime (sec) and Ratio of Output Sizes to Source Table sizes for all methods on all \tpch Benchmarks
    % }
    \caption{Scalability via Average Runtime (sec) and Ratio of Output Sizes to Source Table sizes over %for all methods on 
    \tpch benchmarks.
    }
    \label{fig:scal_plots}
    % \vspace{-.1in}
\end{figure}

Figure~\ref{fig:scal_plots} shows the scalability of \name, \alite, \aliteps, and \aPip across benchmarks as the number and/or size of tables grows. %, from the smallest \tpch benchmark (\tpchSmallBench), to \tpchMedBench tables embedded in a large data lake (\santosLargeBench). %In 
Figure~\ref{fig:scal_plots}(a) %, we 
reports %the 
average runtimes for all methods across all four benchmarks, starting from ingestion of the candidate tables. For \name, this time includes the time it takes to prune the set of candidate tables to a set of originating tables, and integrate it to produce a reclaimed table. For %the 
other %three 
methods, this time only includes integration. %time. 
We find that \aPip only runs on \tpchSmallBench without timing out, and \alite, which performs full disjunction, is exponential in time and times out for the last two benchmarks. 
% Following the timeouts reported by each baseline, we adopt the same ratio of reported timeout to the data lake size, taking the maximum timeout across ratios for each benchmark. 
% We set the timeouts for each method according to the details reported in respective papers and with respect to the data lake size. %, taking the maximum timeout across ratios for each benchmark. 
% For experiments on the \tpchSmallBench benchmark, we use a timeout of 30min. For experiments on \tpchMedBench and \santosLargeBench+\tpchMedBench benchmarks, we use a timeout of 7hrs, and for experiments on \tpchLargeBench benchmark, we use a timeout of 24hrs.
% We set the timeouts for each method according to the details reported in respective papers and with respect to the data lake size. 
We set the timeouts as 30min for \tpchSmallBench, 7hrs for \tpchMedBench and \santosLargeBench+\tpchMedBench and 24hrs for \tpchLargeBench.
% , shown in Table~\ref{tab:timeouts}. 
% The timeout reported for \tpchMedBench also applies to \santosLargeBench+\tpchMedBench. 

% \begin{table}[ht]
%     \small
%     \begin{tabular}{lccc}\\ \toprule
%     & \tpchSmallBench    & \tpchMedBench & \tpchLargeBench\\ \midrule%   & \tdBench  \\ \midrule
%     Timeouts & 30 min & 7 hrs & 24 hrs\\ \bottomrule% & 99 sec \\ \bottomrule
%     \end{tabular}
%     \caption{Timeouts for all methods on \tpch benchmarks.\roee{Trimming suggestion: I think we can blend this in the text.}}
%     \label{tab:timeouts}
% \end{table}

% We can see that 
\name
% , although having an overhead of matrix initialization and traversal,
has a more consistent runtime across all benchmarks compared to all baselines. 
% \name is 2X faster in runtime compared to \aPip on the \tpchSmallBench. 
% On \tpchMedBench, \name is 60X faster than \alite and on the \tpchLargeBench, \name is 7X faster than \aliteps.
\name is 3X faster compared to \aPip on \tpchSmallBench. 
On \tpchMedBench, \name is 40X faster than \alite and on \tpchLargeBench, \name is 5X faster than \aliteps. 
Thus, pruning candidate tables to originating tables seems to cut the cost of integration, a prevalent issue 
as shown by the baselines.

%In 
Figure~\ref{fig:scal_plots}(b) %, we 
reports the average output sizes, or number of cell values in the reclaimed tables, with respect to the average Source Table sizes. % for all methods across all benchmarks. 
As the number and size of tables grows across benchmarks, the output size relative to the size of the Source Table (expected output size) can easily grow at a fast rate if the integration is among more or larger tables, especially if it includes noisy tables from the real data lake (\santosLargeBench). 
% The output sizes for \name remain consistent across benchmarks, being 1.3-1.8X larger than the average Source Table's size. This trend largely accounts for the higher precision of \name, compared to the baselines, since \name produces tables that mostly consist of tuples from the Source Table. 
Output sizes for \name remain consistent across all benchmarks (1.4-4.5X larger than the average Source Table size). This trend largely accounts for the higher precision of \name since its output tables mostly consist of Source tuples. In contrast, output sizes for \alite, \aliteps, and \aPip are 200-300X, 2.5-250X, and 4X larger than Source Tables' sizes, respectively. % for benchmarks on which they do not time out.
Thus, \name's runtimes and output sizes remain consistent across benchmarks of different sizes.
% , even when the expected tables for integration are immersed in a large %, real 
%  data lake.

\subsection{Generalizability}\label{subsec:generalizability}
% \roee{Maybe even more explicitly refer to this case as a real-world scnerio where \textbf{we do not know whether and how} the tables were generated and we are trying to reclaim them using other existing tables?}\grace{resolved}
%In addition to the \tpch benchmarks, w
We also experiment with the \tdBench benchmark to see how well we can apply \name to a real-world scenario with Web Tables. %in which 
In this case, we do not know \textit{whether} or \textit{how} the tables were originally generated. Accordingly, we attempt to reclaim each table using a subset of other tables in the benchmark by iterating through each 515 tables as potential Source Tables. %, and run the methods. %with no groundtruth.
\name successfully reclaims 3 Source Tables from an integration of multiple tables (5-6 tables), such that the outputs have perfect Recall, Precision, Instance-Div., and near-perfect \dkl. %Meanwhile, 
\name also finds duplicate tables for 12 Source Tables, or 6 sets of duplicates. This indicates that we can apply \name in a different domain, even if no sources are known to be reclaimed, and retrieve successful reclamations. %The b
Baseline methods are able to reclaim 12-13 Source Tables, %all of 
which are included in the 15 Source Tables reclaimed by \name. 
% Specific results are given in a technical report~\cite{gen-t_technical_report}.

\begin{table}[ht]
\centering
\scalebox{0.8}{
\begin{tabular}{l|cc|cc}\hline
         Method & Recall & Precision & Inst-Div.& \dkl \\ \hline 
\alite & 0.956 & 0.490 & 0.009 & 0.627 \\
\aliteps & 0.956 & 0.796 & 0.009 & 0.627 \\ 
\aPip & 0.881 & 0.725 & 0.088 & 19.261 \\ \midrule
\name & 0.956 & 1.000 & 0.009 & 0.627\\ \bottomrule
\end{tabular}}
% \caption{33 Sources from \tdBench for which all methods have non-empty output tables, and a data lake with \tdBench tables immersed in the \wdc benchmark. }
\caption{Sources from \tdBench immersed in the \wdc %data lake 
for which all methods have non-empty outputs.}
\label{tab:td_wdc}
% \vspace{-.15in}
\end{table}

We then run experiments on a data lake consisting of both \tdBench and \wdc tables.
This way, we evaluate how well the methods perform when a set of candidate tables returned from Set Similarity may contain irrelevant or misleading tables from the \wdc benchmark. 
Table~\ref{tab:td_wdc} presents the similarity and divergence scores for all methods on 33 of the common sources from \tdBench for which all methods have non-empty, reasonably sized output tables. We can see that \name outperforms the baselines for all measures, even having a precision of 1.0. In contrast, the baseline methods that are given the candidate tables from Set Similarity integrate all candidate tables and produce tables that contain many additional tuples. 

% For 33 of the common sources for which all methods have non-empty, reasonably-sized output tables, \name outperforms the baselines, even having perfect precision of 1.0. In contrast, baselines that integrate all candidate tables produce tables that contain many additional tuples, leading to much lower precision scores (best-performing baseline has precision 0.796). 

%\section{Conclusion}\label{sec:conclusion}
\section{Conclusion and Open Problems}\label{sec:conclusion}

Table Reclamation 
is essential in verifying if a data lake supports the tuples (facts) in a Source table.
Our results show that, despite the large search space, \name can solve the reclamation problem efficiently for source tables with keys.
\revision{In future work, we will relax the key assumption with regard to source tables, and use a fast, approximate instance comparison algorithm to compare instances from a source table and data lake tables~\cite{Glavic2024}.
}
When a table can only be partially reclaimed, we plan to investigate whether the originating tables can be embedded in a new data lake and used to possibly generate a better reclamation.  Alternatively, we plan to investigate if reclamation can be combined with data cleaning (for example, value imputation over missing values or entity resolution) to produce a better reclamation.
%We will also look at combining reclamation with provenance enabling data scientists to ask provenance queries over erroneous or other source values.  This will require understanding provenance over queries containing subsumption and complementation operators.
 % Furthermore, we will look at combining reclamation with provenance enabling data scientists to ask provenance queries over erroneous or other source values.  This will require understanding provenance over queries containing subsumption and complementation operators.
\revision{
In addition, we plan to consider the case in which values from a source table do not syntactically align with values from a data lake, in which case we can explore the semantic similarity of instances.
} 
Table reclamation can also be used to verify the tabular results of generative AI or large language models.  Verifying the output of generative AI using a data management lens is an emerging and important area~\cite{DBLP:journals/corr/abs-2307-02796,DBLP:journals/pvldb/FernandezEFKT23}.  
For example, users who generate summary tables and charts (e.g. Microsoft Copilot~\cite{ms_copilot}) or presentation slides (e.g. SlidesAI~\cite{slidesai}) from input data would find it useful to verify model outputs and examine what data was used to generate them.
Our approach will advance this area by allowing the automatic verification of tables that are created through complex integrations of other tables, something not yet considered in the literature.  

\section*{Acknowledgements}
This work was supported in part by NSF award numbers IIS-2107248, IIS-1956096, and IIS-2325632.

\balance
\bibliographystyle{IEEEtran}
\bibliography{main} 
\pagebreak
\setcounter{page}{1}
\appendix

\subsection{Preliminaries: Table Operators}\label{app:ops_proof}
Suppose we have two tables, $T_1, T_2$, 
    % that both contain join column $a_j$, 
    that share common columns $C$,
    and are in their minimal forms in which there are no duplicates and no tuples that can be subsumed or complemented. 
    We show that for each pairwise table operator, Inner Union, Inner Join, Left Join, Outer Join, Cross Product, there exists an equivalent query consisting of Outer Union and/or unary operators.
    % that produce instance-equivalent tables 
    (SP of SPJU queries are accounted for by the unary operators).

    \begin{lemma}[Inner Union]\label{lemma-union-op}
    Inner Union($\cup$): it is known that if the schemas of two tables are equal, then Inner Union = Outer Union
    \end{lemma}
    \begin{lemma}[Inner Join]\label{lemma-innerjoin-op}
    Inner Join ($\bowtie$):
    \begin{equation}\label{eq:inner_join}
        T_1 \bowtie T_2 = \sigma(T_1.C = T_2.C \neq \bot, \beta(\kappa(T_1 \uplus T_2)))
    \end{equation}
    \end{lemma}
    \begin{lemma}[Left Join]\label{lemma-leftjoin-op}
    Left Join ($\leftouterjoin$)~\cite{DBLP:conf/sigmod/Galindo-Legaria94}:
    \begin{equation}
        T_1 \leftouterjoin T_2 = \beta((T_1 \bowtie T_2) \uplus T_1)
    \end{equation}
    \end{lemma}
    \begin{lemma}[Outer Join]\label{lemma-outerjoin-op}
    Full Outer Join ($\fullouterjoin$)~\cite{DBLP:conf/sigmod/Galindo-Legaria94}:
    \begin{equation}\label{eq:full_outer_join}
        T_1 \fullouterjoin T_2 = \beta(\beta((T_1 \bowtie T_2) \uplus T_1) \uplus T_2)
    \end{equation}
    \end{lemma}
    \begin{lemma}[Cross Product]\label{lemma-crossproduct-op}
    Cross Product($\times$): We denote columns in $T_1$ and $T_2$ as $T_1.C$ and $T_2.C$, respectively. Consider a constant column $c$.
    \begin{equation}
        T_1 \times T_2 = \kappa(\pi((T_1.C, c), T_1) \uplus \pi((T_2.C, c), T_2))
    \end{equation}
    \end{lemma}
    
    Thus, $\uplus,\sigma, \pi, \kappa, \beta$ operators
    form queries that are equivalent to all SPJU queries.

\subsubsection{Proof of Lemma~\ref{lemma-innerjoin-op}[Inner Join]}
Given two tables $T_1, T_2$ that join on a set of common columns $C$, such that $T_1, T_2$ are in their minimal forms in which they contain no duplicate tuples and no tuples can be subsumed or complemented , $T_1 \bowtie T_2$ can be expressed by an equivalent query containing Outer Union, complementation, and subsumption. Specifically, $T_1 \bowtie T_2$ is equivalent to query $\sigma(T_1.C = T_2.C \neq \bot, \beta(\kappa(T_1 \uplus T_2)))$.

\begin{IEEEproof}
    We first prove that all tuples in $T_1 \bowtie T_2$ are contained in $\sigma(T_1.C = T_2.C \neq \bot, \beta(\kappa(T_1 \uplus T_2)))$.
    Let tuple $t \in T_1 \bowtie T_2$, such that join columns $C$'s values in $t$ appear in both $T_1.C$ and $T_2.C$, and are non-null: $t.C \in T_1.C \cap T_2.C$ s.t. $t.C \neq \bot$.

    % \roee{Can we formalize the following textual description?} 
    When applying $\beta(\kappa(T_1 \uplus T_2))$, only tuples with common non-null values $T_1.C_i = T_2.C_i \neq \bot$ in same column(s) $i$ are complemented and subsumed. This is similar to tuple $t$, which is formed by joining on $T_1.C_i = T_2.C_i$. Thus, tuple $t$ is derived by selecting on tuples from  $\beta(\kappa(T_1 \uplus T_2))$ with non-null $C$ values in both $T_1.C$ and $T_2.C$, so $t \in \sigma(T_1.C = T_2.C \neq \bot, \beta(\kappa(T_1 \uplus T_2)))$.

    Next, we show that all tuples in $\sigma(T_1.C = T_2.C \neq \bot, \beta(\kappa(T_1 \uplus T_2)))$ are found in $T_1 \bowtie T_2$. Let tuple $t' \in \sigma(T_1.C = T_2.C \neq \bot, \beta(\kappa(T_1 \uplus T_2)))$. Here, all $C$ values in $t'$ are non-null values found in both $T_1.C$ and $T_2.C$ as a result of selection. From $\beta(\kappa(T_1 \uplus T_2))$, $t'$ contains all values from all columns in $T_1$ and $T_2$ in a single tuple, formed by complementing and subsuming based on common $C$ values. Thus, $t' \in T_1 \bowtie T_2$.

    We have thus shown that all tuples from $T_1 \bowtie T_2$ are found in $\sigma(T_1.C = T_2.C \neq \bot, \beta(\kappa(T_1 \uplus T_2)))$ and vice versa, and so $T_1 \bowtie T_2$ is an equivalent query to $\sigma(T_1.C = T_2.C \neq \bot, \beta(\kappa(T_1 \uplus T_2)))$.
\end{IEEEproof}

\subsubsection{Proof of Lemma~\ref{lemma-leftjoin-op}[Left Join]}
Given two tables $T_1, T_2$ that join on a set of common columns $C$%\roee{can we prove if for a set of join attributes? Also, I think that by our new definition, we should refer to this set as implicit rather than explicit.}\grace{updated}
, such that $T_1, T_2$ are in their minimal forms in which there are no duplicates and no tuples can be subsumed or complemented %\roee{Can't we just say that if there are we simply apply complementation/subsumption? I think this would make the IEEEproof more complete.}\grace{updated}
, $T_1 \leftouterjoin T_2$ can be expressed by an equivalent query containing Outer Union and subsumption. Specifically, $T_1 \leftouterjoin T_2$ is equivalent to query $\beta((T_1 \bowtie T_2) \uplus T_1)$.

\begin{IEEEproof}
    We first prove that the resulting table of $T_1 \leftouterjoin T_2$ is contained in the resulting table of $\beta((T_1 \bowtie T_2) \uplus T_1)$: 

    %\roee{May be try to make this more formal, e.g., let $t\in T_1 \leftouterjoin T_2$... then use $t$ in the two scenarios you described, i.e., $t\in T_1$ and $t\in T_2$ (or $t\in T_1\cap T_2$) then $t\in (T_1 \bowtie T_2)$} \grace{updated}
    Let tuple $t \in T_1 \leftouterjoin T_2$. There are two cases for join column $C$'s values in tuple $t$: $t.C \in T_1.C \cap T_2.C$ (i.e., $t.C$ values are in both $T_1.C$ and in $T_2.C$) 
    % \grace{Should the notation be $t.C$ or $t[C]$?} \roee{I prefer $t[C]$, but I think $t.C$ works as well} 
    and $t.C \in T_1.C \setminus T_2.C$ (i.e., $t.C$ values are only in $T_1.C$ and not in $T_2.C$). Since we are performing left join on $T_1$ and $T_2$, $t.C \notin T_2.C \setminus T_1.C$. %\roee{What about the case where $t \in T_2 \setminus T_1$, since it is a left join this is not possible? Maybe say that explicitly?}\grace{resolved}
    \begin{enumerate}
        \item $t.C\in T_1.C\cap T_2.C \implies t\in(T_1 \bowtie T_2)$. Since $t$ is in the inner join result and contains more non-Null values than other tuples with $C$ values only in $T_1$ or $T_2$, it would not be subsumed when applying $\beta((T_1 \bowtie T_2) \uplus T_1)$.
        \item 
        $t.C \in (T_1.C \setminus T_2.C) \implies t \in \beta((T_1 \bowtie T_2) \uplus T_1)$. Since $T_1$ is in its minimal form, and $t$ does not share any $C$ values with any tuple in $T_2$, it is not subsumed when applying $\beta$ to $(T_1 \bowtie T_2) \uplus T_2$, and thus appear as is in $\beta((T_1 \bowtie T_2) \uplus T_1)$.
    \end{enumerate}
    Thus, all tuples from $T_1 \leftouterjoin T_2$ are contained in the resulting table of $\beta((T_1 \bowtie T_2) \uplus T_1)$.

    Next, we show that the resulting tuples of $\beta((T_1 \bowtie T_2) \uplus T_1)$ are contained in the resulting table of $T_1 \leftouterjoin T_2$.

    % \roee{for the other direction as well, try to make it more formal} \grace{updated}
    Let's consider tuple $t' \in \beta((T_1 \bowtie T_2) \uplus T_1)$. There are two cases for $C$ values in tuple $t'$: $t'.C \in T_1.C \cap T_2.C$ and $t'.C \notin T_1.C \cap T_2.C$.
    
    \begin{enumerate}
        \item $t'.C \in (T_1.C \cap T_2.C) \implies t' \in (T_1 \bowtie T_2)$. Since $(T_1 \bowtie T_2) \subseteq
        (T_1 \leftouterjoin T_2)$, $t' \in (T_1 \leftouterjoin T_2)$.
        \item All tuples in $((T_1 \bowtie T_2) \uplus T_1)$ are either subsumed by tuples from $(T_1 \bowtie T_2)$, or are in $T_1 \setminus (T_1 \bowtie T_2)$. Thus, $t'.C \notin T_1.C \cap T_2.C \implies t' \in T_1 \setminus (T_1 \bowtie T_2) \implies t' \in T_1 \leftouterjoin T_2$.
    \end{enumerate}
    Thus, all tuples from $\beta((T_1 \bowtie T_2) \uplus T_1)$ are contained in the resulting table of $T_1 \leftouterjoin T_2$.

    Now that we have shown that tuples from $T_1 \leftouterjoin T_2$ are contained in the resulting table of $\beta((T_1 \bowtie T_2) \uplus T_1)$ and vice versa, we have shown that $\beta((T_1 \bowtie T_2) \uplus T_1)$ is an equivalent query to $T_1 \leftouterjoin T_2$.
    \end{IEEEproof}

    \subsubsection{Proof of Lemma~\ref{lemma-outerjoin-op}[Outer Join]}
    Given two tables $T_1, T_2$ that join on a set of common columns $C$, such that $T_1, T_2$ are in their minimal forms in which there are no duplicates and no tuples can be subsumed or complemented, $T_1 \fullouterjoin T_2$ can be expressed by an equivalent query containing Outer Union and subsumption. Specifically, $T_1 \fullouterjoin T_2$ is equivalent to query $\beta(\beta((T_1 \bowtie T_2) \uplus T_1) \uplus T_2)$.

    \begin{IEEEproof}
    We first prove that the resulting table of $T_1 \fullouterjoin T_2$ is contained in the resulting table of $\beta(\beta((T_1 \bowtie T_2) \uplus T_1) \uplus T_2)$: 

    Let tuple $t \in T_1 \fullouterjoin T_2$. There are three cases for join column $C$'s values in tuple $t$: $t.C \in T_1.C \cap T_2.C$ (i.e., $t.C$ values are in both $T_1.C$ and in $T_2.C$), $t.C \in T_1.C \setminus T_2.C$ (i.e., $t.C$ values are only in $T_1.C$ and not in $T_2.C$), and $t.C \in T_2.C \setminus T_1.C$ (i.e., $t.C$ values are only in $T_2.C$ and not in $T_1.C$). 
    \begin{enumerate}
        \item $t.C \in T_1.C \cap T_2.C \implies t \in T_1 \bowtie T_2$. Tuple $t$ is a result of inner joining two tuples from $T_1, T_2$ on shared values in common columns $C$. This is similar to taking $T_1 \uplus T_2$, and applying subsumption and complementation on tuples with shared values in $C$ (Lemma~\ref{lemma-innerjoin-op}) to get $t$. Since $t$ does not share any values in $C$ with other tuples, it cannot be subsumed. Thus, $t \in \beta(\beta((T_1 \bowtie T_2) \uplus T_1) \uplus T_2)$.
        \item $t.C \in T_1.C \setminus T_2.C \implies t \in (T_1 \leftouterjoin T_2) \setminus (T_1 \bowtie T_2)$. When we take $(T_1 \bowtie T_1) \uplus T_1$, we append all tuples from $T_1$ to $T_1 \bowtie T_2$. After applying subsumption, all tuples from $T_1$ that are used in $T_1 \bowtie T_2$ are subsumed by tuples from $T_1 \bowtie T_2$ on shared values in $C$. Thus, the only tuples remaining are tuples like $t$ in $(T_1 \leftouterjoin T_2) \setminus (T_1 \bowtie T_2)$. Since $t$ does not share any common values with any tuple in $T_2$, it is not subsumed when taking $\beta(\beta((T_1 \bowtie T_2) \uplus T_1) \uplus T_2)$, and so $t \in \beta(\beta((T_1 \bowtie T_2) \uplus T_1) \uplus T_2)$.

        \item $t.C \in T_2.C \setminus T_1.C \implies t \in (T_2 \leftouterjoin T_1) \setminus (T_1 \bowtie T_2)$. Taking the subsumption of $\beta((T_1 \bowtie T_2) \uplus T_1) \uplus T_2$ removes all tuples from $T_2$ that are subsumed by tuples in $T_1 \bowtie T_2$. Since the remaining tuples in $T_2$ cannot be subsumed by any tuple from $T_1$ not in $T_1 \bowtie T_2$, $t \in (T_2 \leftouterjoin T_1) \setminus (T_1 \bowtie T_2)$. Thus, $t \in \beta(\beta((T_1 \bowtie T_2) \uplus T_1) \uplus T_2)$.
    \end{enumerate}

    Thus, all tuples from $T_1 \fullouterjoin T_2$ are contained in the resulting table of $\beta(\beta((T_1 \bowtie T_2) \uplus T_1) \uplus T_2)$.

    Next, we show that all tuples in $\beta(\beta((T_1 \bowtie T_2) \uplus T_1) \uplus T_2)$ are contained in the resulting table of $T_1 \fullouterjoin T_2$. Let's consider tuple $t' \in \beta(\beta((T_1 \bowtie T_2) \uplus T_1) \uplus T_2)$. There are two cases for $C$ values in tuple $t'$: $t'.C \in T_1.C \cap T_2.C$ and $t'.C \notin T_1.C \cap T_2.C$.
    
    \begin{enumerate}
        \item $t'.C \in (T_1.C \cap T_2.C) \implies t' \in (T_1 \bowtie T_2)$. Since $(T_1 \bowtie T_2) \subseteq
        (T_1 \fullouterjoin T_2)$, $t' \in (T_1 \fullouterjoin T_2)$.
        \item All tuples in $((T_1 \bowtie T_2) \uplus T_1) \uplus T_2$ are either subsumed by tuples from $(T_1 \bowtie T_2)$, are in $T_1 \setminus (T_1 \bowtie T_2)$, or are in $T_2 \setminus (T_1 \bowtie T_2)$. Thus, $t'.C \notin T_1.C \cap T_2.C \implies t' \in (T_1 \uplus T_2) \setminus (T_1 \bowtie T_2) \implies t' \in T_1 \fullouterjoin T_2$.
r    \end{enumerate}
    Thus, all tuples from $\beta(\beta((T_1 \bowtie T_2) \uplus T_1) \uplus T_2)$ are contained in the resulting table of $T_1 \fullouterjoin T_2$.

    Now that we have shown that tuples from $T_1 \fullouterjoin T_2$ are contained in the resulting table of $\beta(\beta((T_1 \bowtie T_2) \uplus T_1) \uplus T_2)$ and vice versa, we have shown that $\beta(\beta((T_1 \bowtie T_2) \uplus T_1) \uplus T_2)$ is an equivalent query to $T_1 \fullouterjoin T_2$.

    \end{IEEEproof}

    \subsubsection{Proof of Lemma~\ref{lemma-crossproduct-op}[Cross Product]}
    
    Given two tables $T_1, T_2$, each with columns $C_{T_1}, C_{T_2}$ respectively and do not share any columns, and a constant column $c$, $T_1 \times T_2$ can be expressed by an equivalent query containing Outer Union, projection, and complementation. Specifically, $T_1 \times T_2$ is equivalent to query $\kappa(\pi((C_{T_1}, c), T_1) \uplus \pi((C_{T_2}, c), T_2))$.
    
    \begin{IEEEproof}
    Since $T_1$ and $T_2$ do not share any columns, the complementation operator cannot be applied to $T_1 \uplus T_2$. Thus, we project on all columns $C_{T_1}$ and constant column $c$ in $T_1$, and columns $C_{T_2}, c$ in $T_2$. This way, $T_1, T_2$ now share all values in $c$ and we can apply complementation on $\pi((C_{T_1}, c), T_1) \uplus \pi((C_{T_2}, c), T_2)$ since $T_1, T_2$. Thus, we iteratively apply complementation on all tuples from $T_1$ on all tuples from $T_2$ to form all tuples in $T_1 \times T_2$. Recall that in every tuple in $T_1 \times T_2$, every value in $t.C_{T_1}$ is from $T_1$ and every value in $t.C_{T_2}$ is from $T_2$.  Therefore, every tuple in $T_1 \times T_2$ is contained in $\kappa(\pi((C_{T_1}, c), T_1) \uplus \pi((C_{T_2}, c), T_2))$ and every tuple in $\kappa(\pi((C_{T_1}, c), T_1) \uplus \pi((C_{T_2}, c), T_2))$ is contained in $T_1 \times T_2$, and so $\kappa(\pi((C_{T_1}, c), T_1) \uplus \pi((C_{T_2}, c), T_2))$ is an equivalent query to $T_1 \times T_2$.
    \end{IEEEproof}

\subsection{Set Similarity}
\begin{algorithm}[ht]
    \setstretch{0.85} % sets line height
    \small
    \SetAlgoLined
    \LinesNumbered
    \textbf{Input}: $\bigT= \{T_1, T_2, \dots T_n\}$: set of data lake tables; $S$: Source Table; $\tau$: Similarity Threshold\label{line:ss_input}\\
    \textbf{Output}: $\bigT' = \{T_1, T_2, \dots T_n\}$: a set of candidate tables with high syntactic overlap with $S$ \label{line:ss_output}\\ 
    $\bigT'_\text{scores} \gets \{\}$ \algocomment{Store a list of scores for each candidate table}\; %\emptyset
    \For {all $S$ columns $c \in C$}{   \label{line:ss_iterateCols}
        $\mathcal{T}_C$, overlapScores $\gets \mathsf{SetOverlap}(\bigT, c, \tau)$\; 
        $\mathcal{T}_C$, diverseOverlapScores $\gets \mathsf{diversifyCandidates}(\mathcal{T}_C, c, \tau)$   \label{line:ss_divCand}\;
		% $\bigT' \gets \bigT' \cup \mathcal{T}_C$;
        \For {all tables $T \in \mathcal{T}_C$}{  
        $\bigT'_\text{scores}[T] $+= diverseOverlapScores[$T$]\;
        }
	} \label{line:ss_endIerateCols}
    Order $\bigT'_\text{scores}$ by average diverseOverlapScores, in descending order\;\label{line:ss_averageScores}
    $\bigT' \gets \text{keys}(\bigT'_\text{scores})$\;
    \For {all tables $T \in \bigT'$}{  \label{line:ss_checkColType}
        alignedTuples $\gets$ tuples in $T$ that contain $S$'s column values\;
        \uIf{set overlap of $T$ values in alignedTuples with $S < \tau$}
        {
            Discard $T$;
        }\label{line:ss_checkAlignedTuples}
        Remove $T$ if its values are contained in another table $T' \in \bigT'$\label{line:ss_removeSubsumedTable}\;
            Rename $T$ columns to aligned $S$ columns\label{line:ss_renameCols}\;  
        }
    \Return $\bigT'$;
    \caption{Set Similarity}\label{alg:set_similarity}
\end{algorithm}

We find candidate tables with values that have high set overlap with those in a Source Table.
As shown in Algorithm~\ref{alg:set_similarity}, we perform Set Similarity with an input set of data lake tables $\bigT$, the Source Table $S$, and a similarity threshold $\tau$ (Line~\ref{line:ss_input}), and output a set of candidate tables (Line~\ref{line:ss_output}). We first find 
%rjm the 
a set of candidate tables, 
%that 
where each table contains a column whose set overlap with a column from $S$ (overlapScore) is above a specified threshold (Lines \ref{line:ss_iterateCols}-\ref{line:ss_endIerateCols}).  This can be done efficiently with a system like JOSIE~\cite{DBLP:conf/sigmod/ZhuDNM19} that computes exact set containment or MATE~\cite{DBLP:journals/pvldb/EsmailoghliQA22} that supports multi-attribute joins. 
In addition, when finding tables with columns that have a high set overlap with 
%rjm each implies all in this context
%each 
columns in $S$, we call $\mathsf{diversifyCandidates()}$ (Line~\ref{line:ss_divCand}) to ensure that each candidate table not only has a high overlap with $S$, but also has minimal overlap with the previous candidates, shown in Diversify Candidates Algorithm~\ref{alg:div_candidates}. 

Formally, given candidate table $T_i \in \bigT$ s.t. $i > 0$, the previous candidate table, $T_{i-1}$, and Source Table $S$, we diversify a set of candidate tables uses the following formula to rank the candidates, in descending order:
\begin{equation}
    \text{diverseOverlapScore} = \frac{|T_i \cap S|}{|S|} - \frac{|T_i \cap T_{i-1}|}{|T_i|} 
    \label{eq:diverseScore}
\end{equation}
When finding diverseOverlapScore, we find the set overlap of $T_i$ with $S$ vs.~the set overlap of $T_i$ with the previous candidate $T_{i-1}$. This way, we arrange the set of candidate tables to ensure diversification of candidates.

\begin{algorithm}[ht]
    \setstretch{0.85} % sets line height
    \small
    \SetAlgoLined
    \LinesNumbered
    \textbf{Input}: $c$: column from Source Table; $\mathcal{T}_C= \{T_1, T_2, \dots T_n\}$: set of candidate tables with columns having high overlap with $c$; $\tau$: Similarity Threshold \label{line:dc_input}\\
    \textbf{Output}: $\mathcal{T}_C'= \{T_1, T_2, \dots T_n\}$: a set of diverse candidate tables \label{line:dc_output}\\ 
    $\mathcal{T}_{\text{scores}} \gets \{\}$\;
    \For {all tables $T \in \mathcal{T}_C$}{   \label{line:dc_iterateTables}
        $C \gets $ column from $T$ with highest set overlap with $c$\;        Ind$_T$ $\gets$ index of $T$ in $\mathcal{T}_C$\; 
        \uIf{Ind$_T$ = 0}{  
			Continue;
		    }
        $C_{\text{prev}} \gets$  column from $\mathcal{T}_C$[Ind$_T$ - 1] with highest set overlap with $c$\algocomment{Get column from previous candidate table with high overlap with $c$}\;
        prevColOverlap $\gets (C \cap C_{\text{prev}}) / |C|$ \algocomment{Set overlap with previous column}\;
        sourceColOverlap $\gets (C \cap c) / |c|$ \algocomment{Set overlap with column from Source table}\;
        \uIf{sourceColOverlap $< \tau$}{  
            Continue;
        }
        overlapScore $\gets$ sourceColOverlap -- prevColOverlap \;
        $\mathcal{T}_{\text{scores}}[T] \gets $ overlapScore\;
    }
    Order $\mathcal{T}_{\text{scores}}$ by values in descending order\;    
    $\mathcal{T}_C' \gets \mathcal{T}_{\text{scores}}$.keys \;
    \Return $\mathcal{T}_C'$;
    \caption{Diversify Candidates}\label{alg:div_candidates}
\end{algorithm}

After we find candidate tables for each column in the Source Table, we average over all overlap scores such that each is for a Source Table's column with which they share many values, and rank them in descending order of averaged scores (Line~\ref{line:ss_averageScores}).
With a set of candidate tables, we find tuples in each candidate table that contain column values from $S$. Within these aligned tuples, we check if each aligned column in a candidate table, with respect to a column in $S$, still has high set overlap (above threshold $\tau$). If not, we remove them (Line~\ref{line:ss_checkAlignedTuples}). Next, we remove any subsumed candidate table, whose columns and column values are all contained in another candidate table (Line~\ref{line:ss_removeSubsumedTable}). We then
% we now check if each table has $S$'s primary key column, $\mathcal{C}_k$. If not, we find a foreign key path such that we can join a candidate table $T$ that does not have $\mathcal{C}_k$ with another table that has $\mathcal{C}_k$ and a foreign key path to $T$ (Line~\ref{line:ss_endJoin}). This way, the joined result has $\mathcal{C}_k$. If no foreign key path can be discovered for a candidate table, we discard it (Line~\ref{line:ss_endFKPath}).
rename each candidate tables' columns to the names of $S$'s columns with which they align (Line~\ref{line:ss_renameCols}), thus implicitly performing schema matching between $S$'s columns and the columns from the candidate tables that have overlapping values with $S$'s columns. 
Finally, we return the set of candidate tables.

\section{Matrix Representations}

\subsection{Expanding Candidate Tables}\label{app:expand}
In order to represent candidate tables as matrices, their tuples need to align with those in the Source Table. However, not all candidate tables may share a key column with the Source Table. Thus, we need to join a given candidate table that does not share a key column with the Source Table with those that do. This way, tuples from all candidate tables can be aligned with tuples from a Source Table using key values.

As illustrated in Expand Algorithm~\ref{alg:join_candidates}, we traverse a graph that consists of candidate tables as nodes and we find a join path between candidate tables that do not have a Source Table's key (start nodes), and candidate tables that do (potential end nodes). If two candidate tables can join on common columns, their nodes are connected with an edge. For each edge, we use standard join-size cardinality estimation to find edge weights~\cite{cowbook}.

After we find a path from a start node to an end node, we iteratively join all tables in the path, resulting in a table that shares a key column with the Source Table. This way, all candidate tables share a key column with the Source Table.
% After joining all tables in the path, we project out any non-key column that was not originally in the candidate table.

\begin{algorithm}
    \caption{Expand}\label{alg:join_candidates}
    \textbf{Input}: $\mathcal{T}= \{T_1, T_2, \dots T_n\}$: set of candidate tables; ${k}$: Source Table's key column(s)\\
    \textbf{Output}: $\mathcal{T}_k= \{T_1, T_2, \dots T_n\}$: set of candidate tables that all now contain Source Table's key column\\
   \algocomment{Initialize Graph}\\
    nodes = candidate tables\;
    edges = tables that have joinable columns\;
    edge Weights = value overlap of joinable columns\;
    $start\_nodes$ = \{candidate tables that do not contain $k$\}\;
    $end\_nodes$ = \{candidate tables that contain $k$\}\;
    \For{each $start$ in $start\_nodes$}{
        \algocomment{Initialize sets and dictionaries}\\
        $visited \gets $ set() \algocomment{visited nodes}\;
        $node\_weights \gets \{\}$ \algocomment{maximum weights before each node}\;
        $descendant \gets \{\}$ \algocomment{best child for each node}\;
        $max\_weight \gets 0$ \algocomment{Maximum weight found so far}\;
        $end\_node \gets $ None \algocomment{end node for a given start node}\;
    
        \algocomment{Initialize $stack$ for DFS}\;
        $stack \gets stack + start$\;
        $visited \gets visited + start$\;

        \While{$stack$ is not empty}{
            $node \gets stack.pop()$\;
            $unvisited\_children \gets$ children of $node$ not in $visited$\;
            \algocomment{Current child's weight is the weight of the path so far, including the edge weight between node and current child}\\
                \For{each $child$ in $unvisited\_children$}{
                    $child\_weight \gets node\_weights[node] + weight(node, child)$\;
                    \algocomment{update descendant if it contains the maximum sum of weights so far}\\
                    \If{$child\_weight > node\_weights[child]$}{
                        $node\_weights[child] \gets child\_weight$\;
                        $descendant[child] \gets node$\;
    
                        \If{$child$ is in $end\_nodes$}{
                            \If{$child\_weight > max\_weight$}{
                                \algocomment{child has $k$ and the maximum weighted path so far}\\
                                $max\_weight \gets child\_weight$\;
                                $end\_node \gets child$\;
                            }
                        }
                        
                        $stack \gets stack + child$\;
                    }
                    $visited \gets visited + child$\;
                    
                }
        }
        
        \If{$end\_node$ is not $null$}{
            \algocomment{reconstruct path with maximum sum of weights by reversing path, starting with found end node}\\
            $path \gets []$\;
            $current\_node \gets end\_node$\;
            \While{$current\_node$ is in $descendant$}{
                $path\gets path + current\_node$\;
                $current\_node \gets descendant[current\_node]$\;
            }
            $path.reverse()$ \;
            $table \gets$ first node in $path$\;
            \For{each $join\_table$ in $path[1:]$}{
                $table \gets join(table, join\_table)$\;
            }
            $\mathcal{T}_k \gets \mathcal{T}_k + table$\;
        }
    }
\end{algorithm}

\subsection{Two-Valued vs. Three-Valued Matrix Representations}
After aligning tuples in candidate tables to Source Table's tuples that share the same key values, we can represent candidate tables as matrices to show how similar the values in candidate tables are to those in the Source Table. These matrices have the same dimensions and indices as the Source Table. First, we consider matrices that consist of binary values, where a 0 in tuple $i$, column $j$ represents a value at index ($i, j$) in the candidate table that is different from the value in the same position in the Source Table, and a 1 represent common values in the same indices.

However, populating alignment matrices with binary values does not fully encode how many values in candidate tables \textit{differ} from those in the Source Table. Specifically, this representation does not distinguish between nullified values in candidate tables (null values in candidate tables at index ($i, j$) where there is a non-null value in the Source Table at the same position), and erroneous values (different non-null values in candidate tables at index ($i, j$) from those in Source Tables at the same position).

Instead, we encode matrix representations using three values, where at a given index in an aligned tuple, there is a 1 for a value shared between a candidate table and the Source Table, 0 for a null value in the candidate table where there is a non-null value in the Source Table, and -1 for a non-null value in the candidate table that differs from the value in the Source Table.

\subsection{Metrics}\label{app:metrics}
\noindent\textbf{Conditional KL-divergence:}
Given column $C$ shared between a Source Table and a 
%rjm resulting 
reclaimed
table $T$, suppose we have probability distributions, $\mathcal{P}$ for $C$ in the Source Table and $\mathcal{Q}$ for $C$ in the reclaimed table. We condition on the key values in key column $K$. The conditional KL-divergence (or conditional relative entropy) between $\mathcal{P}$ and $\mathcal{Q}$ of sample space $X$ of column $C$ conditioned on key $K$ is as follows:
\begin{equation}
    D_{KL}(Q || P) = -\sum_{x \in X, k \in K} P(x|k)\text{log}\left(\dfrac{Q(x|k)(1-Q(\neg x|k)}{P(x|k)}\right)
\end{equation}

% \renee{above I used $n$ as number of non-key attributes, if that is not what you need here, use $n+1$}

Given $n$ non-key columns $\mathcal{C}$ in a Source Table 
% \renee{source table is $S$, why are we calling it $\mathcal{C_S}$ now?} \grace{this is for columns - moved}
we take the average \dkl for each column divided by the probability of a key value in $T$ matching a key value from the Source Table ($Q(K)$) and the number of non-key columns ($n$). % found at 
%rjm - we haven't defined this column index 
% \rjm{key value} $j$ in the resulting table.
Then, the conditional KL-divergence of the reclaimed table is as follows:
\begin{equation}
    \resizebox{.9\hsize}{!}{
    $D_{KL}(T) = \dfrac{D_{KL}(Q_1 || P_1)+D_{KL}(Q_2 || P_2)+\dots+D_{KL}(Q_n || P_n)}{Q(K) * n}$
    }
\end{equation}
%\begin{equation}
%    D_{KL}(T) = \dfrac{\frac{D_{KL}(Q_1 || P_1)}{Q(K)}+\frac{D_{KL}(Q_2 || P_2)}{Q(K)}+\dots+\frac{D_{KL}(Q_n || P_n)}{Q(K)}}{|\mathcal{C_S}|}
%\end{equation}
%rjm changed denominator from $\/|matchcal{C_S}|$ to n
%\renee{what is $Q_j(K)$?  I'm not sure what $j$ is or what K is.}\grace{replaced $Q_j(K)$ with $Q(k)$ - referring to the probability that all key values from the Source Table (K) are found in the reclaimed table}
The conditional KL-divergence of the reclaimed table is a score $\in [0, \infty)$, with 0 being the ideal score. There is no upper limit on this metric since it naturally approaches $\infty$ when no key value from the Source Table is found in the reclaimed table.

\subsection{Effectiveness Experiments}
\noindent\textbf{Effectiveness of LLM Baseline:} 
To further assess the effectiveness of \name, we  compare it to OpenAI's ChatGPT3.5~\cite{chatgpt} as a representative Large Language Model (LLM). In our prompt for ChatGPT, we define the table reclamation problem and input a source table and a set of tables. We ask ChatGPT to return an integration result that reproduces the source table as closely as possible. Due to scalability constraints, we only use the integrating set as input. 

In our effectiveness experiments, we run ChatGPT on \tpchSmallBench benchmark, since it times out for all source tables in the larger benchmarks. We find that ChatGPT achieves a Recall of 0.239 and a precision of 0.256. For divergence measures, ChatGPT achieves an Instance-Divergence of 0.540 and a \dkl of 209.83. ChatGPT returns tables that only contain some source tuples, while containing many non-null erroneous values (high \dkl score).

\begin{figure*}[ht]
\centering
    {
    \centering
    \begin{minipage}[t]{0.25\linewidth}
    \includegraphics[width=\linewidth]{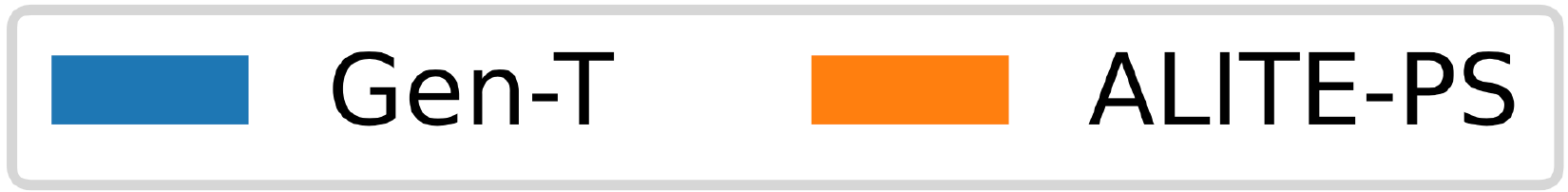}
    \end{minipage}
    }
    \subfloat[Recall]{
    \begin{minipage}[t]{0.33\linewidth}
    \includegraphics[width=\linewidth]{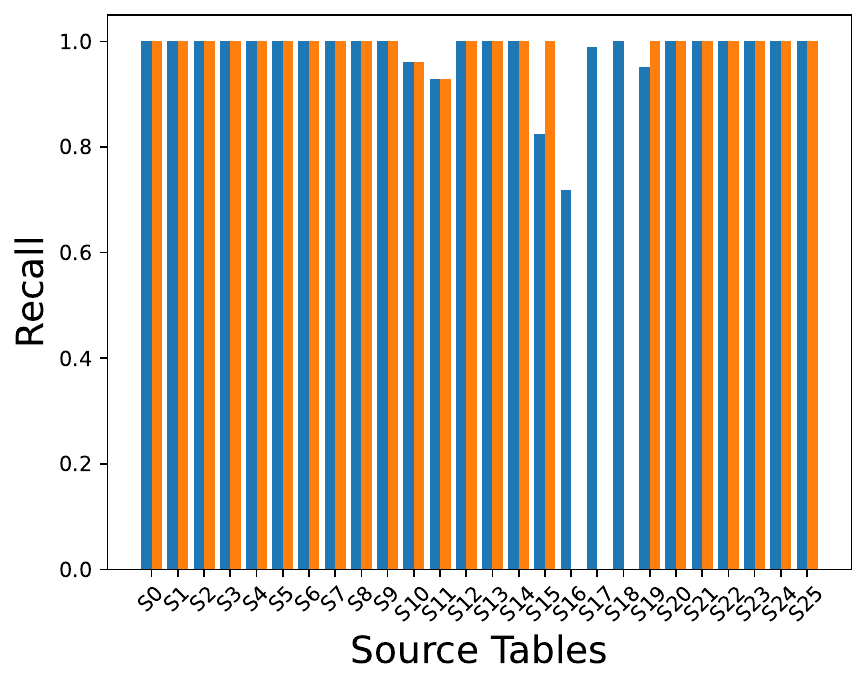}
    \end{minipage}
    }
    \subfloat[Precision]{
    \begin{minipage}[t]{0.33\linewidth}
    \includegraphics[width=\linewidth]{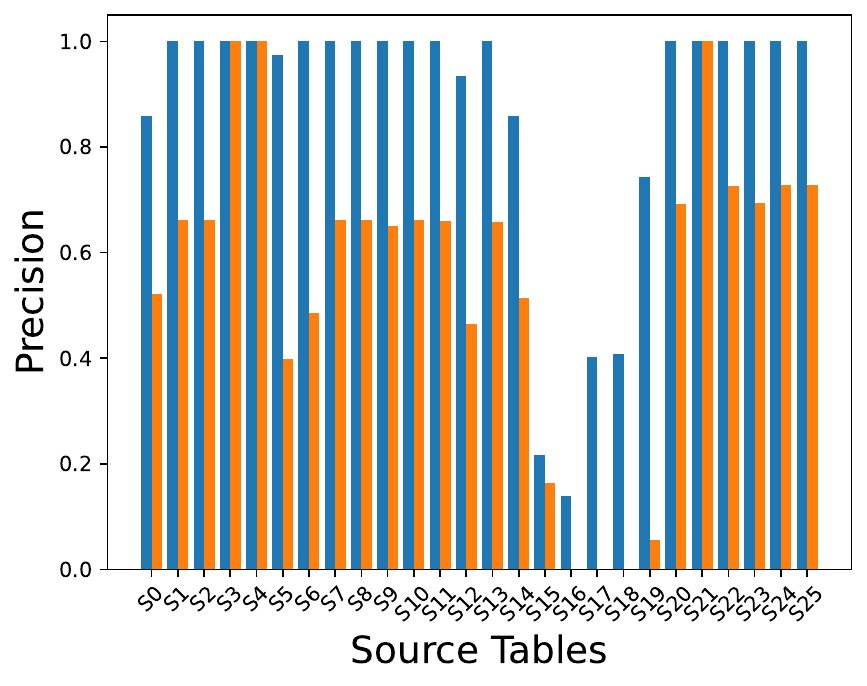}
    \end{minipage}
    }
    \subfloat[F1 Score]{
    \begin{minipage}[t]{0.33\linewidth}
    \includegraphics[width=\linewidth]{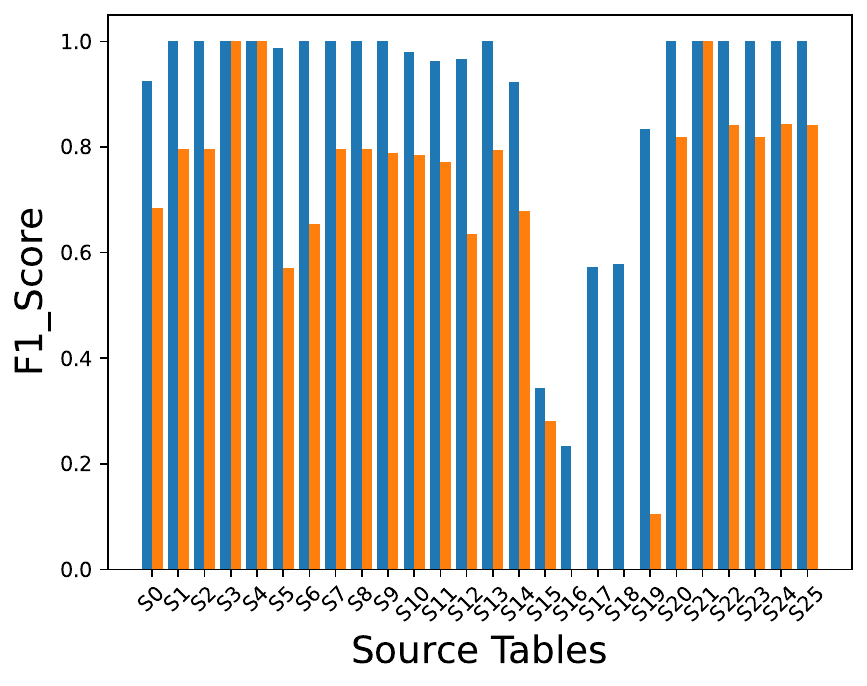}
    \end{minipage}
    }
    \caption{
    Recall, Precision, and F1 Scores of \name and \aliteps for each Source Table in \tpchMedBench benchmark.
    }
    \vspace{128in}
    \label{fig:ind_analysis}
\end{figure*}

\smallskip
\noindent\textbf{Effectiveness of Pruning in \name:} For a more detailed analysis, we analyze Recall, Precision, and F1-Scores of \name and baseline, \aliteps, on \tpchMedBench for each of its 26 Source Tables (Figure~\ref{fig:ind_analysis}). Note that \aliteps directly integrates a set of candidate tables, whereas \name first prunes the set of candidate tables to a set of originating tables before performing table integration. \name outperforms \aliteps in Precision for all Source Tables, and outperforms \aliteps in Recall for 24 of 26 total Source Tables. This shows that \aliteps, which directly integrates candidate tables without pruning, reclaims more Source Tuples than \name, which does include a pruning step, for only a few Source Tables. Also, \name outperforms \aliteps in F1-Score for all Source Tables (Figure~\ref{fig:ind_analysis}(c)), showing that even if \aliteps outperforms \name in Recall, it does not impact the F1-Score. 

\end{document}